\documentclass[final,5p,twocolumn]{elsarticle}

\usepackage{lineno,hyperref}
\usepackage{doi}
\modulolinenumbers[5]
\usepackage{framed}
\newcommand{\thedate}{\today}
\journal{}
\makeatletter
\def\ps@pprintTitle{%
 \let\@oddhead\@empty
 \let\@evenhead\@empty
 \def\@oddfoot{
 \begin{footnotesize}
\begin{tabular}{p{17.8cm}}
 ~ \\
\hline
 \multicolumn{1}{|p{17.8cm}|}{
 \footnotesize This version, created on \@date, is an electronic reprint of the original article published in the Applied Soft Computing Journal, 2017, Vol. 56, pp. 262--285,
 \href{http://dx.doi.org/10.1016/j.asoc.2017.02.024}{doi:10.1016/j.asoc.2017.02.024}. This reprint differs from the original in pagination and typographic detail.} \\
 \hline
\end{tabular}
\end{footnotesize}
 }%
 \let\@evenfoot\@oddfoot
 }
\makeatother

\usepackage{algorithmic}
\usepackage{amssymb}
\usepackage{color}
\usepackage{multirow}

\usepackage{hyperref}
\hypersetup{colorlinks,urlcolor=blue}

\newcommand{\labs}{\texttt{labs}}

\newcommand{\lMAts}{\texttt{lMAts}}
\newcommand{\lssMAts}{\texttt{lssMAts}}
\newcommand{\lssRRts}{\texttt{lssRRts}}
\newcommand{\lssOrel}{\texttt{lssOrel}}

\newcommand{\lssOrelU}{\texttt{lssOrel\_{U}}}
\newcommand{\lssOrelE}{\texttt{lssOrel\_{8}}}
\newcommand{\lssMAtsE}{\texttt{lssMAts\_{8}}}

\newcommand{\CALL}[2]{\texttt{{#1}(#2)}}
\newcommand{\PROCEDURE}[1]{%
	\renewcommand{\algorithmicwhile}{\textbf{procedure}}
	\renewcommand{\algorithmicdo}{ }
	\WHILE{#1}
	\renewcommand{\algorithmicwhile}{\textbf{while}}
	\renewcommand{\algorithmicdo}{\textbf{do}}	
}
\newcommand{\ENDPROCEDURE}{\ENDWHILE}

\definecolor{Gray}{rgb}{0.9,0.9,0.9}

\newcommand{\valueBest}{$\Theta({\underline{\varsigma}}^{*})$}

\newcommand{\valueTarget}{$\Theta^{ub}_{L}$}

\definecolor{shadecolor}{RGB}{220,220,220}

\newcommand{\OMIT}[1]{} 

\newcommand{\defeq}{\stackrel{\textup{\tiny def}}{=}}

\usepackage{lipsum} 
\usepackage{subfig}
\usepackage{graphicx}

\begin{document}

\begin{frontmatter}

\title{Low-Autocorrelation Binary Sequences:\\On Improved Merit Factors and Runtime Predictions to Achieve Them}


\author[feri]{Borko Bo\v{s}kovi\'{c}}
\ead{borko.boskovic@um.si}
\author[ncsu]{Franc~Brglez}
\ead{brglez@ncsu.edu}
\author[feri]{Janez Brest}
\ead{janez.brest@um.si}
\address[feri]{Faculty of Electrical Engineering and Computer Science, University of Maribor, SI-2000 Maribor, Slovenia}
\address[ncsu]{Computer Science, NC State University, Raleigh, NC 27695, USA}

%
%
\begin{abstract}
The
search for binary sequences with a high figure of merit, known as the low autocorrelation binary sequence (\labs) problem, represents a formidable 
computational challenge. To mitigate the computational constraints of the problem, we consider solvers that accept odd values of sequence length
$L$ and return solutions for skew-symmetric binary sequences only -- with the consequence that not all best solutions under this constraint will be
optimal for each $L$. In order to improve both, the search for best merit factor {\em and} \,the asymptotic runtime performance, we instrumented three
stochastic solvers, the first two are state-of-the-art solvers that rely on variants of memetic and tabu search (\lssMAts\ and \lssRRts), the third
solver (\lssOrel) organizes the search as a sequence of independent contiguous self-avoiding walk segments.
By adapting a rigorous statistical methodology to performance testing of all three combinatorial solvers, experiments show that the solver with the
best asymptotic average-case performance, $\lssOrelE = 0.000032*1.1504^L$, has the best chance of finding solutions that  improve, as $L$ increases,
figures of merit reported to date. The same 
methodology can be applied to engineering new \labs\ solvers  that 
may return merit factors
even closer to the conjectured asymptotic value of 12.3248.
\end{abstract}

\begin{keyword}
\texttt{Low-autocorrelation binary sequences}\sep \texttt{self-avoiding walk}\sep \texttt{stochastic combinatorial optimization} \sep \texttt{asymptotic average-case performance}
\end{keyword}

\end{frontmatter}

\section{Introduction}
\label{sec_introduction}
\noindent
The aperiodic low-autocorrelation binary sequence (\labs) problem 
has a simple formulation: take a binary sequence of length $L$,  
$S = s_1 s_2 \ldots s_L$, $s_i \in \{ +1,-1 \}$, the autocorrelation function
$C_k(S) = \sum_{i=1}^{L-k}s_{i}s_{i+k},$ and minimize the energy function:
\vspace*{-0.5ex}
\begin{equation}
E(S) =  \sum_{k=1}^{L-1}C_{k}^{2}(S)
\label{eq_energy}
\vspace*{-0.3ex}
\end{equation}
or alternatively, maximize the 
{\em merit factor F}~\cite{Lib-OPUS-labs-1977-IEEE_TIT-Golay-sieves,
Lib-OPUS-labs-1982-IEEE_TIT-Golay-merit-proof,
Lib-OPUS-labs-1990-IEEE_TIT-Golay-skewsym}:
\vspace*{-0.9ex}
\begin{equation}
F(S) =  \frac{L^2}{{2E(S)}}.
\label{eq_meritFactor}
\vspace*{-0.3ex}
\end{equation}

A binary sequence with the best merit factor has important applications in communication engineering.
These sequences are used for example as modulation pulses in radar and sonar ranging 
\cite{Lib-OPUS-labs-1972-Golay,Lib-OPUS-labs-1985-Phillips-Beenker,Lib-OPUS-labs-2000-Pasha}.
To physicists, the optimum solution of the \labs\ problem corresponds to the ground state of a generalized
one dimensional Ising spin system with long range 4-spin interactions
\cite{Lib-OPUS-labs-1987-JourPhys-Bernasconi},
also known as the Bernasconi model with aperiodic autocorrelation. The \labs\ problem also appears in mathematics  in terms of the
Littlewood problem~\cite{Lib-OPUS-labs-1968-littlewood,Lib-OPUS-labs-2002-Borwein}, in chemistry~\cite{Lib-OPUS-labs-1996-Stadler}, and in
cryptography~\cite{Lib-OPUS-labs-2016-Schmidt, Lib-OPUS-labs-2016-Packebusch}.
Solving the \labs\ problem is clearly important in a number of different areas. For more
details, we refer the reader to surveys~\cite{Lib-OPUS-labs-1994-Marinari,Lib-OPUS-labs-2004-Springer-Jedwab-survey}.

\begin{figure*}[t!]
\centering
\vspace*{-5ex}
\subfloat[]{
\begin{minipage}{0.47\textwidth}
\includegraphics[width=0.98\textwidth]{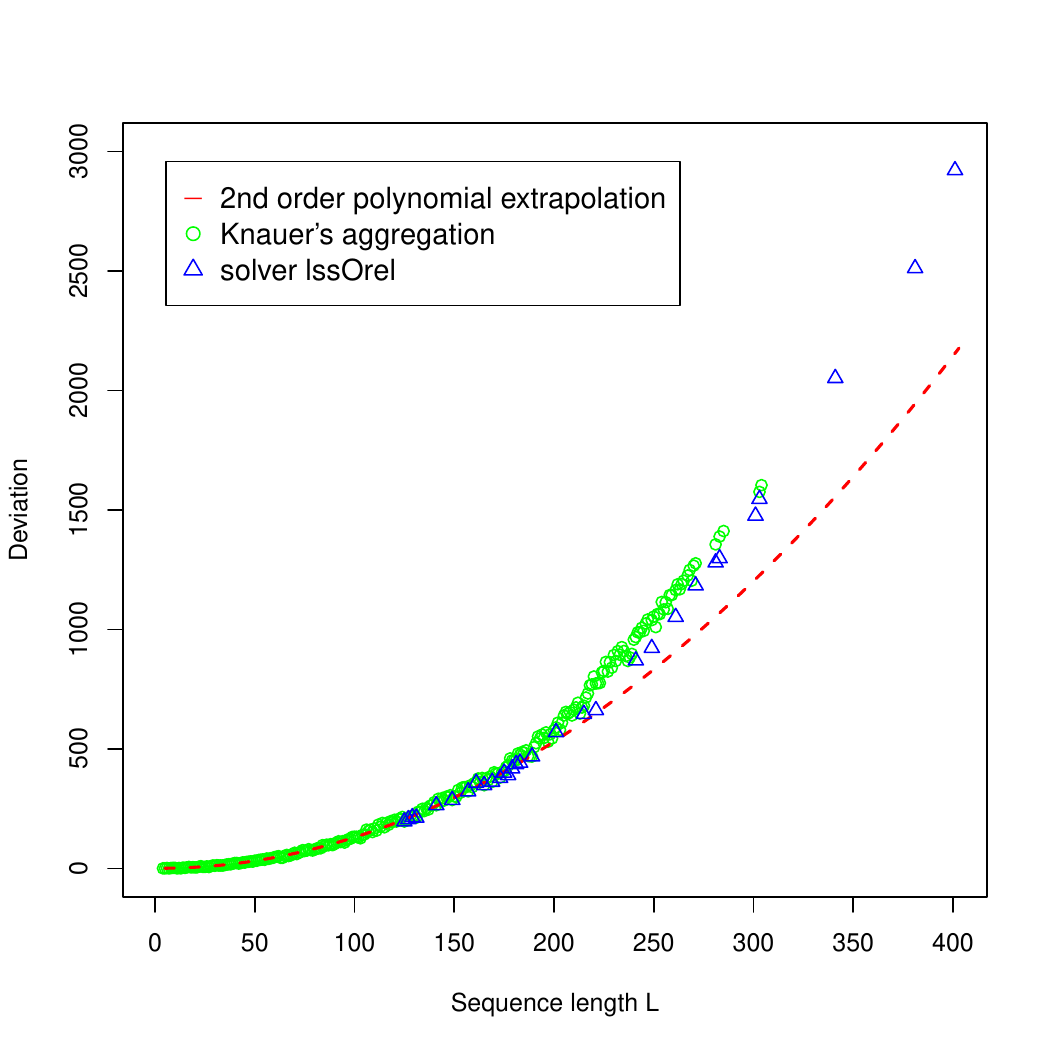}
\vspace*{-1ex}
\end{minipage}
}
\subfloat[]{
\begin{minipage}{0.47\textwidth}
\includegraphics[width=0.98\textwidth]{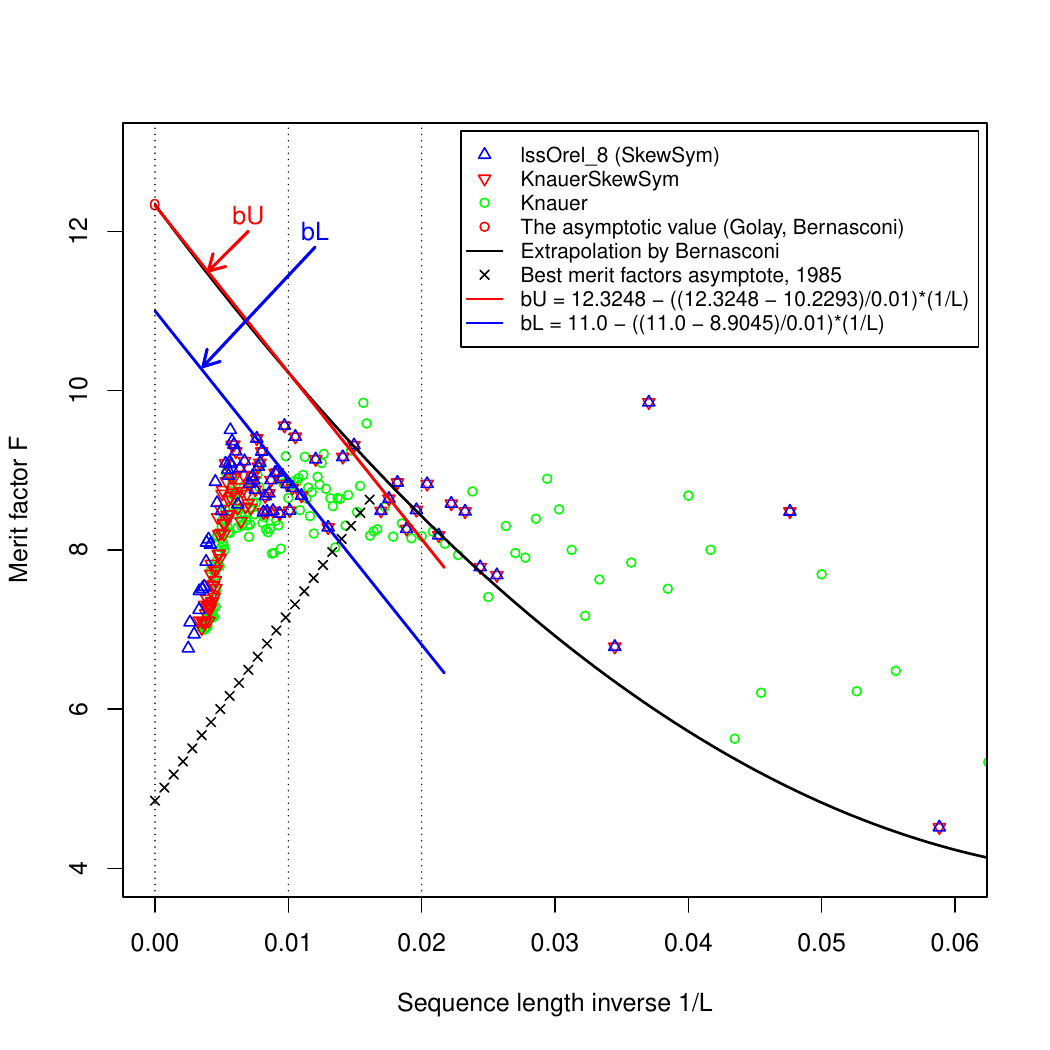}
\vspace*{-1ex}
\end{minipage}
}
\vspace*{-5ex}
\subfloat[]{
\begin{minipage}{0.47\textwidth}
\includegraphics[width=0.98\textwidth]{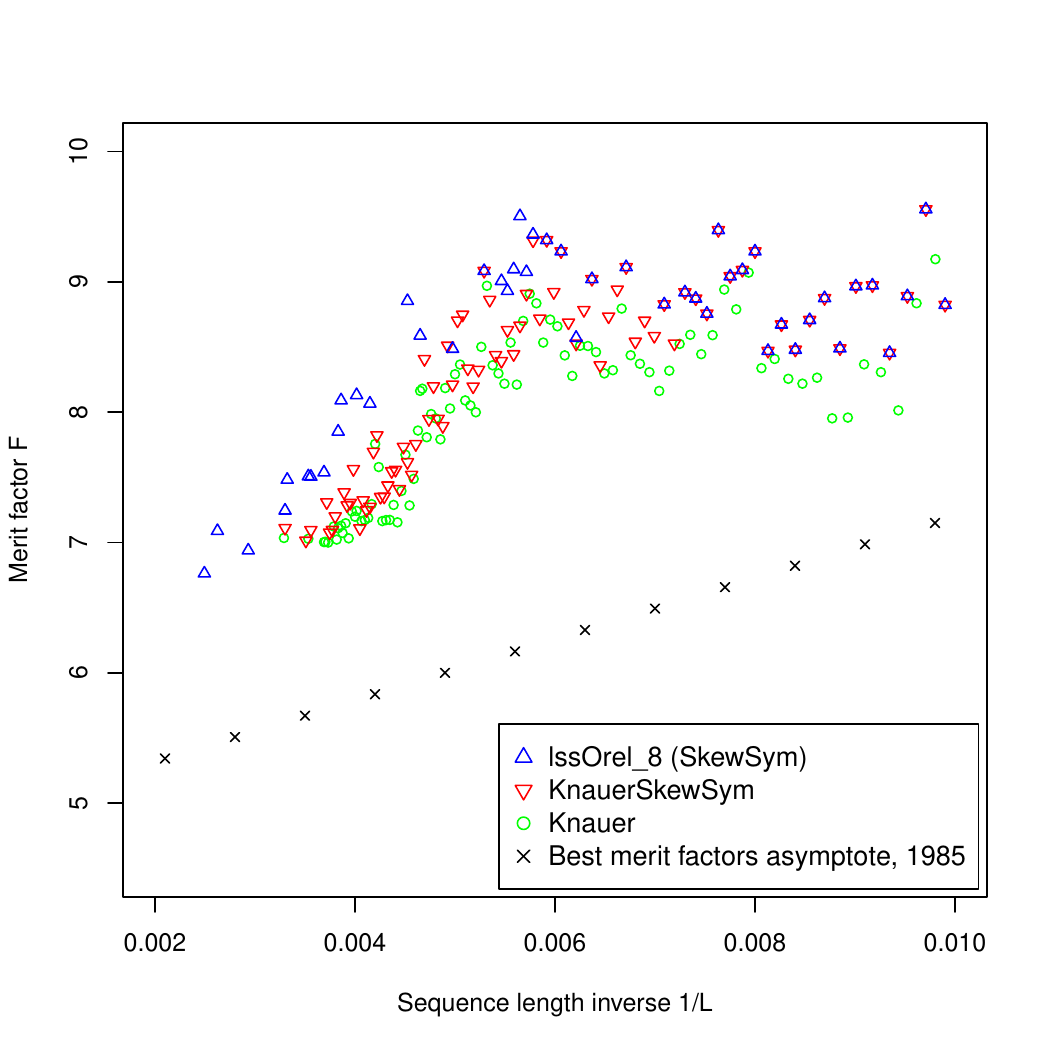}
\vspace*{-1ex}
\end{minipage}
}
\subfloat[]{
\begin{minipage}{0.47\textwidth}
\includegraphics[width=0.98\textwidth]{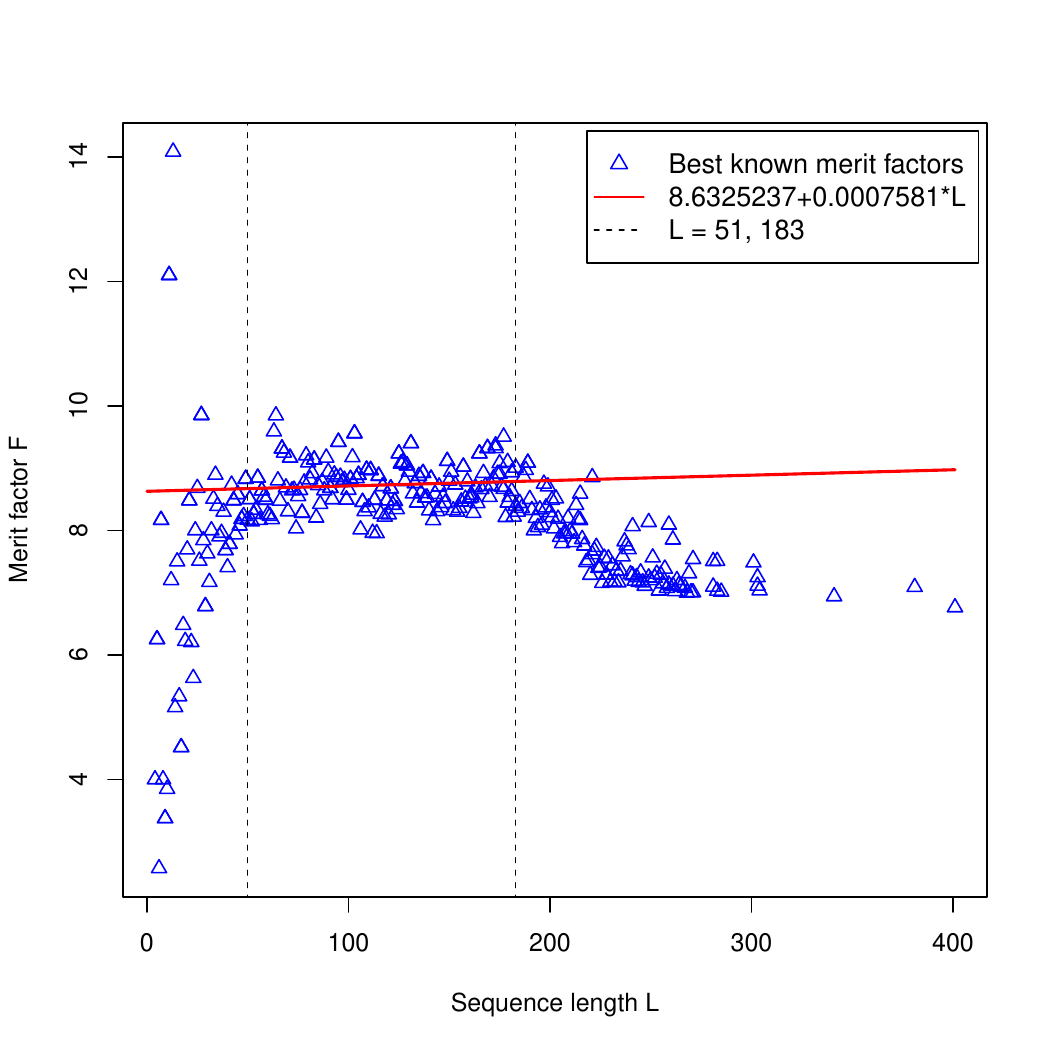}
\vspace*{-1ex}
\end{minipage}
}

\caption{
Asymptotic views of the \labs\ problem best merit factors, aggregated from results with several generations of \labs\ solvers: 
(a) 
A deviation-versus-length plot, introduced in~\cite{Lib-OPUS-labs-2010-Elsevier-Ukil-theory}, fits deviations from known optimal solutions  for $L \le 60$ to a second-order polynomial (the dotted line);
(b) A global view of the merit factor-versus-inverse sequence length, rigorously introduced in~\cite{Lib-OPUS-labs-1987-JourPhys-Bernasconi}, demonstrates that the conjectured asymptotic value of $F=12.3248$  may indeed be 
reachable with sufficient computational 
resources~\cite{Lib-OPUS-labs-1982-IEEE_TIT-Golay-merit-proof,Lib-OPUS-labs-1987-JourPhys-Bernasconi}, i.e. the improvements vis-a-vis the 
`Best merit factors asymptote, 1985'\cite{Lib-OPUS-labs-1985-Phillips-Beenker} are significant~\cite{BernasconiMail};
(c) The expanded view of experimental results, particularly 
for $L > 183$ (or $1/L < 0.00546$) suggests that better merit factors may be found not only by providing massively parallel computational resources
but also by continuing to improve the current generation of \labs\ 
solvers;
(d) 
A direct plot of merit factors $F$ versus sequence length $L$ illustrates that 
for $L \leq 50$ there are large and also {\em asymmetrical} variations in
values of merit factors, ranging from less than 4 to more than 14. 
For $L > 50$ and L $\leq$ 183, i.e. for sequences where \lssOrelE\ 
solutions reports an {\em observed number of hits} larger than 1 (Figure~\ref{fg_merit_factors_and_solvability}), merit factor variations are
not only significantly reduced, they also exhibit an {\em increasing trend}
and lead to a linear predictor model $8.6325237 + 0.0007571*L$.
We demonstrate that, given the state-of-the-art \labs\ solvers, the  downwards trend
in merit factors for $L > 183$ is due to insufficient application of computational resources -- sequences with better merit values than the ones shown here exist and are expected to be found in the future!
}
\label{fg_Deviation-vs-SequenceLength}
\end{figure*}

The asymptotic value for the maximum merit factor $F$, introduced in
\cite{Lib-OPUS-labs-1982-IEEE_TIT-Golay-merit-proof}, has been re-derived
using arguments from statistical mechanics \cite{Lib-OPUS-labs-1987-JourPhys-Bernasconi}:
\vspace*{-0.5ex}
\begin{equation}
{\rm as~~~} L \rightarrow \infty {\rm ,~~~then~~~}F  \rightarrow 12.3248
\label{eq_asymptotic}
\vspace*{-0.3ex}
\end{equation} 
The publication of the asymptotic value in Eq. \ref{eq_asymptotic}
is providing an on-going challenge since no published solutions can yet claim to converge to this value
as the length of the sequence increases. 

Finding the optimum sequence is significantly harder than solving the special cases of the Ising spin-glass problems
with limited interaction and periodic boundary conditions, for example
\cite{Lib-OPUS-labs-2003-GECCO-Goldberg-periodic}. 
While effective methods have been presented to solve the special cases up to $L = 400$ 
\cite{Lib-OPUS-labs-2003-GECCO-Goldberg-periodic}, 
the best merit factors that has also been {\em proven optimal} for the problem as formulated in Eq. \ref{eq_meritFactor}
are presently known for values of $L \le 60$ 
only~\cite{Lib-OPUS-labs-1996-JPhysA-Mertens-BB_solutions}.
A web page of \labs\ best merit factors and solutions, 
up to the sequence length of $L=304$, has been compiled by Joshua Knauer in 2002.
This page is no longer accessible and has now been restored 
under~\cite{Lib-OPUS2-git_labs-Boskovic} next to 
comprehensive tables of {\em best-value solutions}.
These tables contain not only updates on the best known figures of merit but also 
the number of {\em unique}  solutions in {\em canonic form} and the solutions themselves.
Relationships between results reported 
in~\cite{Lib-OPUS-labs-1982-IEEE_TIT-Golay-merit-proof,
Lib-OPUS-labs-1985-Phillips-Beenker,
Lib-OPUS-labs-1987-JourPhys-Bernasconi,
Lib-OPUS-labs-1990-IEEE_TIT-Golay-skewsym}, 
and all subsequent updates under~\cite{Lib-OPUS2-git_labs-Boskovic}
are depicted in four panels in Figure~\ref{fg_Deviation-vs-SequenceLength}.
The latest experimental results support the trend towards the conjectured asymptotic value of $F=12.3248$,
however as we demonstrate later on in the paper, the computational cost to reach this value may well exceed the currently available resources unless a better solver is discovered.

While branch and bound solvers,
for both even and odd sequences~\cite{Lib-OPUS-labs-1996-JPhysA-Mertens-BB_solutions} and for skew-symmetric
sequences only~\cite{Lib-OPUS-labs-2013-arxiv-Prestwich-branch-and-bound-odd} have been 
pursued, they do not scale as well as the stochastic solvers.
Stochastic solvers cannot prove optimality,
they can only be compared on the basis of the {\em best-value solutions}, and to a limited extent, also on
the average runtime needed to find such solutions under a sufficiently large number of repeated  
trials~\cite{Lib-OPUS-labs-1992-Optimization-deGroot, Lib-OPUS-labs-1993-Diploma-Reinholz-genetic-alg,
Lib-OPUS-labs-1996-PRL-Dittes,
Lib-OPUS-labs-1998-IEEE_EC-Militzer,
Lib-OPUS2-labs-2003-FEA-Brglez-short,
Lib-OPUS2-labs-2004-InfoSciences-Brglez-InPress,
Lib-OPUS-labs-2007-Prestwich-local-search,
Lib-OPUS-labs-2008-CP-Halim-tabu,
Lib-OPUS-labs-2008-LMSLNS-Borwein,
Lib-OPUS-labs-2009-ASC-Gallardo-memetic}.
The experimental results obtained with our stochastic solver \lssOrel\ are compared to
the instrumented versions of solvers 
in~\cite{Lib-OPUS-labs-2009-ASC-Gallardo-memetic}.\footnote{The generosity of authors to provide the source code
of  \lMAts, \lssMAts\ and its derivative, \lssRRts, is greatly appreciated.
The robust performance of these solvers
has revealed additional insights about the \labs\ problem.}




Compared to heuristic methods, the body of the theoretical literature 
on the merit factor problem for binary sequences is 
considerable~\cite{Lib-OPUS-labs-2004-Springer-Jedwab-survey}.
By 1983, it has been  established by computation (R. J. Turyn) and made plausible by the ergodicity postulate,
that long Legendre sequences offset by a quarter of their length have an asymptotic merit factor 
of 6~\cite{Lib-OPUS-labs-1983-IEEE_TIT-Golay-merit-legendre}. 
A rigorous proof for the asymptotic merit factor of 6, also based on
the quarter-rotated Legendre sequences, has followed in 
1991~\cite{Lib-OPUS-labs-1991-IEEE_TIT-Jensen-cyclic-difference_sets}.
The current records
for asymptotic merit factors, obtained by {\em various construction techniques}, stand at 
6.3421~\cite{Lib-OPUS-labs-2004-IEEE_TIT-Borwein,Lib-OPUS-labs-2004-IEEE_TIT-Parker-Legendre,Lib-OPUS-labs-2013-Elsevier-Jedwab-merit-factor,Lib-OPUS-labs-2013-AdvMath-Jedwab-Littlewood_Polynomials}
and 6.4382~\cite{Lib-OPUS-labs-2011-IEEE_TIT-Baden-Jacobi}.
Clearly, the challenge of
finding long binary sequences that would converge towards the asymptotic merit factor of $F = 12.3248$, as postulated 
in Eq.~\ref{eq_asymptotic}, remains open for experimentalists as well as for theoreticians.

The paper is organized as follows. Section~\ref{sec_notation}  introduces notation, definitions, and examples that motivate
the approach  taken in this paper. Section~\ref{sec_solvers}  highlights details about 
three \labs\  solvers as they are instrumented for comparative performance experiments to measure, in a platform-independent
manner, solver's asymptotic performance as the size of the \labs\  problem increases.
Section~\ref{sec_experiments}  summarizes results of extensive experiments with these solvers, including bounds and
projections for computational resources needed to increase the likelihood of finding better solutions of the \labs\ problem
for sizes $L > 141$. The section concludes with two views of  merit factors as $L$ increases towards the value of 5000. 
The first view depicts an asymptotic convergence of merit factor towards 6.34, achievable 
by constructing each sequence under runtime polynominal complexity of $O(L^3)$.
The second view outlines challenges for the next generation of labs solvers.
Given current computational resources, the trend of merit factors achieved with \lssOrelE\ points in the right direction;
however, sequences that would converge closer
to the conjectured asymptotic value of 12.3248 are yet to be discovered
and will be the focus of future work.

\section{Notation and Definitions}
\label{sec_notation}
\begin{figure*}[!t]
\vspace*{-4ex}
\centering
\subfloat{
\begin{minipage}[t]{0.46\linewidth}
\includegraphics[width=0.90\textwidth]{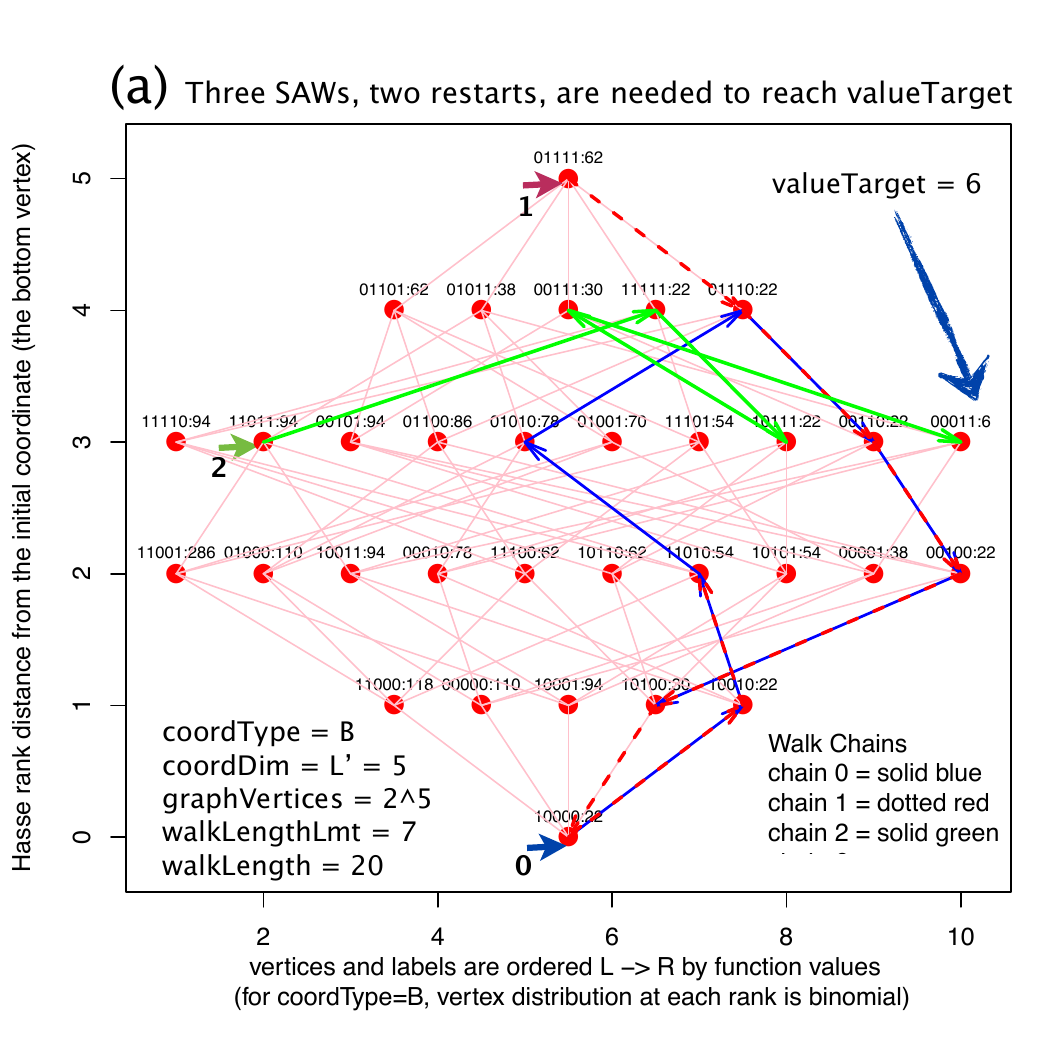}
\label{fg_walks_restarts_vs_saw-a}
\end{minipage}
}
\hfill
\subfloat{
\begin{minipage}[t]{0.46\linewidth}
\includegraphics[width=0.90\textwidth]{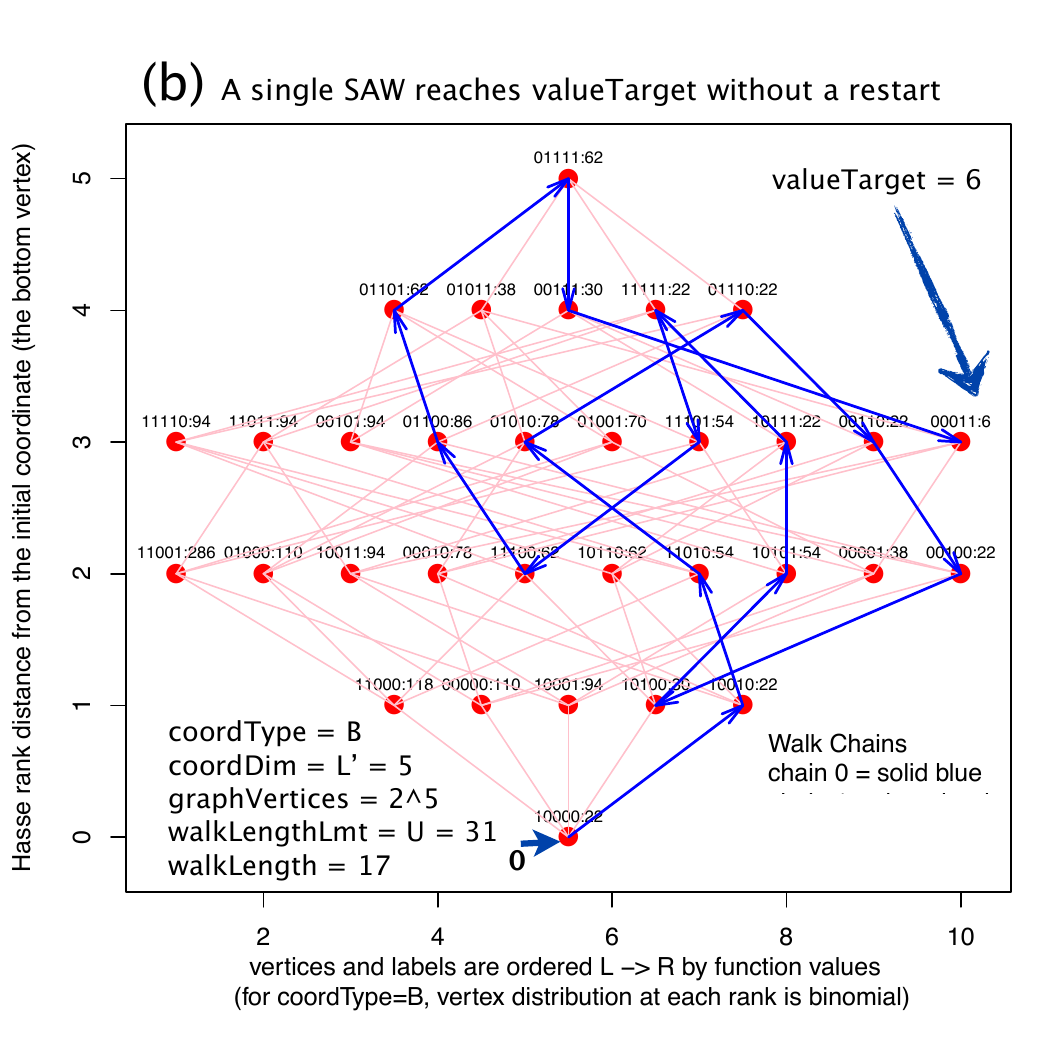}
\label{fg_walks_restarts_vs_saw-b}
\end{minipage}
}
\caption{
Two cases of self-avoiding walks (SAWs): (a) three SAWs where each SAW is limited to 7 or less steps
and is followed by random restarts; (b) a single SAW where the number of steps is limited by the total
number of vertices in the graph. In general, the average walk length required to reach the target value
from any starting point will be less when the walk is self-avoiding and contiguous rather than when the
walk is composed from multiple shorter self-avoiding walk segments.
Nominally, this is an instance of size $L=13$; here we are solving the instance under skew-symmetry,
so $L'=7$. By considering the canonical solutions only (i.e. solutions with the coordinate prefix of {\tt 00}),
we can use a Hasse graph with only $2^5 = 32$ vertices to illustrate such walks.
}
\label{fg_walks_restarts_vs_saw}
\end{figure*}

\noindent
This section follows notation, definitions, and metaphors introduced 
in~\cite{Lib-OPUS2-dice-2011-EV-Brglez} and~\cite{Lib-OPUS2-walk-2013-MIDEM-Brglez}.
The first paper defines Hasse graphs and relates them to 
average-case performance of combinatorial optimization
algorithms, the second paper demonstrates merits of  
{\em long and entirely contiguous self-avoiding walks} which are
searching, under concatenation of binary and ternary coordinates, for the maximum number of bonds in
the 2D protein folding problem.
Combined, these papers also support a simple and intuitive introduction of the
{\em self-avoiding walk segments} as the key component of an effective strategy 
which we apply to finding best solution to instances of the \labs\ problem
in this paper. 
There are two illustrations
of such walks: 
a small one in Figure~\ref{fg_walks_restarts_vs_saw} at the end of this section,
and a larger instance in Section~\ref{sec_solvers} where,
in Figure~\ref{fg_walks_ts_vs_saw},
we compare a self-avoiding walk induced by our solver \lssOrel\ with a walk based on tabu search induced by the solver \lssMAts~\cite{Lib-OPUS-labs-2009-ASC-Gallardo-memetic}.
We proceed with a brief reprise of notation and definitions, some of them extended
to specifics of the \labs\ problem.

\par\vspace*{1ex}\noindent
{\bf Solution as a coordinate-value pair.}
While the energy of the autocorrelation function as defined in Eq.~\ref{eq_energy}
may be simple to interpret in terms of binary symbols $s_i \in \{ +1,-1 \}$,
we define, for the remainder of the paper, any solution of Eq.~\ref{eq_energy}
as a coordinate-value pair in the form
\vspace*{-0.5ex}
\begin{equation}
  {\underline \varsigma}:\Theta({\underline \varsigma})
\label{eq_coord_value_pair}
\vspace*{-0.3ex}
\end{equation} 
where ${\underline \varsigma}$ is a binary string of length $L$, 
also denoted as the {\em coordinate} from $[0, 1]^L$,
and $\Theta({\underline \varsigma})$ is the {\em value} associated with this coordinate
(an integer value denoting energy as shown in Eq. \ref{eq_energy}).

\OMIT{
Examples of coordinate-value pairs are shown in Table~\ref{tb_labs_symmetries}.
Specifically, for instance sizes of $L = 8, 9, 13$, these pairs represent 
two quadrants from the complete sets of 
{\em optimal solutions}, listed in lexicographical order, 
under coordinate prefixes of {\sf `00'} and {\sf `01'}, as the
{\em coordBest} and {\em valueBest} pairs
\begin{equation}
  {\underline \varsigma^*}:\Theta({\underline \varsigma^*})
\label{eq_coord_value_pair_best}
\end{equation}
We could extend the solutions to more quadrants,
under coordinate prefixes of {\sf `11'} and {\sf `10'}, by simply taking a binary complement of
the solutions shown in Table~\ref{tb_labs_symmetries}. 
\begin{table*}[th!]
\begin{center}
\caption{
Optimal solutions for the \labs\ problem for $L = 8, 9$ and 13. 
These solutions are also analyzed with respect to their symmetries and the sizes of their neighborhoods.
The column {\bf sym} reports values of the coordinate symmetry parameter, 
the column {\bf ssym} reports values of the coordinate skew-symmetry parameter.
Solutions under the 'coordinate prefix of 00` are defined as {\bf canonical solutions.}
}
\label{tb_labs_symmetries}
{\small
\begin{tabular}{|c|c|c|c|c||c|c|c|c|}

 \multicolumn{1}{c}{\textbf{}} & \multicolumn{4}{c}{\textbf{coordinate prefix of 00}} & \multicolumn{4}{c}{\textbf{coordinate prefix of 01}} \\ 
 \multicolumn{1}{c}{\textbf{instanceSize    }} &  \multicolumn{1}{c}{\textbf{coordBest}} &  \multicolumn{1}{c}{\textbf{valueBest}} & 
 \multicolumn{1}{c}{\textbf{sym}} & \multicolumn{1}{c}{\textbf{ssym}} & 
 \multicolumn{1}{c}{\textbf{coordBest}} & \multicolumn{1}{c}{\textbf{valueBest}} & 
 \multicolumn{1}{c}{\textbf{sym}} & \multicolumn{1}{c}{\textbf{ssym}} \\ \hline \hline
			& 00001101 & 8 & 0 & NA & 01101000 & 8 & 0 & NA \\ 
L = 8			& 00011010 & 8 & 0 & NA & 01011000 & 8 & 0 & NA \\ 
neighbSize$^\dagger$ = 28		& 00010110 & 8 & 0 & NA & 01001111 & 8 & 0 & NA \\ 
(for each quadrant)	& 00111101 & 8 & 0 & NA & 01000011 & 8 & 0 & NA \\ \hline
\hline
			& 000011010 & 12 & 0 & 2 & 011000010 & 12 & 0 & 0 \\ 
			& 000110010 & 12 & 0 & 1 & 011010111 & 12 & 0 & 0 \\ 
L = 9			& 000101001 & 12 & 0 & 0 & 010110000 & 12 & 0 & 1 \\ 
neighbSize$^\dagger$ = 47		& 000101100 & 12 & 0 & 0 & 010011000 & 12 & 0 & 1 \\ 
(for each quadrant)	& 001101000 & 12 & 0 & 0 & 010000110 & 12 & 0 & 0 \\ 
			& 001111101 & 12 & 0 & 0 & 010000011 & 12 & 0 & 0 \\ \hline
\hline
L = 13			&	    &    &   &   & 	       &    &   & \\ 
neighbSize$^\dagger$ = 13		& 0000011001010 & 6 & 0 & 1 & 0101001100000 & 6 & 0 & 1 \\ 
(for each quadrant)     &	    &    &   &   &        &    &   & \\ \hline\hline
\end{tabular}

\par\vspace*{2.75ex}
\begin{minipage}[t]{0.99\textwidth}
{\em
($\dagger$) The value of $L$ denotes not only the instance size of the \labs~ problem but also 
the size of the binary coordinate. A binary coordinate of size $L$ has $L$ 
immediate neighbors or adjacent coordinates, with distance of 1 from each neighbor. 
The neighborhood size for a set $m_L$ coordinates that represent the number of global minima in a quadrant is computed as 
$L \times m_L - r$
where $r$ is the number of replicated coordinates produced while
generating neighborhood coordinates. 
When evaluating neighborhoods of $L = 8$ and $L = 9$, 
there are 4 and 7 replicated neighborhood coordinates, thus values of neighbSize are 28 and 47, respectively.
}
\end{minipage}
}
\end{center}
\end{table*}
}

\par\vspace*{1ex}\noindent
{\bf Coordinate distance.}
The distance between two binary coordinates {\underline a}
and {\underline b} is defined as the Hamming distance:
\vspace*{-0.5ex}
\begin{equation}
  d({\underline a}, {\underline b}) = \sum_{i=1}^{L} a_{i} \oplus b_{i}
\label{eq_coord_distance}
\vspace*{-0.3ex}
\end{equation}

\noindent
{\bf Coordinate symmetries.}
There are four coordinate transformations that reveal the symmetries of the
\labs~ problem function as formulated in Eq.~\ref{eq_energy}:

\begin{quote} 
{\em complementation:}
For example, the complement of 
0000011001010 is 1111100110101
\par
{\em reversal:}
For example,  the reversal of 
{\sf 0000011001010} is {\sf 0101001100000}
\par
{\em symmetry:}
For example, for $L$ even,  we say that
{\sf 011110} is symmetric compared to  $L/2$ without coordinate complementation.
\par
{\em skew-symmetry:}
Skew-symmetry has been introduced in \cite{Lib-OPUS-labs-1977-IEEE_TIT-Golay-sieves}.
It is defined for odd values of $L$ only and 
the solution of the \labs~ problem can be expressed
with coordinates that are significantly reduced in size: 
\vspace*{-0.5ex}
\[ L' = (L+1)/2 \] 
\vspace*{-0.5ex}
For example, 
{\sf 0000011001010}  is skew-symmetric since $(L-1)/2$ left-most coordinates and
$(L-1)/2$ right-most coordinates are skew-symmetric under coordinate reversal.
Formally, the  skew-symmetry definition in terms of 
binary coordinate components $b_k$ and their complements $\overline{b}_{k}$ is
\vspace*{-0.5ex}
\begin{equation}
  b_{L'+i} = \left\{ \begin{array}{ll}
                      \overline{b}_{L'-i} \mbox{\hspace*{2ex} if~ $i=1, 3, 5, ...$}\\
                                b_{L'-i}  \mbox{\hspace*{2ex} if~ $i=2, 4, 6, ...$}
                      \end {array}
             \right.
\label{eq_skew_symmetry}
\vspace*{-0.3ex}
\end{equation}

\end{quote}

The introduction of skew-symmetry significantly reduces the computational
complexity of the \labs\ problem -- but not every optimal solution for odd values of 
$L$ is also skew-symmetric. Only recently, a branch-and-bound 
solver~\cite{Lib-OPUS-labs-2016-Packebusch,Lib-OPUS-labs-2013-arxiv-Prestwich-branch-and-bound-odd}
extended known optimal solutions for skew-symmetric sequences to length up to 119.
However, even under skew-symmetry, only stochastic solvers have demonstrated the
potential to find improved, if not optimal solutions, as $L$  increases to 201 and beyond.

\par\vspace*{1ex}\noindent
{\bf Canonic solutions and best upper bounds.}
The \labs~ problem symmetries partition solutions into four quadrants with coordinate prefixes of 
{\sf 00}, {\sf 01}, {\sf 10}, and {\sf 11}.
Without loss of generality, we transform coordinates of all optimal or {\em best value} solutions found in quadrants 
{\sf 01}, {\sf 10}, {\sf 11} to the quadrant {\sf 00}  and denote the set of unique optimal 
{\sf coordinate:value} solution pairs in the quadrant {\sf 00} as the {\em canonic solutions set}. For a given $L$, only
the coordinates in this set are unique; the optimum or the best known value, also denoted as {\em the best upper bound}
$\Theta^{ub}_L$ is the same for each coordinate in this set. We say that $C_L$ is the cardinality of canonic solutions
set and that $\Theta^{ub}_L$ is the best upper bound
we associate with the \labs\ instance of size $L$.

\par\vspace*{1ex}\noindent
{\bf Distance=1 neighborhood.}
The distance=1 neighborhood of a coordinate  ${\underline \varsigma}_j$  is a set of coordinates
\begin{equation}
  {\cal N} ({\underline \varsigma}_j) = \{~ {\underline \varsigma}_j^i ~|~ d({\underline \varsigma}_j, {\underline \varsigma}_j^i ) = 1, 
  ~~~i = 1, 2, \dots , L ~\}
\label{eq_neighborhood}
\end{equation}
Informally, a binary coordinate ${\underline \varsigma}_j$ of size $L$, also called a {\em pivot coordinate}, has $L$ 
neighbors, each a distance of 1 from the pivot coordinate. 
\OMIT{
In Table~\ref{tb_labs_symmetries}, instances with $L=7$ and $L=13$ have a single solution in each quadrant,
and hence also neighborhoods of size 7 and 13, respectively.
For instances with the cardinality of canonic solutions $m_L > 1$, some of the neighborhood coordinates may be replicated.
Adjustments are needed when computing the size such neighborhoods:
\[ L \times m_L - r \] 
where $r$ is the number of replicated coordinates produced while
generating neighborhood coordinates. 
For examples of such computations, see Table~\ref{tb_labs_symmetries}.
}

\par\vspace*{1ex}\noindent
{\bf Contiguous walks and pivot coordinates.}
Let the coordinate ${\underline \varsigma}_0$ be the initial coordinate from which the walk takes the first step. Then the sequence  
\begin{equation}
  \{{\underline \varsigma}_0, {\underline \varsigma}_1, {\underline \varsigma}_2, \dots , {\underline \varsigma}_j, \dots ,
  {\underline \varsigma}_\omega\}
\label{eq_coord_pivots}
\end{equation}
is called a {\em walk list} or a {\em walk} of length $\omega$, the coordinates ${\underline \varsigma}_j$ are denoted as
{\em pivot coordinates} and $\Theta({\underline \varsigma}_j)$ are denoted as {\em pivot values}.
Given an instance of size $L$ and its best upper bound $\Theta^{ub}_L$,
we say that the walk {\em reaches} its target value (and stops) when
$\Theta({\underline \varsigma}_\omega) \le \Theta^{ub}_L$.

We say that the walk is contiguous if the distance between adjacent pivots is 1; i.e.,
given Eq.~\ref{eq_coord_distance}, we find
\begin{equation}
  d({\underline \varsigma}_j, {\underline \varsigma}_{j-1}) = 1, ~~~j = 1, 2, ..., \omega
\label{eq_coord_pivots_distance}
\end{equation}

\par\vspace*{1ex}\noindent
{\bf Self-avoiding walks (SAWs).}
We say that the walk is {\em self-avoiding} 
if all pivots in Eq.~\ref{eq_coord_pivots} are unique. 
We say that the walk is composed of two or more {\em walk segments} if the initial pivot of each walk segment has been
induced by a well-defined heuristic such as {\em random restarts},
a heuristic associated also with all solvers described in this paper. Walk segments can be of different lengths and
if viewed independently of other walks,
may be self-avoiding or not. A walk composed of two or more self-avoiding walk segments may 
no longer be a self-avoiding walk, since some of the pivots may overlap and also form cycles.
This is illustrated after we define the Hasse graph below.

\par\vspace*{1ex}\noindent
{\bf Hasse graph.}
Hasse graph has been defined 
in~\cite{Lib-OPUS2-dice-2011-EV-Brglez,
Lib-OPUS2-walk-2013-MIDEM-Brglez}
as a model of hyperhedron (or informally, a dice) based on an
extension of the Hasse diagram. In the case of the \labs\ problem,
Hasse graph is an undirected  {\em labeled graph} with $2^L$ vertices and
$L\times2^{L-1}$ edges; the degree of each vertex is $L$ and the label is the pair
${\underline \varsigma}:\Theta({\underline \varsigma})$ as defined in
Eq.~\ref{eq_coord_value_pair}. 
By projecting this graph  with its labeled vertices  onto a plane, we
can not only illustrate concepts of coordinate/pivot neighborhoods
but also specific walks as a combinatorial search heuristics. 

Figure~\ref{fg_walks_restarts_vs_saw} illustrates not only two Hasse graphs,
with each vertex displaying a {\em coordinate:value}; it also illustrates
that the target value can be reached either by 
a sequence of three shorter SAW segments
(each segment represents a contiguous SAW) or by a single contiguous SAW.
The three contiguous {\em self-avoiding walk segments}
in Figure~\ref{fg_walks_restarts_vs_saw-a} have lengths of 7, 7, and 4,
covering a total of 19 vertices in 18+2=20 steps.
We add two steps since the second and the third walk segment are induced by two restarts. Just as
the pivot ${\underline \varsigma}_0$ is the initial pivot for the first step in the first walk, pivots 
${\underline \varsigma}_7$  and  ${\underline \varsigma}_{15}$ are taken as initial pivots for 
the first step in the second walk and the first step in the third walk, respectively. Here is a linear depiction of the 19
vertices and three walk segments.
\[ \bullet     \underbrace{\bullet \bullet \bullet \bullet \bullet \bullet \bullet}_7      
            ~~~\underbrace{\bullet \bullet \bullet \bullet \bullet \bullet \bullet}_7 
            ~~~\underbrace{\bullet \bullet \bullet ~\bullet}_4 \]
To keep the Hasse graph less cluttered, the steps of the walk that are induced by two restarts are not
shown with additional edges. We denote such steps as {\em jump steps} since the distance between pivots may exceed 1. For example,        
$d({\underline \varsigma}_8, {\underline \varsigma}_7) = d({\sf 10100, 01111}) = 4$
and 
$d({\underline \varsigma}_{16}, {\underline \varsigma}_{15}) = d({\sf 11010, 11011}) = 1$.

A more formal description of SAW as a general purpose combinatorial search
algorithm is given in Section~\ref{sec_solvers}.
A summary of results in Section~\ref{sec_experiments} demonstrates that in 
the asymptotic sense (as $L$ increases), a contiguous SAW has
walk lengths that are on the average shorter than walk lengths achieved
under heuristic which limits the length of each SAW and then, 
with repeated random restarts,
assembles 
the shorter SAWs into a single long walk. In general, the assembled walk
is no longer contiguous, see Figure~\ref{fg_walks_restarts_vs_saw-a}. 

\par\vspace*{1ex}\noindent
{\bf On origins of self-avoiding walks.}
The notion of self-avoiding walks (SAWs) was first introduced by the chemist Paul Flory
in order to model the real-life behavior of chain-like entities such as solvents and polymers, 
whose physical volume prohibits multiple occupation of the same spatial point~\cite{Lib-OPUS-walk-self-avoiding-2014-Wikipedia}.
In mathematics, a SAW lives in the n-dimensional lattice $\mathbb{Z}^n$ which consists
of the points in $\mathbb{R}^n$ whose components are all integers 
\cite{Lib-OPUS-walk-2011-EMS-Slade-self-avoiding,Lib-OPUS-Book-2013-Birkhauser-Madras-The-Self-Avoiding-Walk}.
The challenge of finding the longest self-avoiding walk in multi-dimensional
lattices {\em efficiently} has been and also continues to be of considerable interest in
physics~\cite{Lib-OPUS-walk-2010-StatPhysics-Clisby-pivot-alg-self-avoiding}.

\section{Solver Instrumentation and Solvers}
\label{sec_solvers}
\noindent
%
We have instrumented a total of four solvers to conduct the experiments  
summarized in the follow-up sections. 
Solvers \lMAts\ and \lssMAts, described 
in~\cite{Lib-OPUS-labs-2009-ASC-Gallardo-memetic},
implement a memetic-tabu search strategy. 
The solver \lMAts\ returns solutions for both even and odd 
sequences, the solver \lssMAts\ returns skew-symmetric sequences only. 
The solver \lssRRts, is a special case of \lssMAts. 
Our solver, \lssOrel, implements a self-avoiding walk strategy for 
odd sequences under skew-symmetry so that experimental results 
with \lssOrel\ can be
directly compared with  \lssMAts\ and \lssRRts.

We begin with {\em solver instrumentation}, follow up
with solver pseudo-code descriptions, and conclude with highlights on differences between \lssMAts\ and \lssOrel.

\begin{table*}[t]
\caption{Summary of notation: symbols, names, and descriptions.}
\label{tb_notation_summary}
\centering
\begin{footnotesize}
\begin{tabular}[]{@{}c@{~~}c@{}}
\begin{tabular}[t]{l l l}  
\centering
\bf{symbol}  & \bf{short name} & \bf{brief description} \\ \hline\\[-1.25ex]
$L$ & coordDim & instance size \\[1.25ex]
$L'$ & coordDim'& instance size under  \\[-0.25ex]
      &          &  skew-symmetry   \\[-0.25ex]
      &          &  $(L+1)/2 $ \\[1.25ex]
$\sigma_0$ &  seedInit     &  initial seed integer \\[1.05ex]
${\underline \varsigma}_0$ & coordInit & initial coordinate \\[1.05ex]
$\Theta({\underline \varsigma}_0)$ & valueInit & initial value \\[1.05ex]
${\underline \varsigma}_j$ & coordPivot & pivot coordinate\\[1.75ex]
$\Theta({\underline \varsigma}_j)$ & valuePivot & pivot value \\[1.75ex]
${\underline \varsigma}_j^i$ & coordNeighb & pivot neighbor coord. \\[1.75ex] 
${\cal N} ({\underline \varsigma}_j)$ & coordNeighbSet & full neighborhood set \\[-0.45ex] 
                                      &                & of pivot coordinate \\[1.75ex] 
{${\cal N}_{saw} ({\underline \varsigma}_j)$} & {sawNeighbSet} & {SAW neighborhood set} \\[+1.15ex] 
$\omega_c$     & walkSegmCoef &  walk segment coefficient\\[+01.20ex]  
$\omega_{lmt} = \omega_c \times L' $ & walkSegmLmt & walk segm. length limit\\[1.75ex] 
{$Walk_{\omega} ~= $}  & {~~~} & {~~~} \\[0.3ex]
~~=~$\{{\underline \varsigma}_0, \dots , {\underline \varsigma}_\omega\}$ 
& walkList & walk list after $\omega$ steps \\[0.5ex]
$t$ & runtime & CPU runtime  \\[0.55ex]  
$t_{lmt}$ & runtimeLmt & solver timeout value\\[0.55ex]
$\tau$ & cntProbe & \# of function probes \\[0.65ex] 
$\rho$ & cntRestart & \# of walk restarts  \\[0.65ex]  
\hline
\multicolumn{3}{l}{ }  \\[-1.4ex]
\multicolumn{3}{l}{For \labs\ problems with an odd value of $L$, $L' = (L+1)/2 $}  \\ 
\multicolumn{3}{l}{represents a {\em de facto} instance size under skew-symmetry.}  \\  
\end{tabular}
&%
\begin{tabular}[t]{ l l l} 
\centering
\bf{symbol}  & \bf{short name} & \bf{brief description} \\ \hline\\[-1.0ex]
$ \beta $ & cntTrapped & \# of trapped solutions \\[0.30ex] 
${\underline \varsigma}^*$ & coordBest & best coordinate \\[0.30ex]
$\Theta({\underline \varsigma}^*)$ & valueBest & best value \\[0.30ex]
$\Theta^{ub}_L$ & valueTarget & best upper bound\\[0.30ex]  
~~                 & targetReached & solution status:  \\
~~                 & ~~         & 0 if $\Theta({\underline \varsigma}^*) > \Theta^{ub}_L$,\\[0.30ex]
~~                 & ~~         & 1 if $\Theta({\underline \varsigma}^*) = \Theta^{ub}_L$,\\[0.30ex]
~~                 & ~~         & 2 if $\Theta({\underline \varsigma}^*) < \Theta^{ub}_L$\\[0.30ex]
~~                 & isCensored & solution censure status: 1 if\\[0.30ex]
~~                 & ~~         & $t \ge t_{lmt}$ (targetReached = 0); \\[0.30ex]
~~		   & ~~         & 0 otherwise (targetReached = 1 or 2) \\[0.30ex]
$N$                & sampleSize & \# of instances and initial  \\[-0.2ex]
~~                 &            & seeds in the experiment \\[0.30ex] 
$N_c$                & sampleSizeCrit & $N_c \ge$ 100 to satisfy $\mathit{CI_{0.95}}$  \\[0.3ex] 
~~                 &            & $\approx [0.8 \times \overline{m},~1.2 \times \overline{m}]$ \\[0.30ex] 
$\overline{m}$         & sampleMean        & an estimate of the sample mean \\
$\mathit{hitO}$    & hitsObserved      & \# of observed uncensored \\[-0.15ex]
~~                 &                   &  solutions (hits)\\[0.30ex]
$\mathit{hitO_r}$  & hitRatioObserved  & observed hit ratio, $\mathit{hitO/N}$ \\[0.30ex]  
$\mathit{hitP}$    & hitsPredicted     & \# of predicted uncensored \\[-0.15ex]
~~                 &                   &  solutions (hits)\\[0.30ex]
$\mathit{hitP_r}$  & hitRatioPredicted & predicted hit ratio, $\mathit{hitP/N}$ \\[0.30ex]
$\mathit{solvP}$       & solvPredicted     & predicted solvability or waiting  \\[-0.15ex]
~~                     &                   & time for a single solver\\[0.30ex]
$\mathit{solvP_{ser}}$ & solvPredictedSer  & predicted solvability or waiting  \\[-0.15ex]
~~                     &                   & time for $N$ serial solvers\\[0.30ex]
\hline
\multicolumn{3}{l}{ }  \\[-1.4ex]
\multicolumn{3}{l}{For \labs\ problems, $\Theta({\underline \varsigma}^*)$ represents the {\em minimum energy value}} \\
\multicolumn{3}{l}{returned by the solver.}
\end{tabular}
\end{tabular}
\end{footnotesize}
\end{table*}

\par\vspace*{1ex}\noindent
{\bf Solver instrumentation.}
We argue that in order to design a better
combinatorial solver one also needs to devise an environment 
and a methodology that supports
reproducible and statistically significant computational experiments.
In our case, this environment continues to evolve
under the working name of {\tt xBed}~\cite{Lib-OPUS2-ebook-xBed-2016-Brglez}.
A generic and standardized notation is an important part of this environment;
Table~\ref{tb_notation_summary} summarizes the notation and description
of principal variables in our solver instrumentation which we also use in
our pseudo-code descriptions. 

For example, reporting the {\em runtime} or $t$ 
is not the only performance variable of importance. The most important variable is
the variable named as 
{\em cntProbe} or $\tau$: a variable that counts how many times the solver evaluates the
objective function before completing the run. By keeping track of this variable, we can compare two
solvers regardless of the platform on which experiments have been performed,
and regardless whether the solver represents a much slower scripted implementation
of an early prototype
or the faster compiled-code implementation. 
In our experiments,
the correlation coefficient between {\em runtime} and {\em cntProbe} consistently 
exceeds 0.999.

Given the labs problem of size $L$, the experiment is defined by $N$ runs of the solver where each 
run is considered a sequence of {\em Bernoulli trials} which are probing the objective function: 
the solver stops as soon as -- and only when -- it finds the coordinate ${\underline \varsigma}^*$ with the best known solution value $\Theta({\underline \varsigma}^*) = \Theta^{ub}_L$. 
If the solver does not time-out, the status variables recorded with each run are {\em targetReached = 1} and  {\em isCensored = 0} or {\em targetReached = 2} and  {\em isCensored = 0}.
If the solver returns a solution $\Theta({\underline \varsigma}^*) < \Theta^{ub}_L$,
we record {\em targetReached = 2}, 
declare  $\Theta^{ub}_L = \Theta({\underline \varsigma}^*)$ as the new best known solution value, and initiate a new series of Bernoulli trials with the new value of $\Theta^{ub}_L$.
If the solver terminates the run  before reaching the target value $\Theta^{ub}_L$ (due to a time-out limit), we record {\em targetReached = 0} and  {\em isCensored = 1}. 
Each time {\em targetReached = 1} is recorded, the counter that measures the
{\em observed number of hits}, $\mathit{hitO}$, is incremented:
\vspace*{-0.5ex}
\begin{equation}
\mathit{hitO} \defeq {\rm number~of~successes~or~} \mathit{hits} {\rm ~in~} N {\rm ~runs}.
\label{eq_hitO}
\end{equation}
\vspace*{-1.2ex}
Concurrently with the observed number of hits $\mathit{hitO}$, we also define the 
{\em observed hit ratio}, $\mathit{hitO_r}$:
\vspace*{-0.3ex}
\begin{equation}
\mathit{hitO_r} \defeq hitO/N
\label{eq_hitO_r}
\end{equation}
\vspace*{-1ex}
The {\em sampleSize} of the experiment is thus $N$: the experiment is performed either serially on a single processor or in parallel on a grid of $N$ processors.

Given the probability of success on any {\em single} Bernoulli trial as $p$, then the number of probes $T$ required to obtain the first success has geometric distribution with parameter~$p$.  
Formally, the probability of observing the first success on $\tau$-th probe or before is thus
\vspace*{-0.3ex}
\begin{equation}
 P(T = \tau) = (1 - p)^{\tau - 1} p 
\label{eq_geometric}
\end{equation}
\vspace*{-1ex}
The cumulative distribution function $F_{T}(\tau)$ is then
\vspace*{-0.3ex}
\begin{equation}
P( T \le \tau ) = F_{T}(\tau) = 1 - (1 - p)^{\tau}
\label{eq_geometricCum}
\end{equation}
\vspace*{-1ex}
i.e. the probability that the solver finds the solution before or on probe $\tau$ is the cumulative probability $F_{T}(\tau)$.
Experimentally, the time required until a success occurs is proportional to the number of probes, hence $T$ also denotes the {\em waiting time}, a characteristic value associated with the solver.

%
%
%


Traditionally, the waiting time is considered as a continuous random variable and for $p < 0.01$ we can approximate, with negligible error, the distribution function $F_{T}(\tau)$
with a cumulative exponential distribution function
\vspace*{-0.3ex}
\begin{equation}
P( T < \tau ) = F_{T}(\tau) = 1 - exp(-\tau/\overline{m})
\label{eq_exponentialCum}
\end{equation}
\vspace*{-1ex}
where $\overline{m} = 1/p$. Given the sample size $N$, we estimate the value of $p$ either by computing the mean value of $\overline{m}$ either as
$\mathit{cntProbe}~\tau$ or $\mathit{runtime}~t$ in units of seconds, minutes, hours, days, etc.
In other words, given a solver with an estimated value of $\overline{m} = 10$ hours, the probability that this solver returns the target solution value in 10 hours 
(or less) is 0.632: by waiting for 20, 40, or 80 hours, the probability of a $\mathit{hit}$ increases to 0.865, 0.982, or 0.999, respectively.

The performance experiments  summarized in the next section confirm that the
{\em uncensored} random variables such as {\em runtime} or {\em cntProbe} 
have near-exponential or near-geometric distribution,   
i.e. we observe ~$s \approx \overline{m}$  where ~$s$ denotes the sample standard deviation 
and ~$\overline{m}$ denotes the sample mean. Under such distributions,
given the uncensored sample size of ~$N_c=100$ used in all of our experiments, 
a reliable rule-of-thumb estimate of the 
95\% confidence interval ($\mathit{CI_{0.95}}$) on value of the sample mean ~$\overline{m}$  
is thus 
\vspace*{-0.3ex}
\begin{equation}
\hspace*{-1.9em}
\mathit{CI_{0.95}} \approx [0.8 \times \overline{m},~1.2 \times \overline{m}] 
{\rm ~...~uncensored}~N_c=100
\label{eq_confidenceInterval}
\end{equation}
\vspace*{-1ex}
When a subset of $N_c$ runs is censured, the confidence interval  can increase significantly beyond the one 
in Eq.~\ref{eq_confidenceInterval}.
We argue that reliable estimates of confidence bounds on the mean values
of {\em runtime} or {\em cntProbe} returned by combinatorial solvers under censoring
are a subject best left to 
statisticians~\cite{Lib-OPUS-statistics-1995-MathStat-Sen-censoring}. 

The question now arises whether and how, for a given solver and a labs instance, we can {\em predict} the number of hits, $\mathit{hitP}$, as well as the hit ratio, $\mathit{hitP_r}$.
The answer can be formulated in the context of the solver time-out value or runtime limit
$t_{lmt}$:
\vspace*{-0.3ex}
\begin{equation}
\mathit{hitP_r} \defeq 1 - \mathit{exp}(-t_{lmt}/\overline{m})
\label{eq_hitP_r}
\end{equation}
\vspace*{-1ex}
Thus, the predicted hit ratio $\mathit{hitP_r}$ in Eq.~\ref{eq_hitP_r}
is just an instance of the cumulative exponential  distribution function in 
Eq.~\ref{eq_exponentialCum}. When the experiment is performed serially on a single processor, we assume that 
the solver is the only significant application scheduled to run on the processor, so the value of $t_{lmt}$ represent the true runtime limit. However, when the experiment is performed in parallel on a grid of $N$ processors, each processor is scheduled to run a number of applications and the value of $t_{lmt}$ is reduced by a {\em loadFactor} $>$ 1,
a random variable with an empirical average value of 
2.4 for our environment under~\cite{2014-Web-SLING}. 

Similarly, we define and calculate the {\em predicted the number of hits}, 
$\mathit{hitP}$:
\vspace*{-0.3ex}
\begin{equation}
\mathit{hitP} \defeq \lfloor N \times \mathit{hitP_r} \rfloor 
\label{eq_hitP}
\end{equation}
\vspace*{-1ex}

In Section~\ref{sec_experiments}, we empirically derive the runtime model for the solver \lssOrel\ to predict a runtime mean. Consider a labs instance for $L = 149$.  The model predicts a runtime mean of $0.000032 * 1.1504^{149}/(3600)= 10.34928$ hours. 
For a runtime limit $t_{lmt} = 96$ hours (4 days) and $N = 100$ processors, 
Eqs.~\ref{eq_hitP_r} and~\ref{eq_hitP} predict
$\mathit{hitP_r} = 0.9999055$ and $\mathit{hitP} = 99$ -- under the assumption of
{\em loadFactor = 1}. Under the empirically verified value of 
{\em loadFactor = 2.4} under~\cite{2014-Web-SLING}, our predictions are modified to
 $\mathit{hitP_r} = 0.9789588$ and $\mathit{hitP} = 97$. The  experiments 
 on the grid, summarized in
Figure~{15} illustrate that for $L=149$ the solver \lssOrel\ reports the {\em observed hit ratio} $\mathit{hitO} = 95$.

%
%
%
%
Given that $\mathit{hitO} = 95 < N_c$ 
with $N_c = 100$ as postulated in Eq.~\ref{eq_confidenceInterval},
the question arises how many processors $N$ should be scheduled on the grid, so that
for each value of $L$, we can maintain $\mathit{hitO} \ge N_c = 100$ with the runtime limit of $t_{lmt} = 96$ hours? A quick back-of-the
envelope calculation returns the answer for $L=149$: $N = 103$. More formally, we find the answer as follows:
\vspace*{-0.3ex}
\begin{equation}
N  \defeq \lceil N_c/\mathit{hitP_r} \rceil 
\label{eq_forN}
\end{equation}
\vspace*{-1ex}

Two more examples for the runtime limit of $t_{lmt} = 96$, extrapolated from Figure~{15}: (1) for $L = 165$ we find
$\mathit{hitP_r} = 0.3365782$ and $N = 298$ in order to achieve $N_c = 100$ hits, and
(2) for $L = 179$ we find
$\mathit{hitP_r} = 0.05607801$ and $N = 1784$ in order to achieve $N_c = 100$ hits.

%
%
%
An experiment of a sample size $N$ performed in parallel with $N$ solvers on a grid 
of $N$ processors has a significant 
advantage in terms of ``waiting time'' when compared to an experiment
where the solver is invoked $N$ times on a single processor.
However, serially scheduled experiments are important in this research:
they support not only
accurate runtime comparisons of different solvers but also allow accurate observations of statistically significant differences, if any, between two or more heuristics implemented by the same solver.
Consider again the labs instance for $L = 149$ analyzed after Eq.~\ref{eq_hitP}: the value of
the runtime mean is 10.34928 hours and when invoking the solvers in parallel on $N=100$ processors we predict to reach the hit ratio
$\mathit{hitP_r} > 0.99$ in 96 hours (4 days).
When invoking the solver $N$ times on a single processor,
the sum of the exponential variates returned by each run has {\em gamma distribution}.
For relationships between Poisson's processes,
exponential distributions, and gamma distributions, see~\cite{1980-Macmillan-Olkin}.

Using the notation of R~\cite{WEB-Project-R},
the cumulative gamma function $\mathit{pgamma}$ and its inverse $\mathit{qgamma}$:
%
\begin{equation}
\hspace*{-1.2em}
P( T \le t_{lmt} ) = \mathit{hitP_{ser}}  \defeq  \mathit{pgamma}(t_{lmt}, N, 1/\overline{m})
\label{eq_gammaCum}
\end{equation}
\vspace*{-0.5ex}
In queueing theory,
we associate the inverse of the cumulative gamma function $\mathit{qgamma}$
with the
{\em waiting time}. In this paper, 
given $\mathit{hitP_{ser}}$ as the solver's predicted hit rate of $N$ solvers invoked serially,
the inverse of cumulative gamma function is the metaphor for 
the {\em predicted solvability} $\mathit{solvP_{ser}}$ of 
$N$ serially invoked solvers:
\begin{equation}
\hspace*{-1.2em}
\mathit{solvP_{ser}} \defeq \mathit{qgamma}(hitP_{ser}, N, 1/\overline{m})
\label{eq_solvP_ser}
\end{equation}
\vspace*{-0.5ex}
%
Using Eq.~\ref{eq_solvP_ser} (the inverse of Eq.~\ref{eq_gammaCum}),
we can compute the solvability $\mathit{solvP_{ser}}$ directly.
Given a solver with $\overline{m} = 10.34928$
and the hit rate of 99/100,  we use $\mathit{qgamma}$
and find $\mathit{qgamma}(0.99, 100, 10.34928) = 1290.79$ hours (53.8 days).

In the case of the single solver with predicted hit ratio $\mathit{hitP_r}$, we again use $\mathit{qgamma}$ to compute the {\em predicted solvability} $\mathit{solvP}$ of this single solver:
\begin{equation}
\hspace*{-1.2em}
\mathit{solvP} \defeq \mathit{qgamma}(hitP_r, 1, 1/\overline{m})
\label{eq_solvP}
\end{equation}
\vspace*{-0.5ex}

Table~\ref{tb_gamma} presents values of cumulative gamma function  in the
range of most practical interest for our purposes. Note that for $N=100$,
$\mathit{pgamma}(125,100,1) = 0.9906$ and that for $N=1$, $\mathit{pgamma}(q,1,1)$ is equivalent to the cumulative distribution of the exponential function.\\

\begin{table}[h]
  \caption{A detail about pg(q, r) = pgamma(q, r, 1) in R.}
  \label{tb_gamma}
  \centering
  \begin{small}
 \begin{tabular}{c|c c c|| c | c}
 q & pg(q, 1)  & pg(q, 2) & pg(q, 4) & q & pg(q, 100)   \\
 \hline
 1 & 0.6321 & 0.2642 & 0.0189 & 80  & 0.0171 \\
 2 & 0.8647 & 0.5939 & 0.1428 & 90  & 0.1582 \\
 3 & 0.9502 & 0.8009 & 0.3528 & 100 & 0.5133 \\
 4 & 0.9817 & 0.9084 & 0.5665 & 110 & 0.8417 \\
 5 & 0.9933 & 0.9596 & 0.7349 & 120 & 0.9721 \\
 8 & 0.9997 & 0.9969 & 0.9576 & 125 & 0.9906 \\
10 & 0.9999 & 0.9995 & 0.9897 & 135 & 0.9993 \\
 \end{tabular}
 \end{small}
\end{table}

\begin{figure*}[t!]
\begin{small}
\subfloat{
\begin{minipage}[t]{0.99\linewidth}
\label{fg_global_search_a}
\begin{algorithmic}[1]
\PROCEDURE{\CALL{lssOrel}{$\sigma_0, \Theta^{ub}_L, t_{lmt}, \omega_{lmt}$}}
\STATE {${\underline \varsigma}_0\!:\!\Theta({\underline \varsigma}_0) \gets {\tt coordInit}(\sigma_0)$}
\COMMENT{initial solution}
\STATE $\tau \gets 1 $ 
\COMMENT {initialize cntProbe}
\STATE ${\underline \varsigma^*}\!:\!\Theta({\underline \varsigma^*}) \gets {\underline \varsigma}_0\!:\!\Theta({\underline \varsigma}_0)$ 
\COMMENT{initial best solution} 
\STATE $isCens \gets 0$
\COMMENT{initialize $isCensored$}
\STATE $tgReached \gets 0$                                     
\COMMENT{initialize $targetReached$}
\STATE $\beta \gets 0$                                         
\COMMENT{initialize $cntTrapped$}
\STATE $\omega \gets 0$                                        
\COMMENT{initialize total number of steps}
\WHILE {{\bf true}}
  \STATE $\omega_{s}\!:\!{\underline \varsigma^*}\!:\!\Theta({\underline \varsigma^*}) \gets {\tt walk.saw}({\underline \varsigma}_{0}\!:\!\Theta({\underline \varsigma}_{0}),t_{lmt},\omega_{lmt})$ 
  \COMMENT{return a completed walk segment}
  \STATE $\omega \gets \omega + \omega_{s} $                   
  \COMMENT{update total number of steps}
  \IF{$\Theta({\underline \varsigma^*}) \leq \Theta^{ub}_L$}
    \IF {$\Theta({\underline \varsigma^*}) = \Theta^{ub}_L$}
      \STATE{$tgReached = 1$}                                  
      \COMMENT{upper-bound is reached}
    \ELSE
      \STATE $tgReached = 2$                                   
      \COMMENT{upper-bound is improved}
    \ENDIF
    \STATE {\bf{break}}
  \ENDIF
  \IF{$t \geq t_{lmt}$}
    \STATE $isCens \gets 1$                                    
    \COMMENT{return solution as ``censored''}
    \STATE {\bf{break}}
  \ENDIF
  \STATE ${\underline \varsigma}_0\!:\!\Theta({\underline \varsigma}_0) \gets {\tt coordInit}()$ 
  \COMMENT{initialize a new walk segment}
  \STATE $\tau \gets \tau + 1 $                               
  \COMMENT{update $cntProbe$}
  \STATE $\omega \gets \omega + 1 $                           
  \COMMENT{update total number of steps}
\ENDWHILE
\STATE $Table \gets (\sigma_0, {\underline \varsigma^*}, \Theta({\underline \varsigma^*}), \omega, \tau, t, isCens, tgReached)$
\ENDPROCEDURE
\end{algorithmic}
\end{minipage}
}

\vspace{0.45cm}

\subfloat{
\begin{minipage}[t]{0.49\linewidth}
\label{fg_global_search_b}
\begin{algorithmic}[1]
\PROCEDURE{\CALL{walk.saw}{${\underline \varsigma}_{0}\!:\!\Theta({\underline \varsigma}_{0}),t_{lmt}, \omega_{lmt}$}}
\IF {$\Theta({\underline \varsigma}_{0}) \le \Theta({\underline \varsigma^*})$}
    \STATE ${\underline \varsigma^*}\!:\!\Theta({\underline \varsigma^*}) \gets {\underline \varsigma}_0\!:\!\Theta({\underline \varsigma}_0)$
    \COMMENT{new best solution}
\ENDIF
\STATE $\omega_{s} \gets 0$     
\COMMENT{walk segment length}                                   
\STATE $Walk_{0} \gets \{ {\underline \varsigma}_0 \}$
\COMMENT{new walk segment}
\WHILE {$\Theta({\underline \varsigma^*})  > \Theta^{ub}_L$ \textbf{and} $\omega_{s} < \omega_{lmt}$}
    \IF [timeout]{$t \geq t_{lmt}$}
        \STATE {\bf break}
    \ENDIF
    \STATE $\omega_{s}=\omega_{s}+1$
    \COMMENT{{\bf a new step!}}
    \STATE $Walk_{\omega_{s}}\!:\!{\underline \varsigma}_{\omega_{s}}\!:\!\Theta({\underline \varsigma}_{\omega_{s}}) \gets$
    \STATE { \hspace{0.5cm} $ \gets {\tt newPivot.saw}({\underline \varsigma}_{\omega_{s} - 1}, Walk_{\omega_{s} - 1})$}
     \IF { $\Theta({\underline \varsigma}_{\omega_{s}}) \le \Theta({\underline \varsigma^*})$} 
       \STATE ${\underline \varsigma^*}\!:\!\Theta({\underline \varsigma^*}) \gets {\underline \varsigma_{\omega_{s}}}\!:\!\Theta({\underline \varsigma}_{\omega_{s}})$
     \ENDIF
\ENDWHILE
\STATE {\bf return} $\omega_{s}\!:\!{\underline \varsigma^*}\!:\!\Theta({\underline \varsigma^*})$
\ENDPROCEDURE
\end{algorithmic}
\end{minipage}
}
\subfloat{
\begin{minipage}[t]{0.49\linewidth}
\label{fg_global_search_c}
\begin{algorithmic}[1]
\PROCEDURE{\CALL{newPivot.saw}{${\underline \varsigma}_{\omega_{s} -1},Walk_{\omega_{s} -1}$}}
\STATE $\mathbb{Z} \gets i = 1,2, \ldots ,L $ 
\STATE $\mathbb{Z}_p \gets permute(\mathbb{Z}) $ 
\STATE ${\cal N} ({\underline \varsigma}_{\omega_{s} -1}) \gets  \{{\underline \varsigma}_{\omega_{s} -1}^i | d({\underline \varsigma}_{\omega_{s} -1}, {\underline \varsigma}_{\omega_{s} -1}^i ) = 1, i \in \mathbb{Z}_p\}$ 
\STATE ${\cal N}_{saw} ({\underline \varsigma}_{\omega_{s} -1}) \gets  \{ {\cal N} ({\underline \varsigma}_{\omega -1}) |
                      {\underline \varsigma}_{\omega_{s} -1}^i \not\in  Walk_{\omega_{s} -1}\}$
   \IF {\strut  ${\cal N}_{saw}  ({\underline \varsigma}_{\omega_{s} -1}) \not= \emptyset$}  
        \STATE ${\underline \varsigma}_{\omega_{s}}\!:\!\Theta({\underline \varsigma}_{\omega_{s}}) \gets {\tt bestNeighbor}({\cal N}_{saw} ({\underline \varsigma}_{\omega_{s} -1}))$
        \STATE $Walk_{\omega_{s}} \gets Walk_{\omega_{s}-1} \cup \{{\underline \varsigma}_{\omega_{s}}\}$
        \STATE $\tau \gets \tau + |~{\cal N}_{saw} ({\underline \varsigma}_{\omega_{s} -1})~| $ 
        \COMMENT{update $cntProbe$}
   \ELSE[\textbf{deal with a trapped pivot}]
        \STATE $\beta = \beta + 1$ 
        \STATE ${\underline \varsigma}_{\omega_{s}}\!:\!\Theta({\underline \varsigma}_{\omega_{s}}) \gets {\tt coordInit}()$ 
        \COMMENT{re-initialize}
        \STATE $Walk_{\omega_{s}} \gets \{ {\underline \varsigma}_{\omega_{s}} \}$
        \STATE $\tau \gets \tau + 1 $ 
        \COMMENT{update $cntProbe$}
    \ENDIF
\STATE \textbf{return} $Walk_{\omega_{s}}\!:\!{\underline \varsigma}_{\omega_{s}}\!:\!\Theta({\underline \varsigma}_{\omega_{s}})$ 
\ENDPROCEDURE 
\end{algorithmic}
\end{minipage}
}
\caption{A fully instrumented version of solver \lssOrel\  and two supporting procedures: {\tt walk.saw} and {\tt newPivot.saw}.
Procedure {\bf \tt lssOrel}:~as a single contiguous self-avoiding walk or as a sequence of contiguous self-avoiding walk segments. 
Procedure {\bf \tt walk.saw}:~Self-avoiding walk with a sequence of best pivot coordinates.
Procedure {\bf \tt newPivot.saw}:~Searching for the best new pivot under restrictions of SAW.}
\label{fg_global_search2}
\end{small}
\end{figure*}

\begin{figure*}[t!]
\begin{small}
\subfloat[\lssMAts\ solver, based on $\mathit{MA_{TS}}$ in~\cite{Lib-OPUS-labs-2009-ASC-Gallardo-memetic}.]{
\begin{minipage}[t]{0.49\linewidth}
\centering
\begin{algorithmic}[1]
\PROCEDURE{\CALL{lssMAts}{$\Theta^{ub}_L, t_{lmt}$}}
\FOR{$i\gets 1$ {\bf to} $popsize$} \label{alg_lssMAts_For1}
    \STATE $pop_i\gets$ \CALL{RandomBinarySequence}{$L$}
    \STATE \CALL{Evaluate}{$pop_i$}
\ENDFOR \label{alg_lssMAts_For2}
\STATE \colorbox{Gray}{\valueBest $\gets \CALL{ValueBest}{pop}$}
\WHILE {\colorbox{Gray}{ $t < t_{lmt}$ {\bf ~and~}  \valueBest $\,>\,$ \valueTarget} }
    \FOR{$i=1$ {\bf to} $\mathit{offsize}$}     
        \IF {recombination is performed ($p_X$)} \label{alg_lssMAts_Rec1}
            \STATE $parent_1 \gets $\CALL{Select}{$pop$}
            \STATE $parent_2 \gets $\CALL{Select}{$pop$}
            \STATE $\mathit{offspring_i}\gets$\CALL{Recombine}{$parent_1, parent_2$}
        \ELSE
            \STATE $\mathit{offspring}_i \gets $\CALL{Select}{$pop$}
        \ENDIF \label{alg_lssMAts_Rec2}
        \IF {mutation is performed ($p_m$)} \label{alg_lssMAts_Mut1}
            \STATE $\mathit{offspring}_i \gets $\CALL{Mutate}{$\mathit{offspring}_i$}
        \ENDIF \label{alg_lssMAts_Mut2}
        \STATE $\mathit{offspring}_i \gets$\CALL{TabuSearch}{$\mathit{offspring}_i$}
        \STATE \CALL{Evaluate}{$\mathit{offspring}_i$}
    \ENDFOR
    \STATE $pop \gets$\CALL{Replace}{$pop$, $\mathit{offspring}$} \label{alg_lssMAts_Sel}
    \STATE \colorbox{Gray}{\valueBest $\gets \CALL{ValueBest}{pop}$}
\ENDWHILE
\ENDPROCEDURE 
\end{algorithmic}
\label{alg_lssMAts_lssRRts-a}
\end{minipage}
}
\subfloat[\lssRRts\ solver, based on reduction of \lssMAts.]{
\begin{minipage}[t]{0.49\linewidth}
{\footnotesize	
The procedure \lssMAts\ on the left is an instrumented versions of the 
\labs\ solver named as $MA_{TS}$
in~\cite{Lib-OPUS-labs-2009-ASC-Gallardo-memetic}.
Settings of all parameters,
used also in our experiments, are described
in~\cite{Lib-OPUS-labs-2009-ASC-Gallardo-memetic}.
See a concise reprise below.
\par\vspace*{2ex}
\begin{tabular}{l l}
 {\bf setting} & {\bf value} \\
 \hline
 population size: & 100 \\
 mutation probability: & $2/(L+1)$ \\
 crossover probability: & 0.9 \\
 tournament selection size: & 2 \\
 crossover: & uniform \\
 tabu search walk length: & a random choice  \\
 ~~              &  from the range\\
 ~~              & {\small [$\frac{L}{2}, \frac{3 L}{2}$]}
\end{tabular}
}
\vspace{0.6cm}
\centering
\begin{algorithmic}[1]
\PROCEDURE{\CALL{lssRRts}{$\Theta^{ub}_L, t_{lmt}$}}
\STATE $pop_1\gets$ \CALL{RandomBinarySequence}{$L$}
\STATE \CALL{Evaluate}{$pop_1$}
\STATE \colorbox{Gray} {\valueBest $\gets \CALL{ValueBest}{pop}$}
\WHILE {\colorbox{Gray}{ $t < t_{lmt}$ {\bf ~and~}  \valueBest $\,>\,$ \valueTarget} }
        \STATE $\mathit{pop_1} \gets $\CALL{RandomBinarySequence}{$L$}
        \STATE $\mathit{pop_1} \gets $\CALL{TabuSearch}{$\mathit{pop_1}$}
        \STATE \CALL{Evaluate}{$\mathit{pop_1}$}
        \STATE \colorbox{Gray}{\valueBest $\gets \CALL{ValueBest}{pop}$}
\ENDWHILE
\ENDPROCEDURE 
\end{algorithmic}
\vspace{0.18cm}
\label{alg_lssMAts_lssRRts-b}
\end{minipage}
}
\caption{We illustrate two instrumented versions of the \labs\ solver named as $\mathit{MA_{TS}}$
in~\cite{Lib-OPUS-labs-2009-ASC-Gallardo-memetic}  
under the caption ``Pseudo code of the memetic algorithm'', Figure 5.
The instrumentation, highlighted with gray background, uses variable names defined in
Table~\ref{tb_notation_summary} and fits unobtrusively within the context of the original
pseudo code: it is designed to control solver termination not only with a runtime limit but also
by monitoring the solution quality in terms of a pre-specified upper bound. The procedure \lssMAts\ on the left describes the actual solver in our
experiments, adapted to solve instances of the \labs\ problem only for odd values of
$L$ under skew-symmetry. A more general \labs\ solver, for both even/odd values of $L$
and also described by the same pseudo code, has been named \lMAts.
The procedure \lssRRts\ on the right describes an alternative solver to \lssMAts\ -- it is also the name of the solver used in our experiments. While the
solver on the left relies on an established evolutionary algorithm  to initialize the tabu search, the modified solver uses randomly generated coordinates
to initialize the tabu search. This set-up allows us to investigate the performance of the tabu search alone.
}
\label{alg_lssMAts_lssRRts}
\end{small}
\end{figure*}

\par\vspace*{4ex}\noindent
{\bf Solver \lssOrel.}
A self-avoiding walk strategy implemented by \lssOrel\ 
follows a few simple rules:
(1) initialize a coordinate ${\underline \varsigma}$ and mark it as the `initial pivot';
(2) probe all unmarked adjacent coordinates, then select and mark the coordinate with the 'best value' as the new pivot.
In the case when more than one adjacent coordinates have ’best value’, one of them is randomly selected as a new pivot; 
(3) continue the walk until either 
$\Theta({\underline \varsigma}^*)  \le \Theta^{ub}_L$ or the walk is blocked by adjacent coordinates that are already pivots;
(4) if the walk is blocked or its length exceeds a threshold, initialize a new coordinate ${\underline \varsigma}^{i}$ as a `new initial pivot'
and restart a new walk segment;
(5) manage the memory constraints with  
an efficient data structure such as a hash table.
For examples of contiguous and non-contiguous self-avoiding walks, see
Figure~\ref{fg_walks_restarts_vs_saw}.

In Figure~\ref{fg_global_search2}
we present 
the fully instrumented pseudo code of solver \lssOrel.
The main procedure {\tt lssOrel} invokes 
{\tt walk.saw} which in turn invokes {\tt newPivot.saw}.
Depending on initial parameters, the procedure {\tt lssOrel}  returns the best solution from a 
single contiguous self-avoiding walk or a sequence of contiguous self-avoiding walk segments.
The procedure {\tt walk.saw} makes a contiguous self-avoiding walk segment
as a sequence of best pivot coordinates, 
an arrangement formalized in Eq.~\ref{eq_coord_pivots}.
The procedure {\tt newPivot.saw} searches 
the distance=1 neighborhood as defined in Eq.~\ref{eq_neighborhood}
for the best new pivot under the
self-avoiding walk restrictions. 

\vspace{-0.02cm}
Since procedure {\tt newPivot.saw}
is the computationally most critical part of the solver,
we provide additional details.
The neighborhood search 
proceeds in randomized order (Step~3) to avoid inducing bias
in the order of best pivot selection. 
The Step 5 eliminates all adjacent coordinates that may have been 
used as pivots already and returns a neighborhood subset
${\cal N}_{saw}({\underline \varsigma}_{\omega -1})$.
To manage this search efficiently,
we use a hash table to store 
pivot coordinates $Walk_\omega$.
If the neighborhood subset is not empty, the procedure 
{\tt bestNeighbor} in Step 7 probes all coordinates in the subset and returns
the new pivot,
updates 
the walk list to $Walk_{\omega}$ in Step 8, and exits on Step 16.
An empty neighborhood implies that the self-avoiding walk is {\em trapped}, 
i.e.~all adjacent coordinates are already pivots -- an event not yet observed.
We complete the procedure with Steps 11,~12,~13.

\begin{figure*}[t!]
\vspace*{-6ex}
\centering
\subfloat{
\begin{minipage}[t]{0.45\linewidth}
\includegraphics[width=0.99\textwidth]{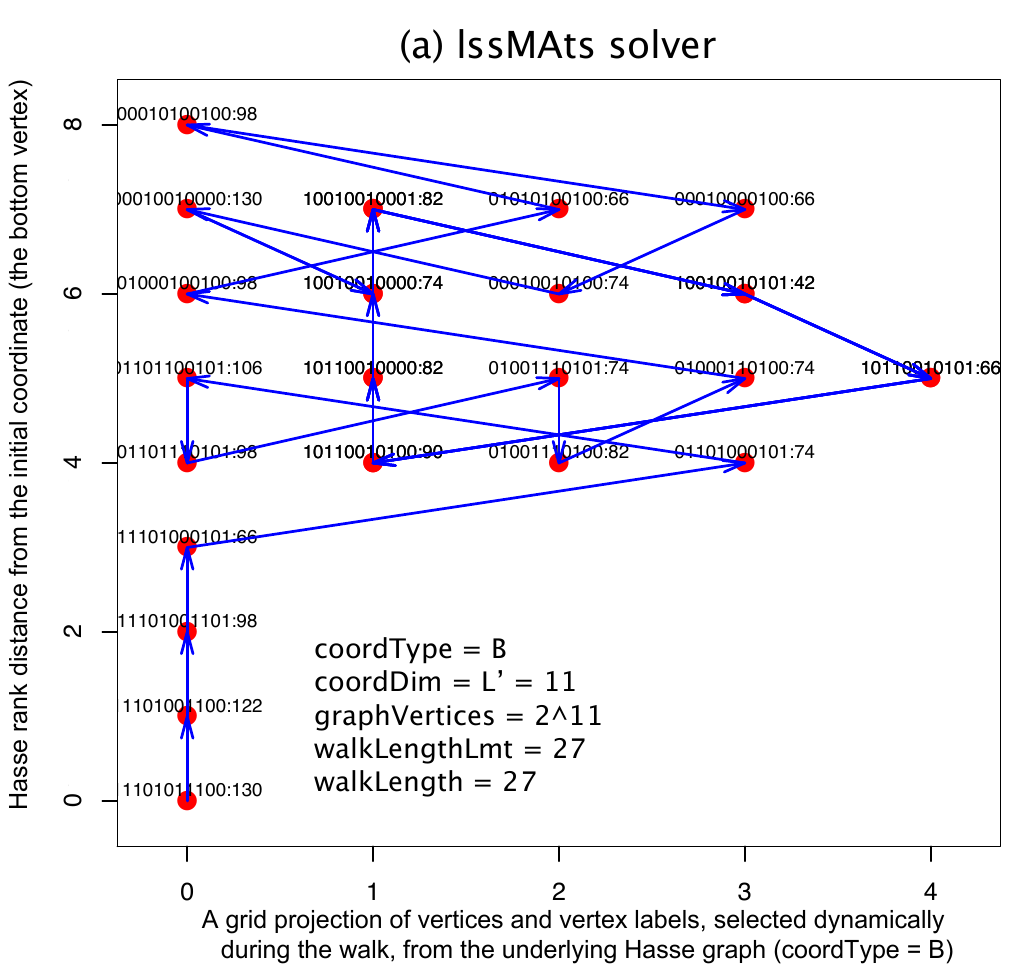}
\label{fg_walks_ts_vs_saw-a}
\end{minipage}
}
\subfloat{
\begin{minipage}[t]{0.45\linewidth}
\includegraphics[width=0.99\textwidth]{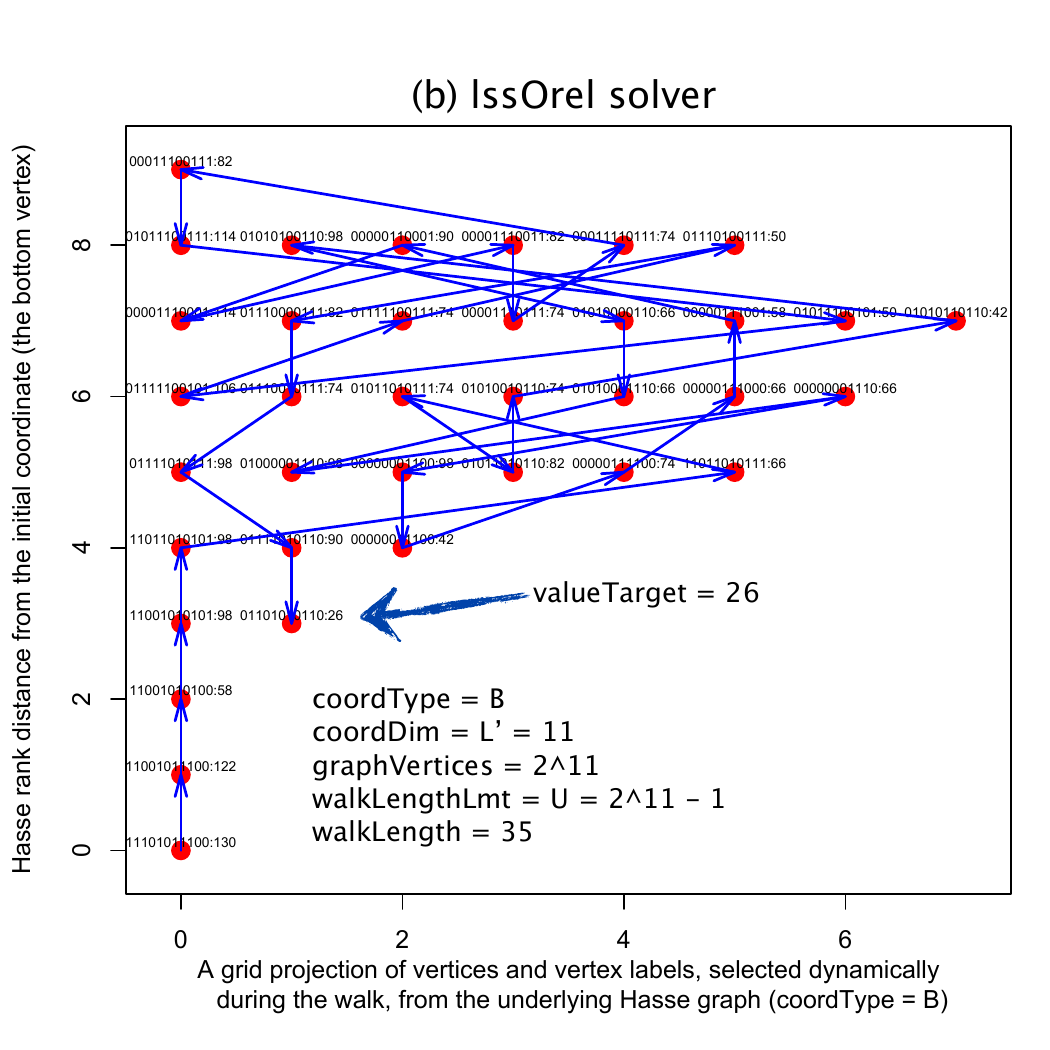}
\label{fg_walks_ts_vs_saw-b}
\end{minipage}
}
\caption{Two instances of walks created by two solvers, \lssMAts\ and \lssOrel.}
\label{fg_walks_ts_vs_saw}
\vspace*{-2ex}
\end{figure*}

\par\vspace*{2ex}\noindent
{\bf Solvers \lMAts, \lssMAts\ and \lssRRts.}
Both solvers \lMAts\ and \lssMAts\ (Figure~\ref{alg_lssMAts_lssRRts-a})
are instrumented versions of the \labs\ solver named as $\mathit{MA_{TS}}$ 
in~\cite{Lib-OPUS-labs-2009-ASC-Gallardo-memetic}.
These solvers, their pseudo code, and associated experiments and results, are 
described in~\cite{Lib-OPUS-labs-2009-ASC-Gallardo-memetic}.
Setting of control parameters 
in our experiments are identical to ones used in~\cite{Lib-OPUS-labs-2009-ASC-Gallardo-memetic}; 
a consise reprise of these setting is shown in the top-right part of
Figure~\ref{alg_lssMAts_lssRRts}.
Our instrumentation
is  highlighted in gray. We also added the {\em cntProbe} variable
which is not shown. 

\par\vspace*{1ex}
The solver \lssRRts\ (Figure~\ref{alg_lssMAts_lssRRts-b}) is a derivative of \lssMAts; we devised it
as a separate solver so we could
investigate the performance of the tabu search, as implemented in \lssMAts, without
its evolutionary component.

\par\vspace*{1ex}\noindent
{\bf Differences in \lssMAts\ and \lssOrel.}
A series of comprehensive
experiments in the next section reveals significant differences
between some of the solvers. Comparisons of most interest are  between 
\lssMAts\ and \lssOrel. We conclude this section with an illustrative example 
which
provides a modicum of explanation why such differences impact the asymptotic performance of both solvers.

Consider an instance of a \labs\ problem for $L = 21$ where we  take advantage of skew-symmetry to reduce the problem size to
$L' = 0.5*(L + 1) = 11$. The corresponding Hasse graph now has $2^{11} = 2048$ vertices and is too large to plot and trace
edges from each vertex to 11 of its neighbor vertices and their labels directly. However, when walk lengths are on the order
of 30--50 steps, we can project vertices and labels that have been visited in the underlying Hasse graph onto a uniform grid. 
In Figure~\ref{fg_walks_ts_vs_saw} 
we display two instances of such projections, based on two different walks returned by two solvers, \lssMAts\ and \lssOrel:
one walk terminates without finding the optimum solution, the other terminates upon finding the optimum solution, the pair
{\sf 01101010110:26}.

Both solvers start the respective walks from the same initial coordinate {\sf 11101011100}, a substring of length $L' = 11$,
which under rules of skew-symmetry expands into the full initial coordinate {\sf 111010111001101111101} of length $L = 21$ and
\labs\ energy value of 130. The labels associated with the initial vertex for each walk are given as the pair {\sf substring:value},
starting with {\sf 11101011100:130}. Both walks are shown in two grids: each grid represents a projection of vertices and vertex
labels, selected dynamically during the walk, from the underlying Hasse graph. The length of the walk is prescribed by the solver.

Under Case (a),  \lssMAts~selects the walk length {\em randomly} from the range  
$[L/2,~ 3L/2] = [\lfloor(L' - 0.5)\rfloor,~ \lfloor(3L' - 1.5)\rfloor] = [10,~ 31]$, 
and for the instance shown, the value of 27 has been selected. Under Case (b),  
\lssOrel\ walk is limited only by the upper bound $2^{L'} - 1$.
For this instance, \lssMAts\ terminates the walk after step 27 without finding the solution target value and therefore needs to
repeat the search from another coordinate. Moreover, the walk in \lssMAts~uses a tabu search strategy and is not self-avoiding in
this instance: six vertices form a cycle {\sf 10010010000:74, 10010010001:82, 10010010101:42, 10110010101:66, 10110010100:90,
10110010000:82, and 10010010000:74}. On the other hand, the self-avoiding walk in \lssOrel~continues for 35 steps and stops only
upon finding the solution target value: {\sf 01101010110:26}. 

In each case, the walk length depends not only on the initial coordinate but also on the initial randomly selected seed. 
With \lssMAts\ and the initial coordinate {\sf 11101011100}, runs with 32 random seeds  return walks of lengths in the range of [10, 31]
where only 14 walks terminate at the target solution value of 26. With \lssOrel~and the initial coordinate {\sf 11101011100},
runs with 32 random seeds  return walks of lengths in the range of [4, 226] where {\em all} 32 walks terminate at the target
solution value of 26.

Additional experiments can determine the more likely walk length means
of each solver, with each reaching the same target value. 
Given one thousand randomly selected initial coordinates and random seeds for $L=21$, the mean value of total walk length
returned by \lssMAts\ is 232.6 with the 95\% confidence interval [214.1, 251.1]. 
This statistics has considerable bias since \lssMAts\ has an advantage by relying on
population of 100 randomly initialized solutions before proceeding with the search proper, 
thereby finding 167 solutions that reach the target value of 26 with walk length of 0 and 833 solutions that reach the target value
with walk length $>$ 0. Now, the mean value of total walk length returned by \lssMAts\ based on 833 runs with walk length $>$ 0
and  under {\em multiple restarts},
is 279.2 with the 95\% confidence interval [258.4, 300.1].

\begin{figure*}[t!]
\vspace*{-6ex}
\centering
\subfloat{
\begin{minipage}[t]{0.46\linewidth}
\includegraphics[width=0.99\textwidth]{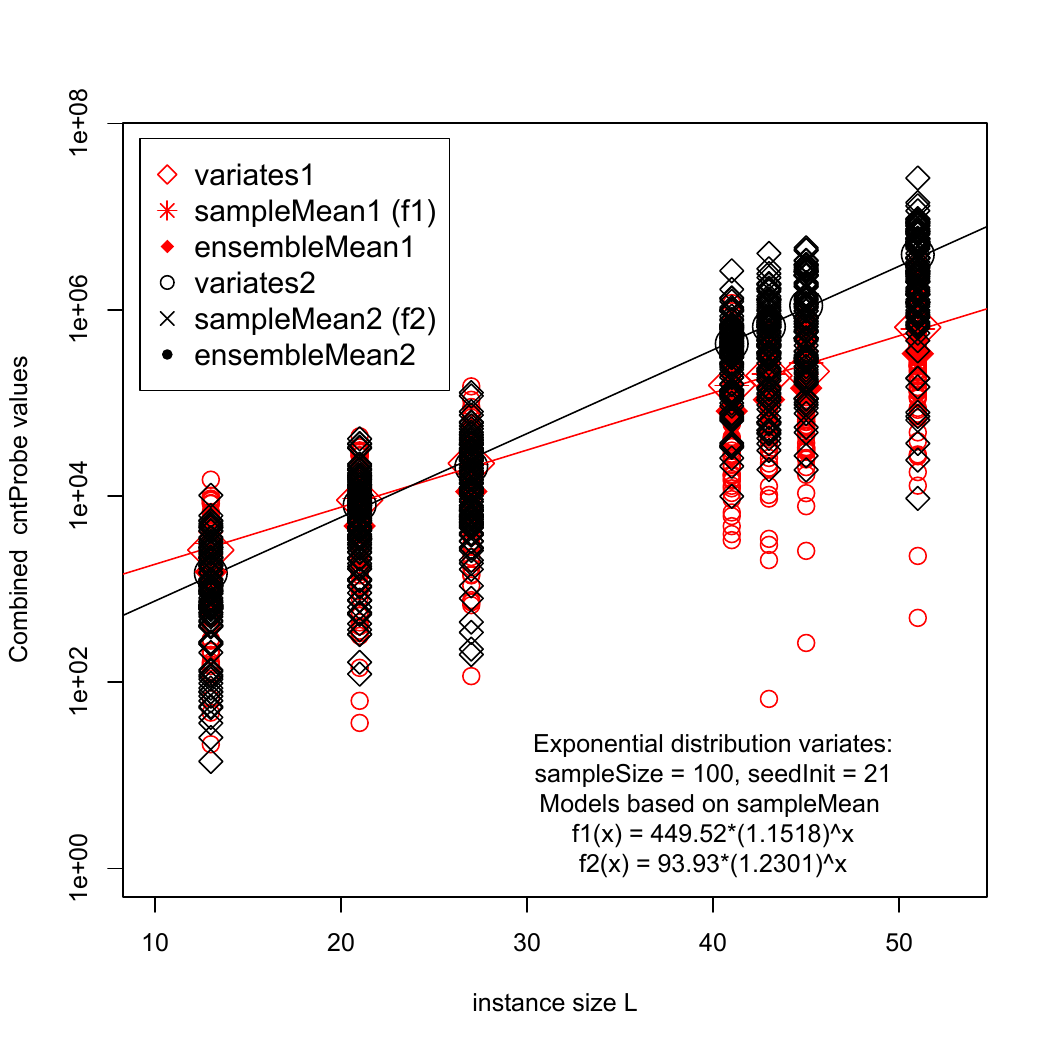}
\end{minipage}
}
\subfloat{
\begin{minipage}[t]{0.46\linewidth}
\includegraphics[width=0.99\textwidth]{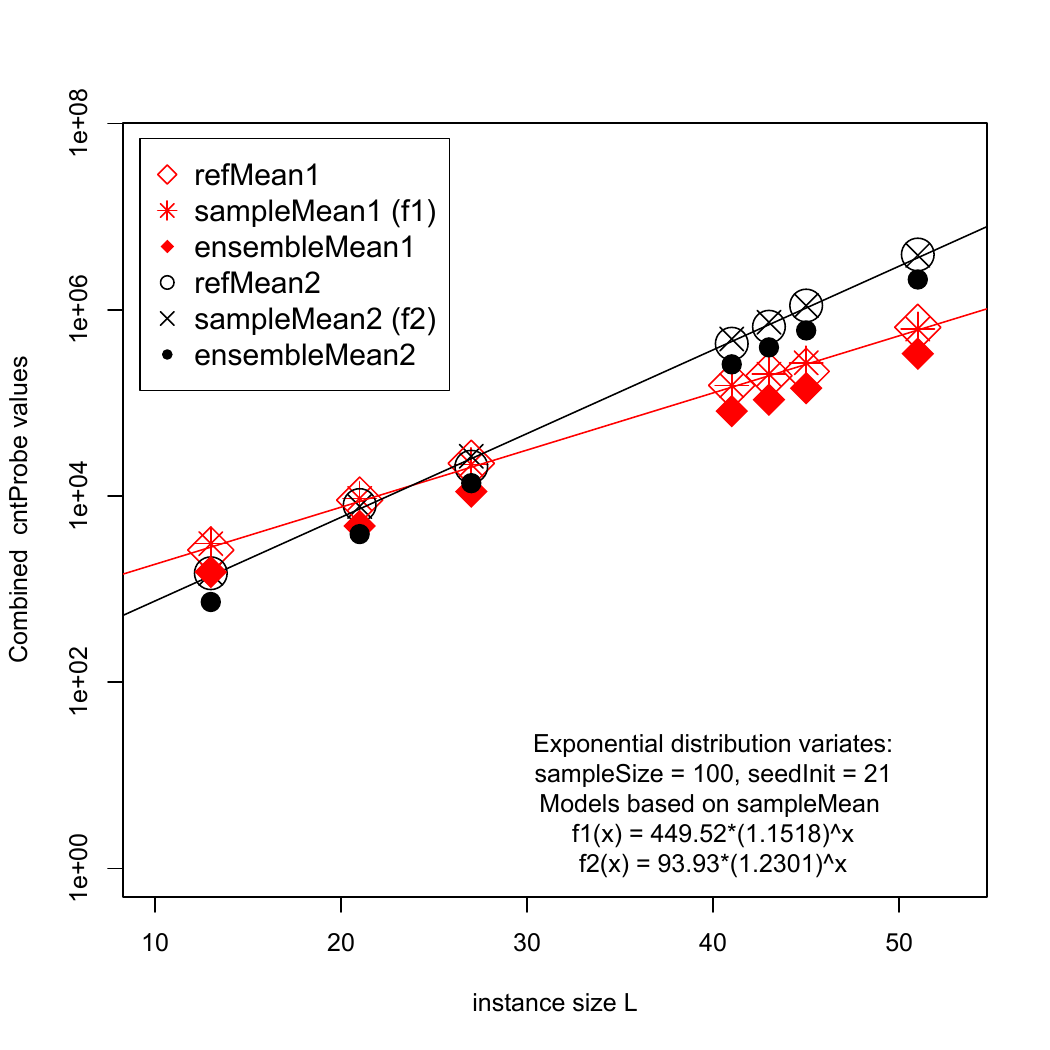}
\end{minipage}
}
\vspace*{-2ex}
\caption{A simulated asymptotic performance evaluation of two \labs\ solvers, given two reference models 
of {\em cntProbe} variable for the instance set $L = \{13, 21, 27, 41, 43, 45, 51\}$:
${\rm ref1} = 500*1.150^L$ and ${\rm ref2} = 100*1.230^L$.
There are $N = 100$ variates generated by each model for each $L$, all variates have exponential distribution. 
Sample means are based on 100 variates for each $L$, values reported for ensemble means are based on the best-fit model
with respect to all variates in the instance set.   
Due to  exponential distribution of variates, sample means are not equivalent to ensemble means which 
we would expect under normal or uniform variate distribution. In all of our experiments that follow, we shall use
sample means to model the asymptotic performance of various solvers. This experiment also demonstrates that
with the sample size of 100, we can reliably model the asymptotic performance differences of two \labs\ solvers.}
\label{fg_asymptotic_sim}
\vspace*{-2ex}
\end{figure*}

In comparison, when the same tests are applied to solver
\lssOrel, each of the one thousand walks terminate at the target value of 26 without a single restart: the  mean value of walk length is 97.3 with the 95\% confidence interval [93.6,~ 100.9].  
This mean value is significantly better than the mean value of the Hamiltonian (self-avoiding) walk.
Given that this instance has 4 minima, the  mean value of the Hamiltonian walk is $(1/8)2^{11} = 256$.
In conclusion, experiments with the instance shown in
Figure~\ref{fg_walks_ts_vs_saw} demonstrate that 
the solver which reduces  the repetition of coordinates during the stochastic 
search more effectively also achieves a  better average case performance. 
In comparison to \lssMAts, \lssOrel\ does  requires more memory.
In \lssMAts, memory is required only to define the tabu data structure
within the solution, the self-avoiding walk stores a path of already visited solutions. The experiments with \lssOrel\ in the next section demonstrate a restart strategy with larger instances so that \lssOrel\  memory remains fixed as well. However, as we show in our experiments, the self-avoiding walk outperforms
the tabu search consistently and the gap widens as the instance size increases.

\section{Summary of Experiments}
\label{sec_experiments}
\noindent
We use the {\em asymptotic performance experiment} 
-- as defined with a simulated
experiment in Figure~\ref{fg_asymptotic_sim} -- to  reliably compare 
the performance of two \labs\ solvers. 
By generating the asymptotic model for each solver, we not only readily compare
the two solvers, we use the model also {\em to predict} computational requirements
for maintaining {\em uncensored experiments}
as the instances size increases -- using also the metrics such as the
{\em  observed hit ratio} (Eq.~\ref{eq_hitO_r}).
We follow this methodology consistently, under the given computational resources and 
time constraints~\cite{Lib-OPUS2-ebook-xBed-2016-Brglez}.

We arrange our experiments into several groups.
Given that the hardest-to-solve instances have only 4 minima --
or equivalently a single canonic solution -- when $L$ is odd,
we restrict the asymptotic performance experiments to this subset of  $L$ only. If during the
experiment we find out that an instance has more than 4 minima, we exclude it from the test set. Similarly, when $L$ is even, 
the hardest-to-solve instances have only two canonic solutions;
we restrict the asymptotic performance experiments to this subset of  $L$ only.
Given the available resources, including runtime, experiments
that do not exploit the skew-symmetry of the labs function have been
limited to $L \le 87$. However, with solvers that do exploit the 
skew-symmetry, we could extend the experiments to $L \le 401$.

\begin{figure*}[t!]
\vspace*{-3ex}
\centering
\subfloat[branch-and-bound solver~\cite{Lib-OPUS-labs-2013-arxiv-Prestwich-branch-and-bound-odd}]{ 
\begin{minipage}{0.50\textwidth}
\includegraphics[width=0.99\textwidth]{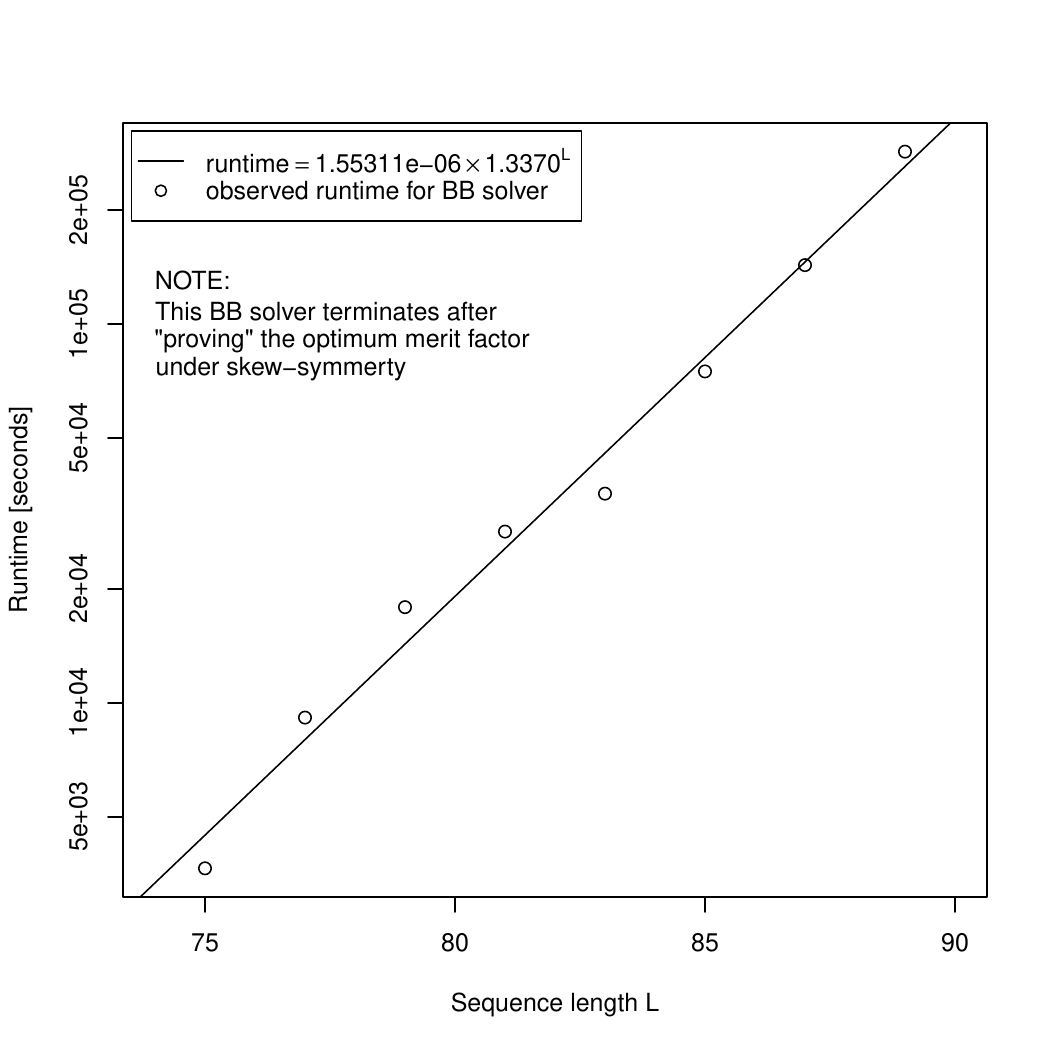}
\label{fg_BB-vs-others-asymptotes_a}
\vspace*{-1ex}
\end{minipage}
}
\subfloat[\lssMAts\ and \lssOrelE\ solvers]{
\begin{minipage}{0.50\textwidth}
\includegraphics[width=0.99\textwidth]{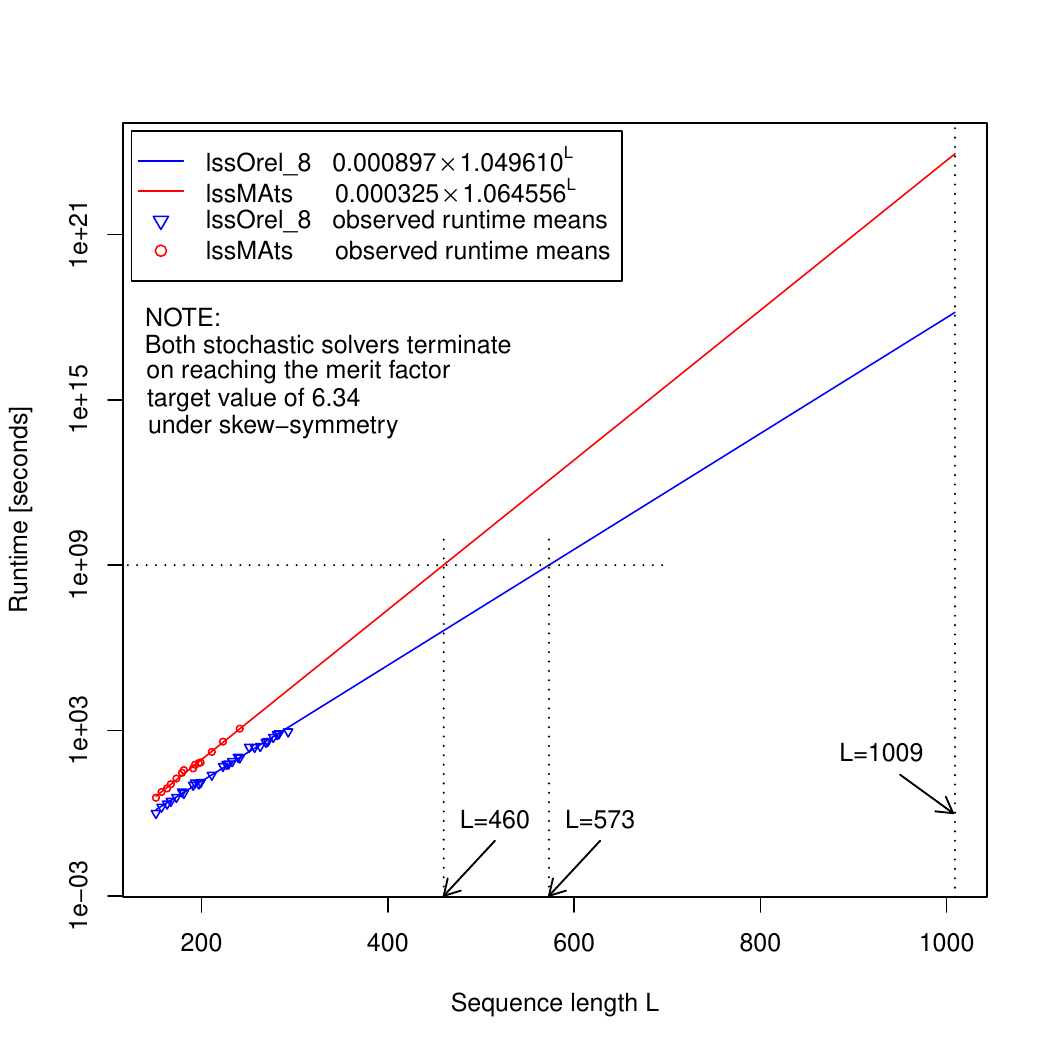}
\label{fg_BB-vs-others-asymptotes_b}
\vspace*{-1ex}
\end{minipage}
}
\vspace*{1ex}
\caption{
A runtime asymptotic comparison of the observed and projected performance of the state-of-the-art branch-and-bound \labs\ solver
under skew-symmetry~\cite{Lib-OPUS-labs-2013-arxiv-Prestwich-branch-and-bound-odd} versus the observed and projected
runtime of two stochastic solvers under skew-symmetry: \lssOrelE\ and \lssMAts. In (a), the objective of the BB solver is to find
(and prove) the optimal merit factor (under skew-symmetry) for increasing values of $L$;  the solution for $L=89$ reports
a runtime of 285326 CPU seconds whereas solvers \lssOrelE\ and \lssMAts\ reach the same merit factor in 
1.6869 and 3.2774 seconds on the average, with each average based on 100 {\em uncensored} trials. 
%
The primary objective in (b) is to determine the asymptotic runtime performance of the two stochastic solvers
while searching for sequences with the merit factor of 6.34:
such sequences can be readily found experimentally with 
constructive methods~\cite{Lib-OPUS-labs-2004-IEEE_TIT-Borwein,Lib-OPUS-labs-2004-IEEE_TIT-Parker-Legendre}.
Experimental results in Figure~\ref{fg_asymptotics}a demonstrate that for  $L = 1009$, the merit factor of approximately 6.34 can be reached in less than 1 CPU second while for $L = 4021$, a 
solution with comparable merit factor takes about 72 CPU seconds.
With stochastic solvers, computational bottlenecks  are observed already for $L > 400$: the {\em observed average runtimes} (based on 100 uncensored trials) 
with \lssOrelE\ rise from 103 CPU seconds at $L=241$ to 910 CPU seconds for $L = 293$ 
while \lssMAts\ requires 1142 CPU seconds on the average to reach 6.34 at $L=241$. 
By extrapolation, finding solutions with merit factor 6.34 for $L =  573$ requires $O(10^9)$ seconds
or around 32 years with \lssOrelE,
whereas with \lssMAts\ solutions of comparable quality are expected for $L =  460$ in the same time frame.
For the large value of $L=1009$,  the average runtime prediction to reach the merit factor of 6.34 with 
for \lssOrelE\ is 46774481153 years -- which exceeds the current estimate about the age of
the universe by a factor of 3.4.
This explains why the best reported merit factor with solver \lssOrelE, valued at
8.0668 for $L=241$ in Figure~\ref{fg_histogram}, is almost surely not optimal. 
}
\label{fg_BB-vs-others-asymptotes}
\end{figure*}

When measuring {\em runtime} precisely is important, we perform experiments
either on a PC or on a cluster of 22 processors~\cite{2013-Web-NCSU-HPC}, running under linux. 
In particular, we run the solver \lMAts\ on the cluster where we control the processor load by
running each solver instance serially -- while also scheduling the runs on 22 processors in parallel.
However, experiments with solvers \lssMAts, \lssRRts, and \lssOrel\ on largest instances are scheduled in
parallel and automatically on the grid with 100 processors, each solving an instance size of $L$ under
different random seeds and a {\em runtimeLmt} of 96 hours (4 days) for each instance.
The PC has an Intel processor i7, clock speed of 2.93 GHz, cache of 8 MB, and main memory of 8 GB.
The grid is a configuration of AMD Opteron processors 6272, clock speed of 2.1 GHz, cache of 2 MB, and main memory of
128 GB assigned to 64 cores~\cite{2014-Web-SLING}. When scheduled on the grid, processors run under variable load
factors and direct comparisons of solver  {\em runtime} are no longer possible. However, by instrumenting each solver with
the counter such as {\em cntProbe}, solver performance comparisons remain platform-independent. Note that this is only
the case when the cost of the evaluation function is the same for all solvers and the evaluation function is the most
compute-intensive method. Both of these criteria are satisfied in our experiments.

Before proceeding to details of experiments about individual solvers, we 
pause to make a realistic assessment about the runtime complexity of the \labs\ problem
in terms of observed and extrapolated experimental results in 
Figure~\ref{fg_BB-vs-others-asymptotes}. 
In contrast to the branch-and-bound solver 
searching for an optimal merit factor, the stochastic solvers \lssOrelE\ (number 8 determines
the maximum length of the self-avoiding walk segment $\omega_{lmt} = \omega_c * \frac{L+1}{2}$, where $\omega_c=8$) and \lssMAts\
search for 
sequences with the merit factor of 6.34:
such sequences can be produced in polynomial time with 
constructive methods~\cite{Lib-OPUS-labs-2004-IEEE_TIT-Borwein,Lib-OPUS-labs-2004-IEEE_TIT-Parker-Legendre}. The most important observations to infer from 
Figure~\ref{fg_BB-vs-others-asymptotes} are:
(1) runtime performance of branch-and-bound solver 
is inadequate to solve instances that are of current interest;
(2) both stochastic solvers start exhibiting computational bottlenecks for $L > 400$ even when the merit factor target values are relaxed and far from the best values.  
Neither \lssOrelE\ 
nor \lssMAts\ and least of all, the branch-and-bound solver, can be expected to find the conjectured optimal merit factors without faster computers and massive parallelism.

The best course of action at this time is to systematically learn 
about limitations of current solvers and to continue
with summaries of the following experiments:
(1) solver \lMAts\ for $L_{odd}$ and $L_{even}$, 
(2) best upper bounds of $L_{odd}$ without skew-symmetry,
(3) solver \lssOrelU\ (each solution is based on a single segment contiguous walk),
(4) solver \lssOrel\ with limited walk length,
(5) solver \lssOrelU\ versus \lssOrelE,
(6) solver \lssMAts\ versus \lssRRts,
(7) solver \lssOrelE\ versus \lssMAts, including asymptotic predictions and hit ratios, and
(8) solver \lssOrelE\ versus best known merit factors in the literature and
 {\em new best-value solutions} of the \labs\ problem,
(9) challenges for the next generation of labs solvers, the asymptotic view.\\

\begin{figure*}[t!]

\centering
\vspace*{-6ex}
\subfloat[The sample mean asymptotic model for {\em cntProbe}.]{
\begin{minipage}{0.46\textwidth}
\includegraphics[width=0.99\textwidth]{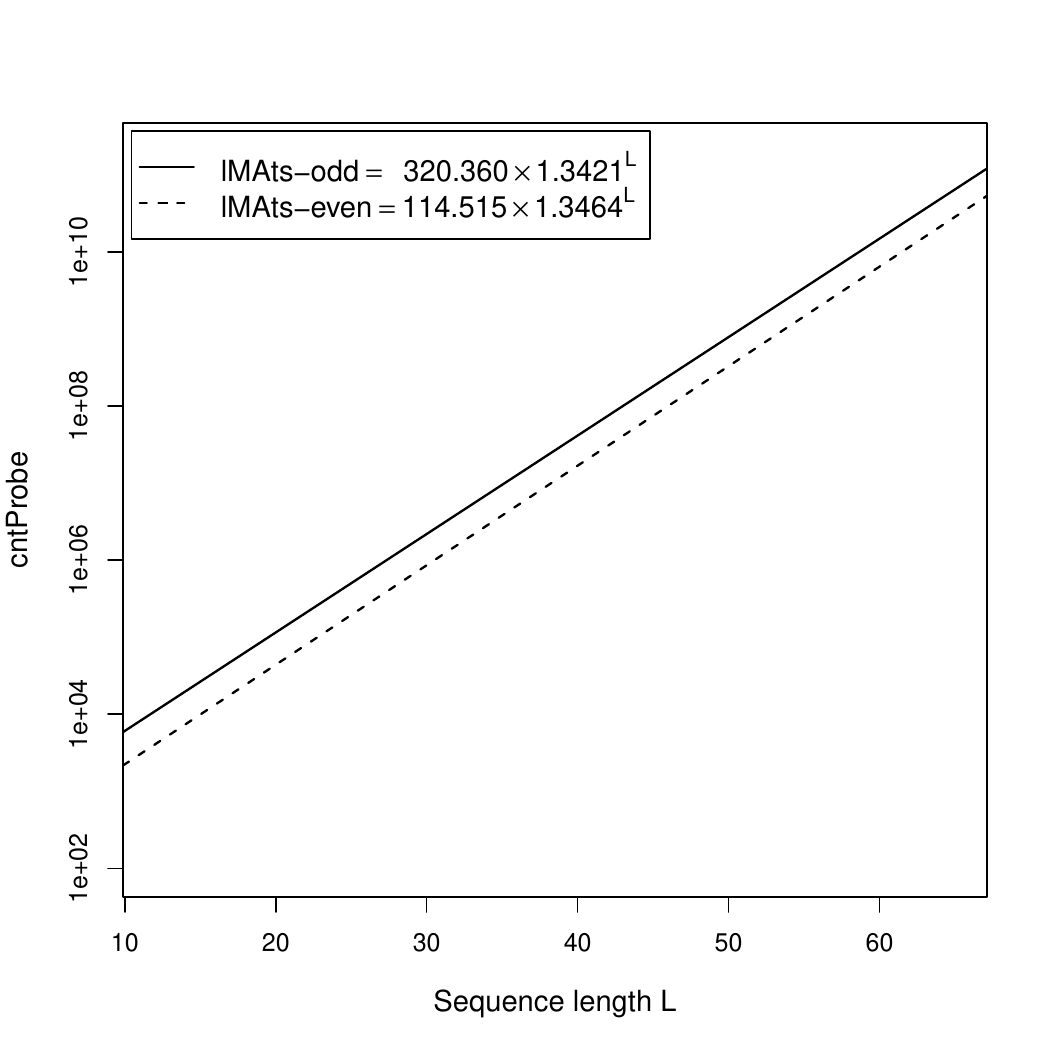}
\label{fg_lMAts-asymptotes_a}
\vspace*{-2ex}
\end{minipage}
}
\subfloat[The sample mean asymptotic model for {\em runtime}.]{
\begin{minipage}{0.46\textwidth}
\includegraphics[width=0.99\textwidth]{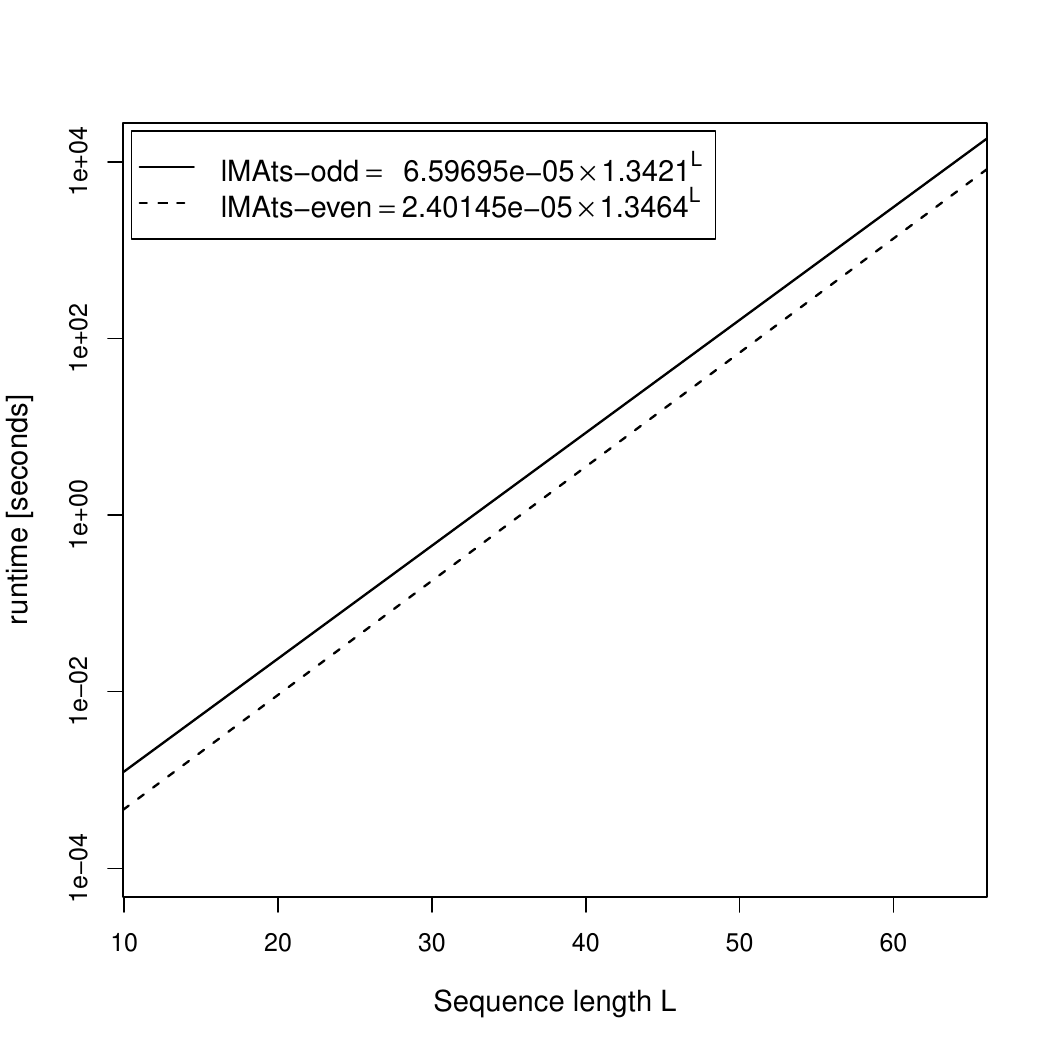}
\label{fg_lMAts-asymptotes_b}
\vspace*{-2ex}
\end{minipage}
}

\vspace*{-4ex}

\subfloat[For runtimLmt=10 hours and $L \le 64$, hitRatio=100\%.]{
\begin{minipage}{0.46\textwidth}
\includegraphics[width=0.99\textwidth]{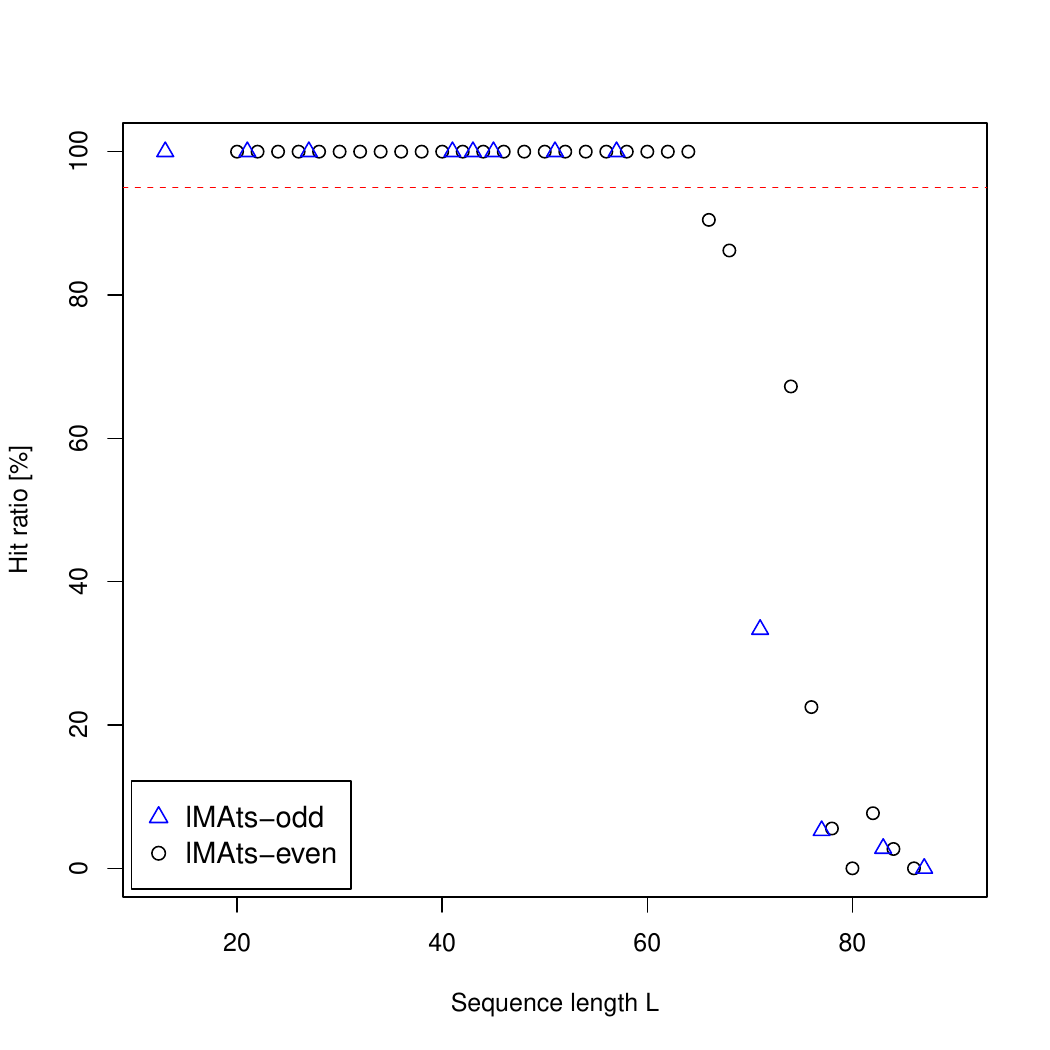}
\label{fg_lMAts-asymptotes_c}
\vspace*{-2ex}
\end{minipage}
}
\subfloat[Asymptotic comparisons to earlier solvers~\cite{Lib-OPUS2-labs-2003-FEA-Brglez-short,Lib-OPUS2-labs-2004-InfoSciences-Brglez-InPress}.]{
\begin{minipage}{0.46\textwidth}
\includegraphics[width=0.99\textwidth]{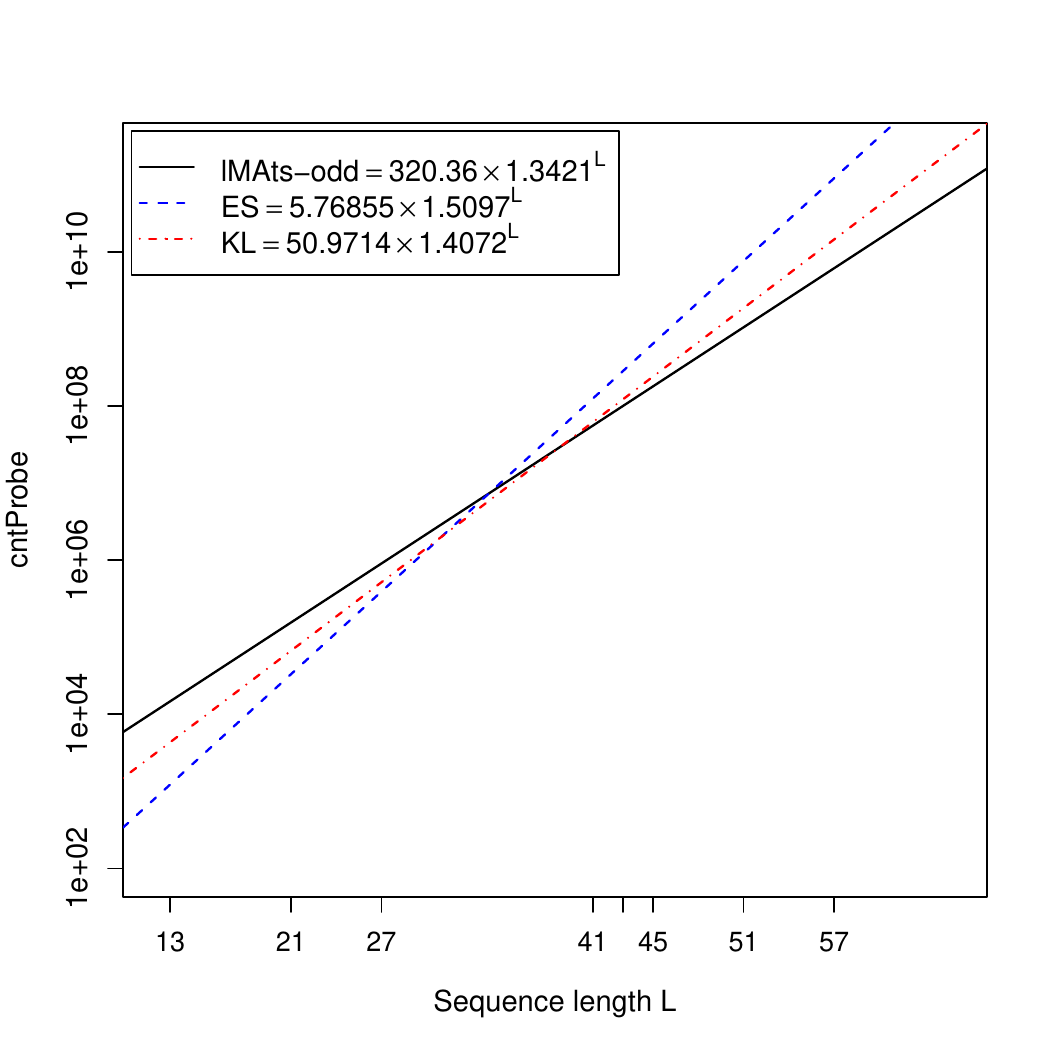}
\label{fg_lMAts-asymptotes_d}
\vspace*{-2ex}
\end{minipage}
}
\vspace*{2ex}
\caption{The asymptotic performance of solver \lMAts\ for sequence lengths $L$
from the subsets $L_{odd}$ and $L_{even}$.
As expected, instances from $L_{odd}$  take significantly more time to solve than
instances from  $L_{even}$. The sample mean asymptotic models
are based on sample size $N=516$ and rely only on runtime samples of $L$ where runtime is not censored,
i.e. for hitRatio = 100\%. For consistency with observed hitRatio in
Table~\ref{tb_lMAts_solvability}, the plot of hitRatio in this Figure also relies on the sample size of 100.
Finally, the solver \lMAts\ significantly outperforms the two earlier stochastic solvers,
{\tt ES} and {\tt KL}
from~\cite{Lib-OPUS2-labs-2003-FEA-Brglez-short,Lib-OPUS2-labs-2004-InfoSciences-Brglez-InPress}.
Here, the performance is measured not in {\em runtime} but the observed values of {\em cntProbe} which are
platform-independent and as such, still relevant 11 years after the initial experiments.
}
\label{fg_lMAts-asymptotes}
\end{figure*}

\par\vspace*{-3ex}\noindent
{\bf (1) Experiments with \lMAts.}
Experiments with solver \lMAts\ have been designed to illustrate its
asymptotic performance;
we summarize it  in 
Figure~\ref{fg_lMAts-asymptotes} and in
Table~\ref{tb_lMAts_solvability}.
We consider two specific subsets of sequence lengths $L$:
\begin{eqnarray}
L_{odd} = \{13, 21, 27, 41, 43, 45, 51, 57, 71, 77, 83, 87 \}
\label{eq_lMAts_odd}
\end{eqnarray}
\begin{eqnarray}
L_{even} =\{20, 24, 28, 32, 34, 36, 38, 40, 42, 44, 48, 52,\\
         54, 58, 60, 62, 64, 66, 68, 74, 76, 78, 80, 82,84, 86\}  \nonumber 
\label{eq_lMAts_even}
\end{eqnarray}

\noindent
For values of $L$ in the subset $L_{odd}$ there are only four optimal solutions, which reduce to a {\em single canonic solution} -- making these instances hardest-to-solve.
For values of $L$ in the subset $L_{even}$ there are only eight optimal solutions, which reduce to a {\em canonic solution pair} -- making these instances hardest-to-solve for even
values of $L$. 
There are four plots in Figure~\ref{fg_lMAts-asymptotes}:\\

\vspace*{-2ex}
\noindent
(a) Predictor models for observed sample mean of {\em cntProbe}
\begin{eqnarray}
{\mathit cntProbe(\lMAts)}_{odd}  &=& 320.360 * 1.3421^L \\
{\mathit cntProbe(\lMAts)}_{even} &=& 114.515 * 1.3464^L    \nonumber 
\label{eq_lMAts_cntProbe}
\end{eqnarray}\\

\begin{table*}[t!]
  \caption{Predictions versus observations from experiments with \lMAts,
  under the constraint of {\em runtimeLmt} of 10 hours for the listed
  values of $L$, each with the sample size of 100.
  For $L$ odd, we only consider  hardest-to-solve
  instances, each having a single canonic solution.
  For $L$ even, we also consider  the hardest-to-solve
  instances, each having two canonic solutions.
  The observed mean
  represents the sample mean based on the value of observed solvability,
  i.e. the sum total of runtimes of each instance. 
  The {\em observed number of hits}, $\mathit{hitO} = 100$, 
  signifies that none of the solutions have been censored. 
  The model mean values are
  computed from the two predictors based on empirical data described in
  Figure~\ref{fg_lMAts-asymptotes}. 
  The values of predicted solvability, 
  computed from Eq.~\ref{eq_solvP_ser}, are waiting times
  to reach {\em hitRatio} of 100\% with probability of 0.99 -- provided 
  (1) the solver has been scheduled  on a single processor to invoke on each instance serially,
  and (2) the solution produced by the solver has not been censored.
}
\label{tb_lMAts_solvability}
\begin{footnotesize}
\subfloat[L is odd]{
\begin{minipage}[t!]{0.49\textwidth}
 \begin{tabular}{c | c c | c c | c }
 ~  & model              & observed & predicted            & observed    & \\[-0.45ex]
 L  & mean$^*$           & mean     & solvability$^\dagger$ & solvability & $\mathit{hitO}^\ddagger$ \\
\hline
&&&& \\[-2ex] 
 57 & 0.3521 & 0.3734 & 43.92    & 37.34 & 100  \\
 71 & 21.66  & 8.345  & 2702     & 834.5 & 33  \\
 77 & 126.6  & 9.686  & 15791    & 968.6 & 5  \\
 83 & 739.9  & 9.887  & 92284    & 988.7 & 3  \\
 87 & 2400   & 10.0   & 299412   & 1000  & 0  \\ \hline 
 \end{tabular}
 \par\vspace*{1.5ex}
 \raggedright
 $^*$ 
 ${\rm runtime(\lMAts)}_{odd} = 1.832 * 10^{-8} * 1.3421^L$\\
 $^\dagger$ 
in hours, using Eq. \ref{eq_solvP_ser}; $N = 100, p = 0.99$
\\
$^\ddagger$
using Eq. \ref{eq_hitO}
\vspace{1.88cm}
\end{minipage}
}
\subfloat[L is even]{
\begin{minipage}{0.49\textwidth}
 \begin{tabular}{c | c c | c c | c }
 ~  & model              & observed & predicted            & observed    & \\[-0.45ex]
 L  & mean$^*$           & mean     & solvability$^\dagger$ & solvability & $\mathit{hitO}^\ddagger$ \\
\hline
&&&& \\[-2ex] 
 58 & 0.2071 & 0.2993 & 25.83  & 29.93 & 100  \\  
 60 & 0.3755 & 0.2656 & 46.83  & 26.56 & 100  \\
 62 & 0.6807 & 0.4877 & 84.90  & 48.77 & 100  \\
 64 & 1.233  & 1.167  & 153.9  & 116.7 & 100  \\
 66 & 2.236  & 3.502  & 279.0  & 350.2 & 90  \\
 68 & 4.055  & 4.264  & 505.7  & 426.4 & 86  \\
 74 & 24.15  & 6.355  & 3013   & 635.5 & 67  \\
 76 & 43.79  & 9.242  & 5461   & 924.2 & 23  \\
 78 & 79.38  & 9.829  & 9901   & 982.9 & 5  \\
 80 & 143.9  & 10.0   & 17949  & 1000  & 0  \\
 82 & 260.8  & 9.475  & 32538  & 947.5 & 7  \\
 84 & 472.9  & 9.817  & 58984  & 981.7 & 3  \\
 86 & 857.3  & 10.0   & 106927 & 1000  & 0  \\ \hline 
 \end{tabular}
 \par\vspace*{1.5ex}
 \raggedright
 $^*$ 
${\rm runtime(\lMAts)}_{even} = 6.671 * 10^{-9} * 1.3464^L$\\
\end{minipage}
}
\end{footnotesize}
\vspace*{-2ex}
\end{table*}

\par\vspace*{-5.5ex}\noindent
(b) Predictor models for observed sample mean of {\em runtime} (converted from seconds in the figure to hours here)
\begin{eqnarray} 
{\mathit runtime(\lMAts)}_{odd} &=& 1.832 * 10^{-8} * 1.3421^L~~~~ \\
{\mathit runtime(\lMAts)}_{even}&=& 6.671 * 10^{-9} * 1.3464^L~~~~   \nonumber   
\label{eq_lMAts_runtime}
\end{eqnarray}
\noindent
(c) Observed {\em hit ratio} as defined in Eq.~\ref{eq_hitO_r}. For {\em runtimeLmt} = 10 hours,
we can observe the {\em hit ratio} of 100\% up to $L=57$ for odd values of $L$ and up to $L=64$ for even
values of $L$. This implies that under current {\em runtimeLmt} we can not reliably measure average-case performance of solver \lMAts\ for larger values of $L$.\\ 
(d) The solver \lMAts\ significantly outperforms the two earlier stochastic solvers,
{\tt ES} and {\tt KL}~\cite{Lib-OPUS2-labs-2003-FEA-Brglez-short,Lib-OPUS2-labs-2004-InfoSciences-Brglez-InPress}.
Note that the observed values of {\em cntProbe}, which are
platform-independent, are still relevant 11 years after the initial experiments:
${\mathit cntProbe({\tt ES})}_{odd} =  5.76855 * 1.5097^L$
and
${\mathit cntProbe({\tt KL})}_{odd} =  50.9714 * 1.4072^L$.
Our current estimates of asymptotic performance are more accurate, since we 
process observations not as single ensemble but as two ensembles, one for odd  and the other for even values of $L$.

\par\vspace*{1ex}\noindent
{\em More about Figure~\ref{fg_lMAts-asymptotes}.} Predictor models for {\em cntProbe} and {\em runtime}
are based on a sample size of 516. The model mean is only an approximate predictor of the observed
sample mean -- it can underestimate as well as overestimate. 
For $L=57$, the observed runtimes range from 2 seconds to slightly more than 2 hours,
with the sample mean of 1340.4 seconds. However, when we report on sample means over five consecutive intervals, with 100 samples
in each interval, sample means range from 1155.3 seconds to 1624.9 seconds -- as anticipated in Eq.~\ref{eq_confidenceInterval}. 

For this series of experiments
we had access to  a cluster of 22 {\em unloaded processors} and could schedule executions in
parallel while still measuring runtimes that would be consistent with runtimes
we would observe serially on a single unloaded processor. Since runtime measurements under 1 seconds are not precise even for an unloaded processor,
we rely on near 100\% correlation with {\em cntProbe} and compute  {\em runtime} indirectly
for all values of $L$. 

\par\vspace*{-0.2ex}\noindent
{\em Predictions and observations in Table~\ref{tb_lMAts_solvability}.}
~We relate observations from experiments with \lMAts\ to
{\em observed hit ratio}, {\em predicted hit ratio}, and runtime models as defined by
Eqs.~\ref{eq_hitO},~\ref{eq_hitP_r}, and~\ref{eq_lMAts_runtime}.
The rapid decline in observed {\em hit ratio}, under the constraint of {\em runtimeLmt} of 10 hours 
can also be observed/predicted in this table and in Figure~\ref{fg_lMAts-asymptotes_c}.
The experiments with  solver \lMAts\ define
the  methodology when focusing on the performance of  solvers 
\lssOrel\ and \lssMAts\ under skew-symmetry. Again, we  define groups for  $L_{odd}$ 
to arrange the sequence of our experiments.
\begin{table*}[!t]
\caption{Pairs of best upper bound values  on \labs\  energies for a  subset of odd values of $L$.
The first number is the best value achieved under coordinates with skew-symmetry;
the adjacent number in brackets gives the number of canonic solutions under skew-symmetry.
The second number is the best value achieved with coordinates that are not skew-symmetric;
the adjacent number in brackets gives the number of canonic solutions that are not skew-symmetric.
The significance 
is this:
solutions under skew-symmetry can be equivalent to solutions without skew-symmetry.
Entries such as {\bf 15}(2)/--, {\bf 26}(1)/--, etc. imply that for $L=15$, $L=21$, etc.
{\em all} solutions are  skew-symmetric; the ones of most interest in this group are 
solutions where the number of canonic solutions is 1, forming {\em a primary group}:
$L = 5, 7, 11, 13, 21, 27, \dots$.
For $L \ge 101$, 
all solution values represent 
the 'best-known values under skew-symmetry'. Values marked with *
are an improvement on values posted by Knauer in 2002, now accessible
under~\cite{Lib-OPUS2-labs-2014-homepage-Knauer}.
We make no attempt to find new and improved solutions for all values of L
in~\cite{Lib-OPUS2-git_labs-Boskovic},
except  to show that 
solver \lssOrel\ consistently returns improved skew-symmetric solutions  
vis-\`{a}-vis Knauer's solutions without skew-symmetry.
For details, see 
Table \ref{tb_new_records} in Section \ref{sec_experiments}.
Entries shown as `?' imply that no `best solutions' have been reported at this time.
}
\label{tb_L_asymptotic}
\begin{footnotesize}
\begin{center}
\begin{tabular}{@{}p{4.5in}@{}p{2.6in}@{}} 
\begin{tabular}{@{}p{4.5in}@{}}
\begin{scriptsize}
\begin{tabular}{@{}c | c c c c c@{}}
\bf{L}    & \bf{1} & \bf{3} & \bf{5} & \bf{7} & \bf{9} \\ \hline
\bf{0}   & -- & -- & {\bf 2}(1)/-- & {\bf 3}(1)/--& {\bf 12}(2)/12(4)	 \\
\bf{10}	 & {\bf 5}(1)--  & {\bf 6}(1)/--  & {\bf 15}(2)/-- & {\bf 32}(1)/32(10) & 33(2)/29(2)	 \\		
\bf{20}	 & {\bf 26}(1)/-- & 51(4)/47(6) & 52(1)/36(2) & {\bf 37}(1)/-- & {\bf 62}(2)/-- \\
\bf{30}	 & 79(1)/67(2) & 88(2)/64(2) & 89(2)/73(2) & 106(1)/86(2) & {\bf 99}(2)/--	 \\
\bf{40}	 & {\bf 108}(1)/-- & {\bf 109}(1)/-- & {\bf 118}(1)/-- & {\bf 135}(5)/135(2) & {\bf 136}(1)/136(2) \\	
\bf{50}	 & {\bf 153}(1)/-- & {\bf 170}(1)/170(1)	 & {\bf 171}(2)/--	& {\bf 188}(1)/-- & {\bf 205}(1)/205(2)	 \\	
\bf{60}	 & 230(1)/226(2)	 & 271(3)/207(2)	 & 272(4)/240(2) & {\bf 241}(3)/241(2) & 282(1)/274(1)		 \\			
\bf{70}	 & {\bf 275}(1)/- & 348(2)/308(2) & 341(1)/329(2) & {\bf 358}(1)/-- & 407(5)/339(1)		 \\	
\bf{80}	 & 400(1)/372(1)	 & {\bf 377}(1)/-- &  442(1)/414(1) &  451(1)/431(1)	 &  484(2)/432(1)		 \\			
\bf{90}	 & {\bf 477}(1)/-- & 502(3)/486(1) & {\bf 479}(1)/--  & {\bf 536}(1)/536(1)  & {\bf 577}(1)/--  \\
\bf{100} & 578(1) & 555(1) & 620(1)  & 677(1)  & 662(1)  \\			
\bf{110}	 & 687(1) & 752(2) & 745(1)  & 786(1)  & 835(1)  \\
\bf{120}	 & 844(1) & 893(1) & 846(1)  & 887(1)  & 920(1)  \\
\bf{130}	 & 913(1) & 1010(1)   & 1027(1)    & 1052(1)    & 1133(1)   \\
\bf{140}	 & 1126(2) & 1191(1)   & 1208(1)    & 1265(2)    & 1218(1)  \\
\bf{150}	 & 1275(4) & 1340(1) & 1437(1) & 1366(1) & 1439(1)	 \\
\bf{160}	 & 1512(2)$^*$ & 1529(1) & 1474(1) & 1563(1) & 1532(1)	 \\
\bf{170}	 & 1677(1) & 1598(1)* & 1687(1)* & 1648(1)* & 1761(1)*	 \\
\bf{180}	 & 1834(1)$^*$ & 1859(2)* & 2028(1) & 1973(1) & 1966(1)	 \\	
\bf{190}	 & 2191(1) & 2272(1) & 2281(1) & 2218(1) & 2275(1)	 \\	
\bf{200}	 & 2380(1)$^*$ & 2421(1) & 2662(1) & 2695(1) & 2664(1) \\								
\bf{210}	 & 2801(1) & 2698(1) & 2691(1)$^*$ & 3036(1) & 3189(1) \\					
\bf{220}	 & 2758(1)$^*$ & 3215(1) & 3416(1) & 3409(1) & 3474(1)	 \\	
\bf{230}	 & 3587(1) & 3692(1) & 3757(1) & 3590(1) & 3711(1)	 \\					
\bf{240}	 & 3600(1)$^*$ & 4073(1) & 4098(1) & 4291(1) & 3812(1)$^*$ \\	
\bf{250}	 & 4165(1) & 4382(1) & 4463(1) & 4472(1) & 4145(1)$^*$ \\					
\bf{260}	 & 4338(1)$^*$ & 4803(1) & 4948(1) & 5037(1) & 4950(1)	 \\					
\bf{270}	 & 4871(1)$^*$ & ? & ? & ? & ? \\
\bf{280}	 &  5260(1)$^*$ & 5333(1)$^*$ & 5790(1) & ? & ?  \\	
\bf{300}	 & 6054(1) & 6335(1)$^*$ & ? & ? & ?	 \\	
\bf{340}	 & 8378(1) & ? & ? & ? & ?	 \\					
\bf{380}	 & 10238(1) & ? & ? & ? & ?	 \\		
\bf{400}	 & 11888(1) & ? & ? & ? & ?	 \\					
\end{tabular}
\end{scriptsize}
\end{tabular}
&
\mbox{{
\begin{tabular}{@{}p{2.6in}@{}}
The {\em primary group} of the hardest-to-solve instances includes the following values of $L$: 
\{$5^{\dagger}$, $7^{\dagger}$, $11^{\dagger}$, $13^{\dagger}$, 21, 27, $41^{\dagger}$, $43^{\dagger}$, 45, 51, 57, $71^{\dagger}$, 77, $83^{\dagger}$, 91, 95, $97^{\dagger}$, 99, $101^{\dagger}$, $103^{\dagger}$, 105\}.
The instance $L = 105$ is the largest instance where the solver
\lssOrel~ does not exceed the maximum memory limit of 8 GB and completes as 
{\em uncensored}  a self-avoiding walk without a single restart from each randomly assigned initial coordinate.
\par\vspace*{1ex}
The {\em secondary group} of the hardest-to-solve instances includes the following values of $L$: 
\{$71^{\dagger}$, 77, $83^{\dagger}$, 91, 95, $97^{\dagger}$, 99, $101^{\dagger}$, $103^{\dagger}$, 105, $107^{\dagger}$, $109^{\dagger}$, 111, 115, 117, 119, 121, 123, 125, $127^{\dagger}$\}. 
The instance $L = 127$ is the largest instance where
the solver \lssOrel\ completes 100 {\em uncensored} performance evaluations (initialized with 100 random seeds)  within 2 days on our PC.
\par\vspace*{1ex}
The {\em tertiary group} in our testing includes instances with the following values of $L$: 
\{141, $151^{\dagger}$, 161, $181^{\dagger}$, 201, 215, 221, $241^{\dagger}$, 249, 259, 261, $271^{\dagger}$, $281^{\dagger}$, $283^{\dagger}$, 301, 303, 341, 381, $401^{\dagger}$\}.
Here, the solver \lssOrel~ reports on the best bound returned in 4 days of computing under random initial seeds on 100 CPU units.
\par\vspace*{1ex}
Values marked with ${\dagger}$ are prime numbers.
\par\vspace*{1ex}
Finally, not all solutions with coordinates that are skew-symmetric are also optimal: such solutions
have been observed for $L =$ 19, 23, 25, 31, 33, 35, 37, 61, 63, 65, 69, 73, 75, 79, 81, 85, 87, 89 and 93.
\end{tabular}
}}
\end{tabular}
\end{center}
\end{footnotesize}
\vspace*{-2ex}
\end{table*}

\par\vspace*{+0.5ex}\noindent
{\bf (2) Best upper bounds pairs for $L_{odd}$.}
The experiments with \lMAts\ show the importance of (a) separating  instances for $L_{odd}$ from  instances for $L_{even}$, and (b),
separating instances in both $L_{odd}$ and $L_{even}$ into additional groups: a group with single canonic solution for  $L_{odd}$,
a group with a single pair canonic solutions for $L_{even}$, and groups with more canonic solutions. For experiments that follow
for $L_{odd}$, we again define the hardest-to-solve instances as having a single canonic solution, and divide them into two groups,
primary and secondary:
\begin{eqnarray}
L_{prim} = \{5, 7, 11, 13, 21, 27, 41, 43, 45, 51, 57,\nonumber \\
             71,77, 83, 91, 95, 97, 99, 101, 103, 105\}
\label{eq_L_primary}
\end{eqnarray}
\begin{eqnarray}
& L_{secd} = \{71, 77, 83, 91, 95, 97, 99, 101, 103, 105,  \nonumber \\
          & 107,109, 111, 115, 117, 119, 121,123, 125, 127\}                  
\label{eq_L_secd}
\end{eqnarray}
The instance $L_{prim} = 105$ is the largest instance where the solver
\lssOrel\ does not exceed the maximum memory limit of 8 GB on our PC and completes as 
{\em uncensored}  a self-avoiding walk without a single restart from each randomly assigned initial coordinate. We observe a single canonic solution for each value of $L$ in this group,
so we can formulate an asymptotic predictor model based on sample means, similarly to
Eqs.~\ref{eq_lMAts_cntProbe} and~\ref{eq_lMAts_runtime}.
The instance $L_{secd} = 127$ is the largest instance where
the solver \lssOrel~ completes 100 {\em uncensored} performance evaluations (initialized with 100 random seeds) 
within 2 days ($\sum_{i=1}^{100} runtime_{i} < 2~days$) on our PC; i.e. observing a {\em hit ratio of 100\%} and returning a mean value for the sample size of 100. Again, we observe a single canonic solution for each value of $L$ in this group,
so again we can formulate an asymptotic predictor model based on sample means.

For the remainder, we consider the tertiary group, with all experiments performed on the grid~\cite{2014-Web-SLING}.

\begin{eqnarray}
\hspace*{-1.0em}
L_{tert} = \{141, 151, 161, 181, 201, 215, 221, 241, 249, \nonumber \\
            259, 261, 271, 281, 283, 301, 303, 341, 381, 401\}
\label{eq_L_tertiary}
\end{eqnarray}

\noindent
We place $L_{tert} = 141$ into the tertiary group since the number of observered canonical solutions is greater than~1.
The instance $L_{tert} = 151$ is the smallest instance where we no longer observe a 
hit ratio of 100\% with the solver \lssOrel\ within {\em runtimeLmt = 4 days}.
With exception of $L_{tert} = 141$, solutions in this group associate with 
{\em a distribution of merit factors} rather than a single best value; see 
Table~\ref{tb_lssOrel_lssMAts_cntProbe_solvability} and
Figure~\ref{fg_merit_factors_and_solvability} later in this section.
For these instances, we can only report the best figure of merit;
the probability that the associated sequence is either optimal or near-optimal is
almost 0 as the instance size increases.

See Table~\ref{tb_L_asymptotic} for a summary of best known upper bound values on \labs\  energies for given subsets of $L_{odd}$. Up to $L = 99$, these energies are listed as pairs: 
the first number represents the best value achieved under coordinates with skew-symmetry,
the adjacent number in brackets gives the number of canonic solutions under skew-symmetry.
The second number represents the best value achieved with coordinates that are not skew-symmetric;
the adjacent number in brackets gives the number of canonic solutions that are not skew-symmetric.
In 2002, Knauer posted the `best-value solutions' for $L > 101$ 
without the restriction of skew-symmetry~\cite{Lib-OPUS2-git_labs-Boskovic};
results in Table~\ref{tb_L_asymptotic} show that
our skew-symmetry solver \lssOrel\  consistently returns improved skew-symmetric solutions  
vis-\`{a}-vis Knauer's solutions without skew-symmetry, and then a few more.

\begin{figure*}[t!]
\centering
\vspace*{-7ex}
\subfloat[$L$ versus {\em cntProbe}]{
\begin{minipage}{0.40\textwidth}
\includegraphics[width=0.99\textwidth]{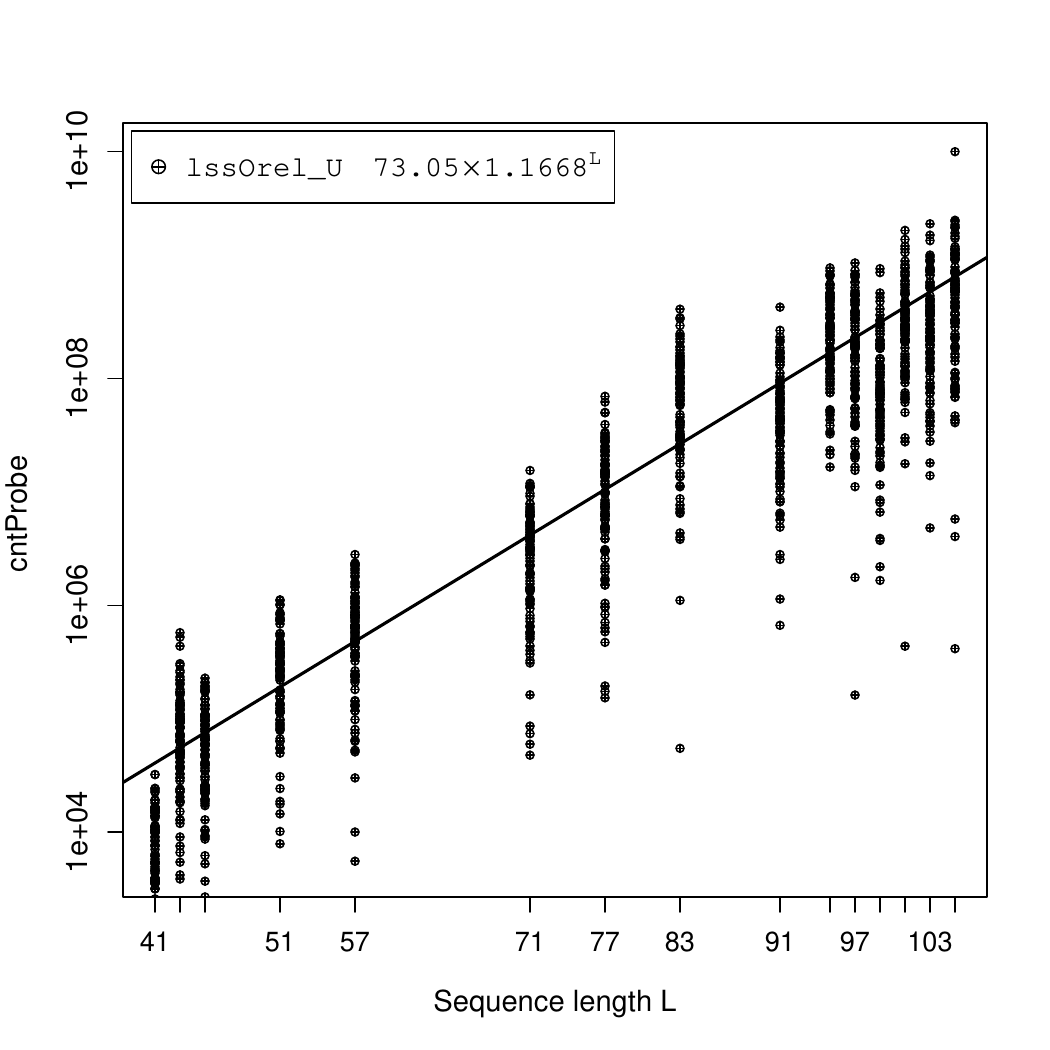}
\label{fg_model_mean-a}
\vspace*{-2ex}
\end{minipage}
}
\subfloat[Correlating {\em runtime} to {cntProbe}]{
\begin{minipage}{0.40\textwidth}
\includegraphics[width=0.99\textwidth]{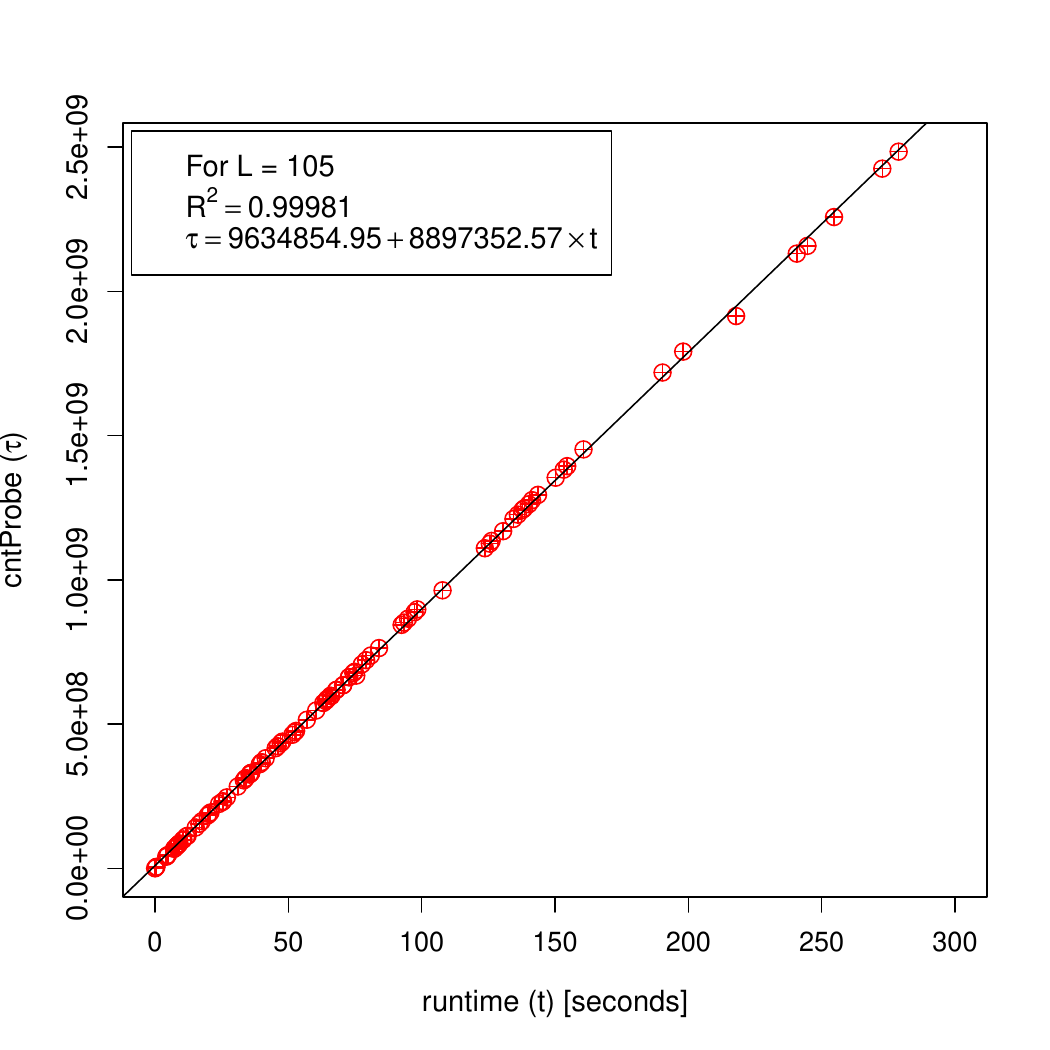}
\label{fg_model_mean-b}
\vspace*{-2ex}
\end{minipage}
}

\vspace*{-5ex}
\subfloat[$L$ versus {\em walkLength}]{
\begin{minipage}{0.40\textwidth}
\includegraphics[width=0.99\textwidth]{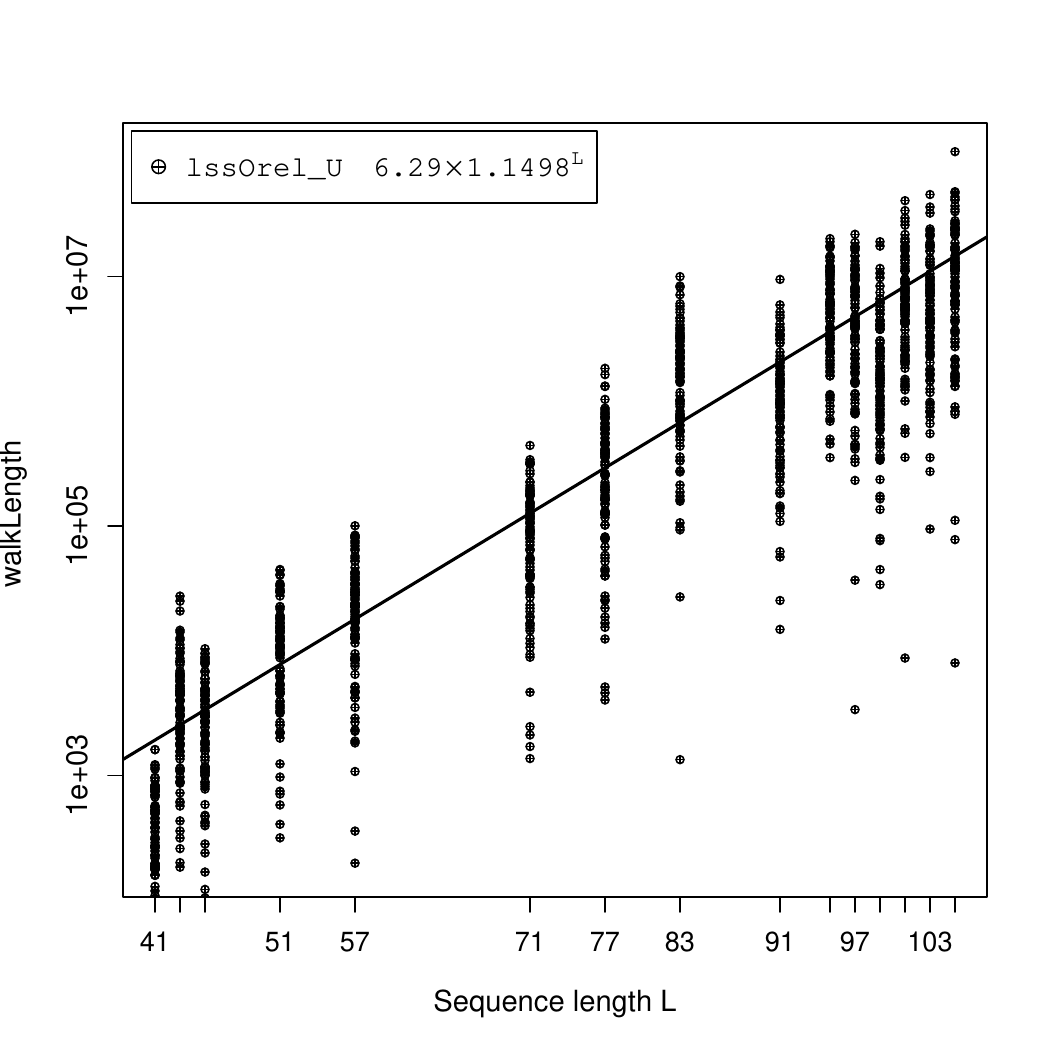}
\label{fg_model_mean-c}
\vspace*{-2ex}
\end{minipage}
}
\subfloat[Correlating {\em cntProbe} to {walkLength}]{
\begin{minipage}{0.40\textwidth}
\includegraphics[width=0.99\textwidth]{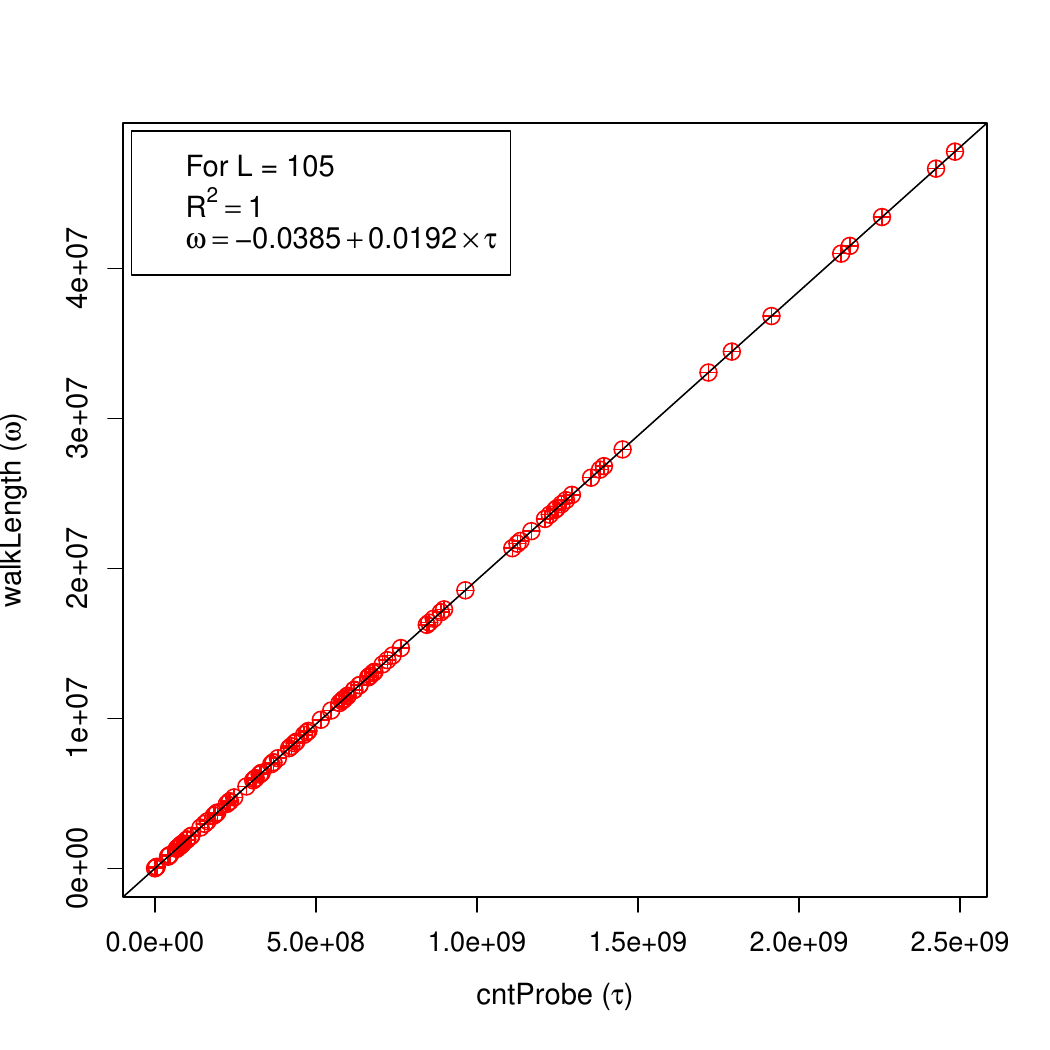}
\label{fg_model_mean-d}
\vspace*{-2ex}
\end{minipage}
}

\vspace*{-5ex}
\subfloat[Results for $L=41$ to $L=105$]{
\begin{minipage}{0.40\textwidth}
\includegraphics[width=0.99\textwidth]{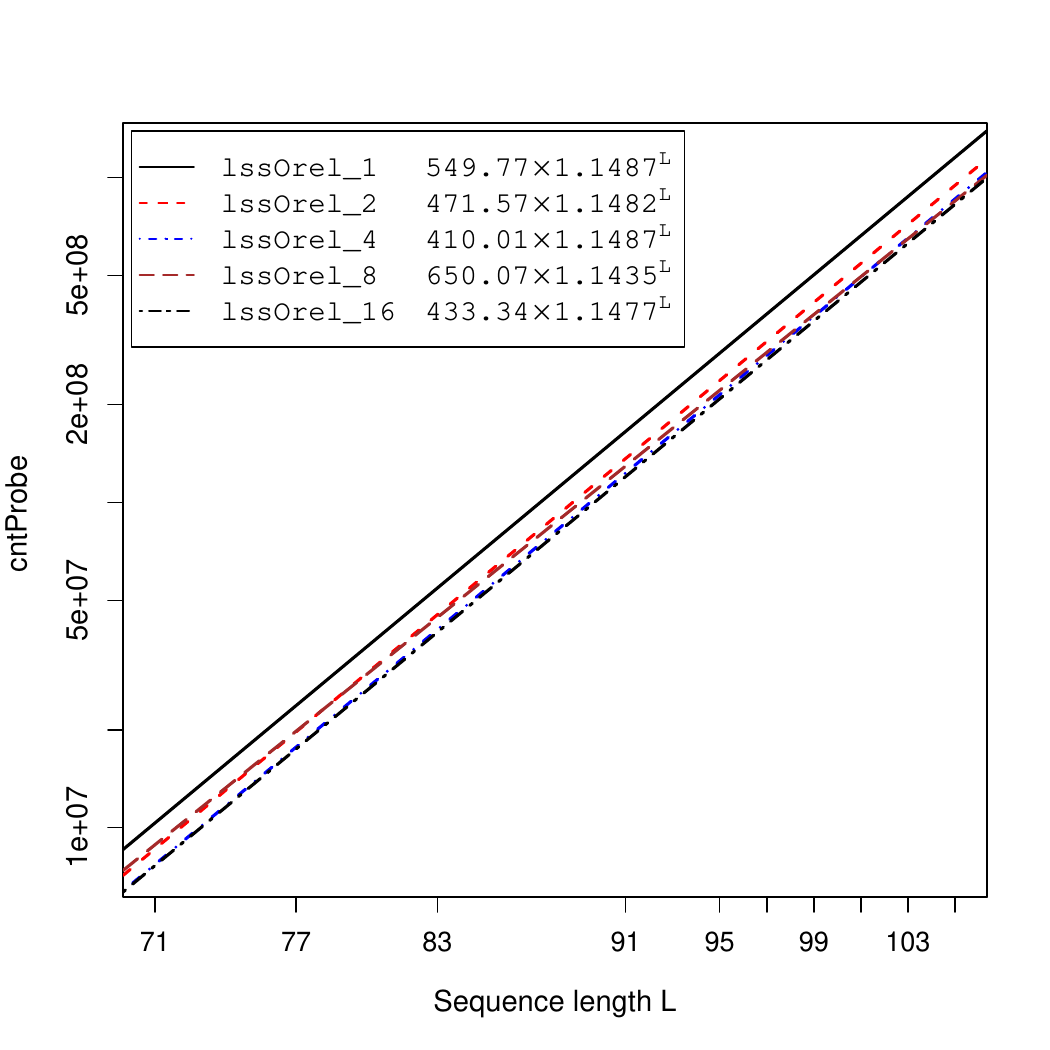}
\label{fg_model_mean-e}
\vspace*{-2ex}
\end{minipage}
}
\subfloat[Results for $L=107$ to $L=127$]{
\begin{minipage}{0.40\textwidth}
\includegraphics[width=0.99\textwidth]{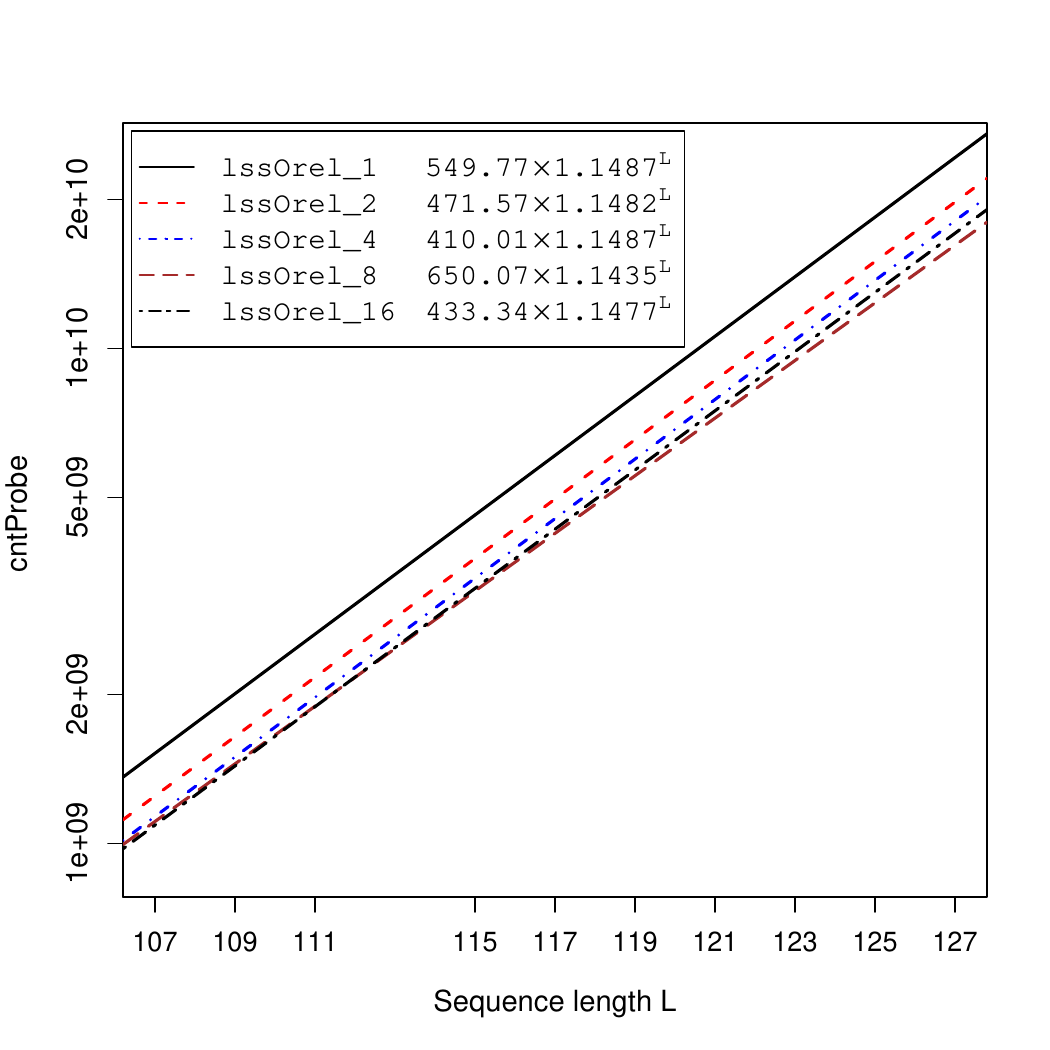}
\label{fg_model_mean-f}
\vspace*{-2ex}
\end{minipage}
}

\caption{
Asymptotic performance of solvers \lssOrelU\ (a-d) and  {\tt lssOrel\_}${\omega_c}$ 
for walk length coefficients $\omega_c \in \{1, 2, 4, 8, 16\}$ (e-f).
The best solver performance is achieved for $\omega_c = 8$.
The values of $L = \{41, 43, 45, 51, 57, 71, 77, 83, 91, 95, 97, 99, 101, 103, 105,
107, 109, 111, 115, 117, \\ 119, 121, 123, 125, 127 \}$ represent the subset of hardest-to-solve instances
of the \labs\ problem; see Eq.~\ref{eq_L_primary} and Table~\ref{tb_L_asymptotic}.
All  predictor models in the form of ($a*b^{L}$) 
are based on a sample size of $N=100$.
}
\label{fg_model_mean}
\end{figure*}

\par\vspace*{1ex}\noindent
{\bf (3) Experiments with \lssOrelU.}
The experiments with \lssOrelU\ for 14 hardest-to-solve instances are summarized in
subfigures ~\ref{fg_model_mean}a-\ref{fg_model_mean}-d; 
they range from $L=41$ to $L=105$. Note the high correlation of 
{\em cntProbe} versus {\em runtime} and {\em walkLength} versus {\em cntProbe}: 
for all values of $L$, this correlation
remains at about 99\%.

The letter {\tt U} in the solver name  
is a parameter that stands for {\em unlimited length of the self-avoiding walk segment}
in contrast to \lssOrelE\ where {\tt 8} stands for the value of walk segment coefficient $\omega_c$ that
determines the maximum length of the self-avoiding walk segment
 $\omega_{lmt} = \omega_c * \frac{L+1}{2}$, already defined in 
Table~\ref{tb_notation_summary};
\lssOrelE\ is discussed in more details later.

\begin{figure*}[t!]
\centering
\subfloat[$cntProbe$ for $L = 51$ up to $L = 105$.]{
\begin{minipage}{0.49\textwidth}
\includegraphics[width=0.99\textwidth]{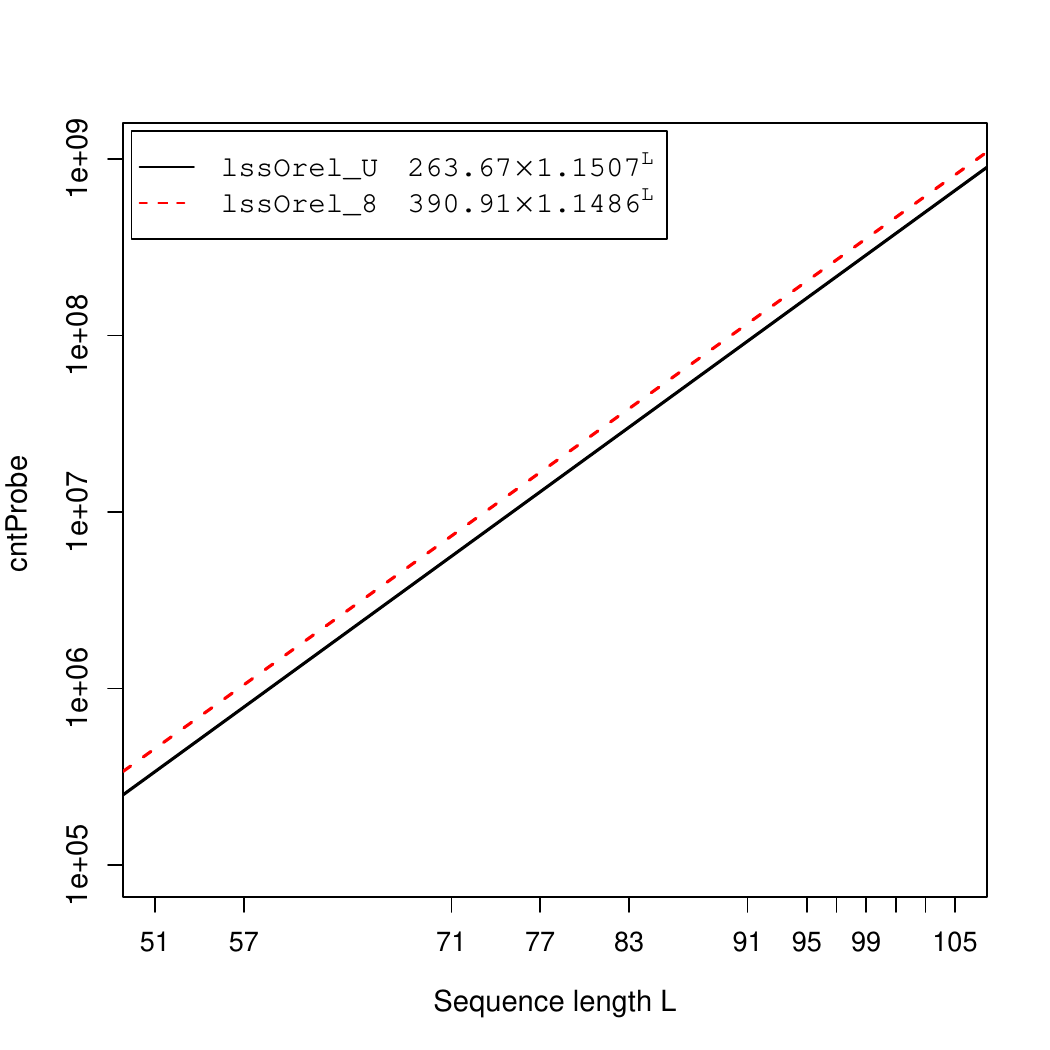}
\label{fg_unlimited_walk_a}
\vspace*{-1ex}
\end{minipage}
}
\subfloat[$runtime$ in seconds for $L = 51$ up to $L = 105$.]{
\begin{minipage}{0.49\textwidth}
\includegraphics[width=0.99\textwidth]{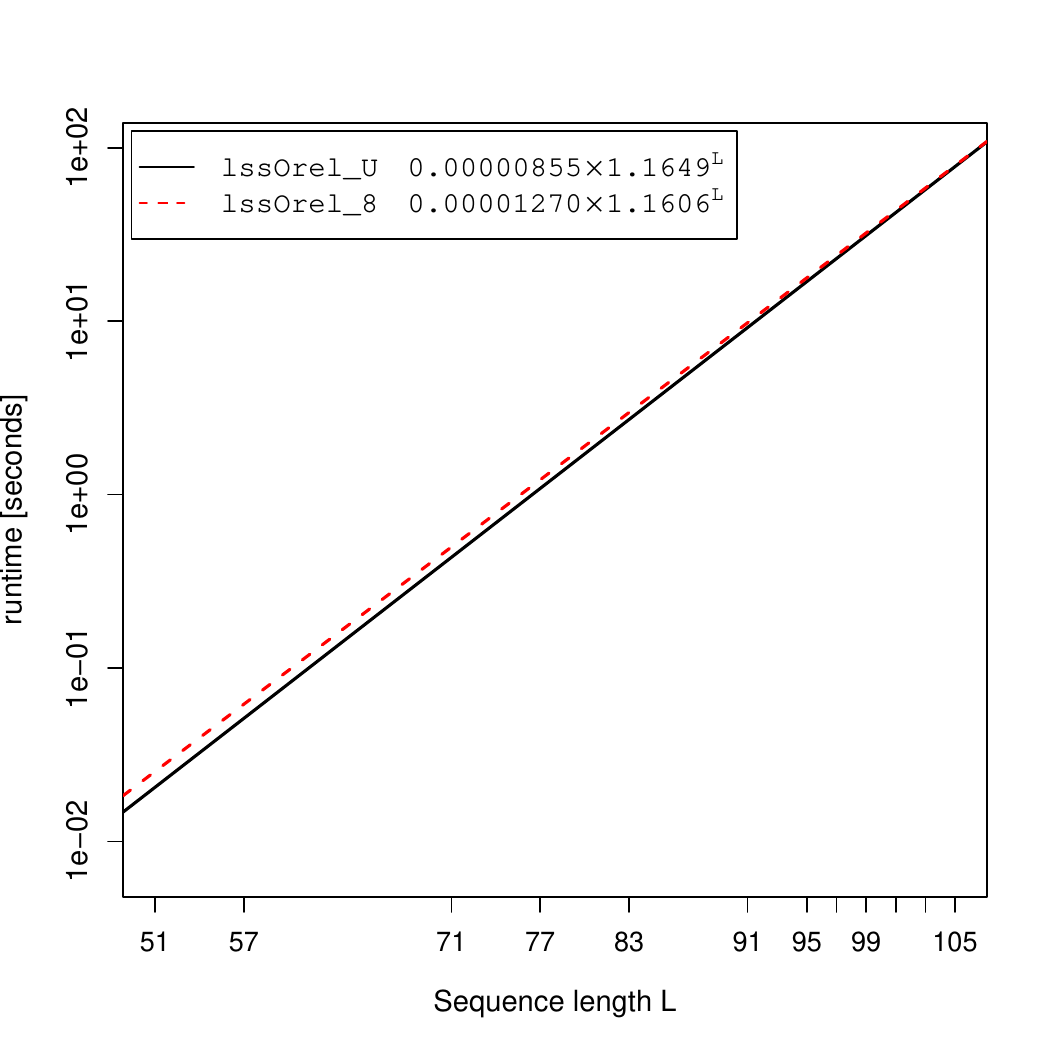}
\label{fg_unlimited_walk_b}
\vspace*{-1ex}
\end{minipage}
}

\subfloat[$speed$ for $L = 71$ up to $L = 105$.]{
\begin{minipage}{0.49\textwidth}
\includegraphics[width=0.99\textwidth]{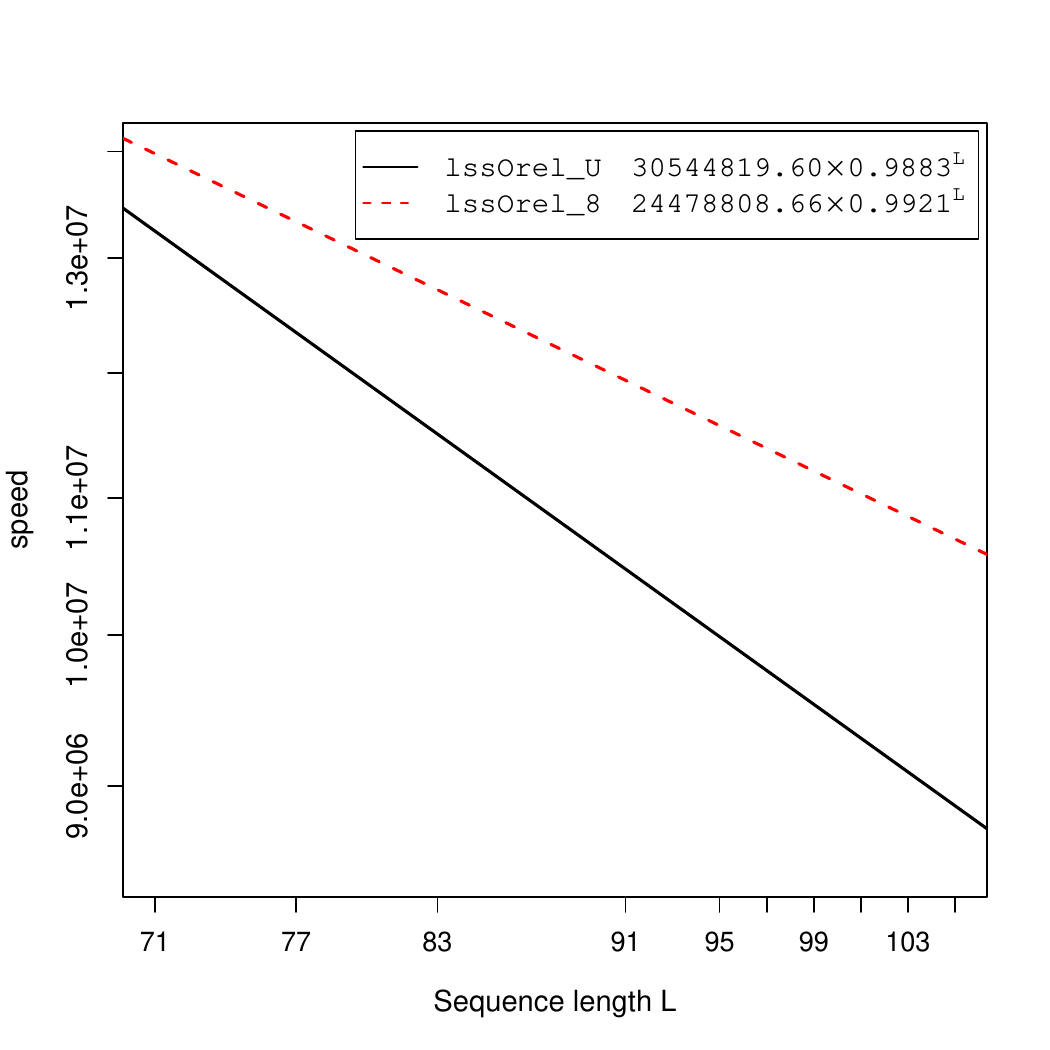}
\label{fg_unlimited_walk_c}
\vspace*{-1ex}
\end{minipage}
}
\subfloat[Maximum memory usage for $L = 51$ up to $L = 105$.]{
\begin{minipage}{0.49\textwidth}
\includegraphics[width=0.99\textwidth]{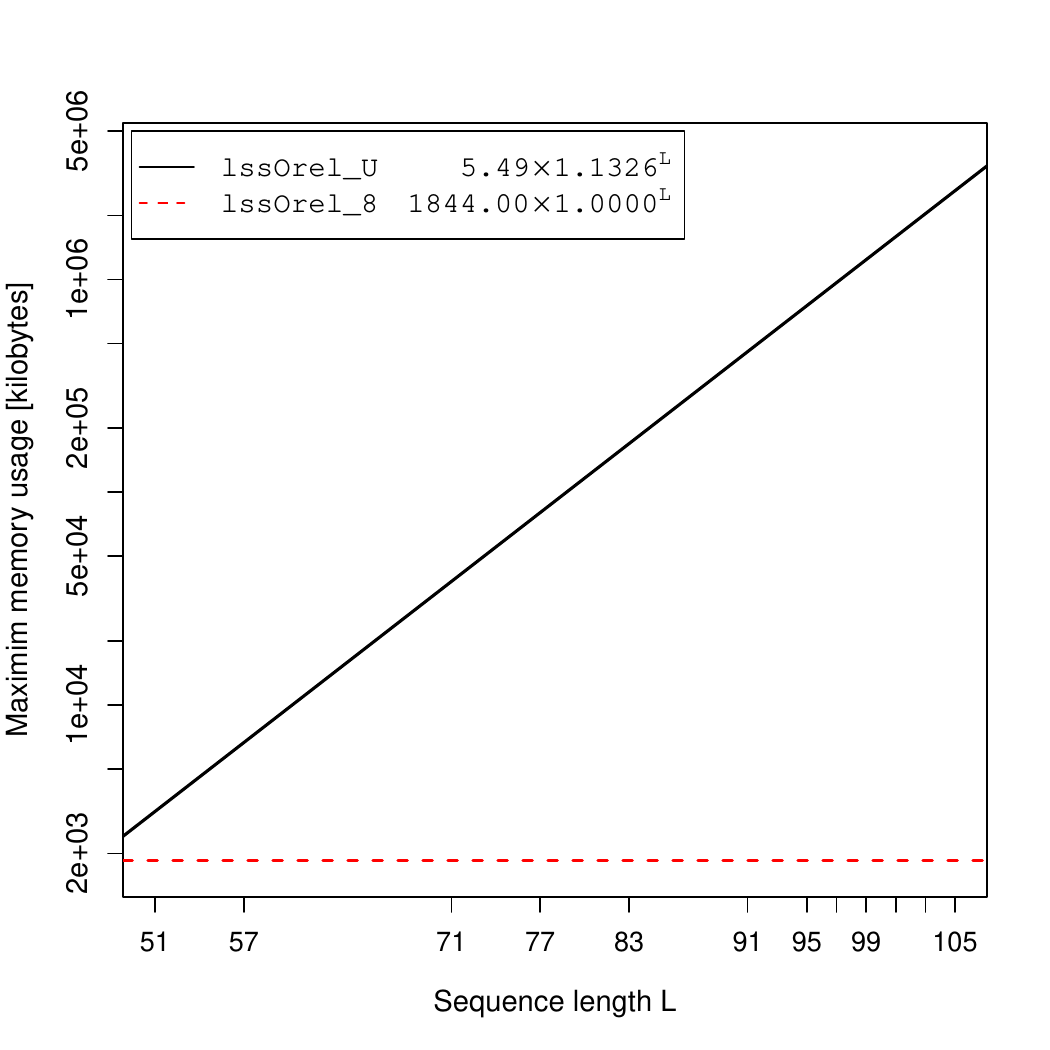}
\label{fg_unlimited_walk_d}
\vspace*{-1ex}
\end{minipage}
}
\vspace*{2ex}
\caption{Asymptotic comparison between two solvers:
\lssOrelU, based on a single segment self-avoiding walk and
\lssOrelE\ with walk composed of several self-avoiding segments, each of  length fixed at $\omega_{lmt} = 8 * \frac{L+1}{2}$.
As expected, the solver \lssOrelU\ has a slight advantage over
\lssOrelE\ when we observe $cntProbe$ only. 
However, the 
probability of a hash collision to maintain a self-avoiding walk under a fixed  memory limit 
also increases with increasing $L$ -- which accounts for the observed reduction in speed of the solver \lssOrelU,
and the approaching crossover in $runtime$ when compared to  \lssOrelE.
A compromise solver, with only a modest memory requirement, such as 
\lssOrelE\ is needed for solving larger instance sizes. 
}
\label{fg_unlimited_walk}
\end{figure*}

Under the walk segment coefficient value of~{\tt U}, solver \lssOrel\
invokes the procedure \textsc{walk.saw} in Figure~\ref{fg_global_search2} only once;
the walk is contiguous and terminates as a single segment only upon reaching
the upper bound $\Theta^{ub}_L$. For $L = 105$, the largest instance reported in this group of experiments,  we have $\Theta^{ub}_{105} = 620$.
In our experiments with $L = 105$ we record instances of 100 distinct single-segment contiguous walks,
each starting at a different randomly selected coordinate and random seed, with each walk terminating at one of the four solution coordinates with the best-known value of 620.
We have not observered a single instance of a trapped pivot that would induce a restart of another walk segment.
The {\em runtime}, {\em cntProbe} and and {\em memory footprint} range from 0.04 to 278.87 seconds, $2^{18.66}$ to $2^{31.21}$ probes and 2.2 MB to 1.97 GB, respectively. The averages for {\em runtime}, {\em cntProbe}, {\em walkLength}, 
and {\em memory footprint}
are 72.48 seconds, $2^{29.28}$ probes, $2^{23.59}$ steps, and 0.516 GB, respectively. 
We could not run instances of size $L = 107$ without a single restart due the 8 GB memory limit of our PC.

In lattices, with grid structures that are simpler when compared to our Hasse graphs,
physicists continue to push
the envelope on the maximum length of self-avoiding walks:
experiments with {\em longest walks under 64 GB of memory} are reported as
having maximum lengths of $2^{28} - 1$ in a 3D lattice and  
$2^{25} - 1$ in a 4D lattice~\cite{Lib-OPUS-walk-2010-StatPhysics-Clisby-pivot-alg-self-avoiding}.

The predictor models for observed sample means of 
{\em cntProbe} and {\em runtime}  (in seconds) are
\begin{eqnarray}
{\mathit cntProbe(\lssOrelU)} &=& 73.05 * 1.1668^L \\
{\mathit runtime(\lssOrelU)}  &=& 1.8 * 10^{-6} * 1.1846^L \hspace{1cm}
\label{eq_lssOrel_cntProbe}
\end{eqnarray}
Of these two models, only the predictor for {\em cntProbe} is platform independent,
the predictor for {\em runtime} (in seconds) is valid for the specified PC only.
Similarly to  Eq.~\ref{eq_lMAts_runtime}, we compute  coefficients 
in ${\mathit runtime(\lssOrelU)}$ indirectly by taking advantage of the 
high correlation between {\em cntProbe} versus {\em runtime}.
 
Results obtained with \lssOrelU\ provide the baseline for all experiments that follow.

\par\vspace*{1ex}\noindent
{\bf (4) Solver \lssOrel\ under limited walk length.}
The length of the self-avoiding walk segment is controlled by the walk segment coefficient $\omega_c$: $\omega_{lmt} = \omega_c * \frac{L+1}{2}$,
see Table \ref{tb_notation_summary}.
The value of $\omega_c$ extends the name of the solver \lssOrel,
for example \lssOrelE\ can be interpreted, in the case of $L=105$, as
limiting the contiguous walk length to a maximum of
$\omega_{lmt}=8 * 53 = 424$ steps.

Under the limited walk length, solver \lssOrel\ invokes the procedure \textsc{walk.saw} in Figure~\ref{fg_global_search2} with a randomly selected initial
coordinate a number of times,
creating the walk as a sequence of contiguous self-avoiding walk segments. However, since
each walk segment is independent, there is no need to store
the previous walk segments.
Thus, the walk segment coefficient determines not only the maximum walk length of the contiguous
self-avoiding walk segment but also the amount of memory needed to store 
the current segment.

To find out the effect of the limited walk length on solver, we ran
experiments with the secondary group of the
hardest-to-solve instances 
(Eq.~\ref{eq_L_secd}, see also Table~\ref{tb_L_asymptotic}) with walk segment coefficient 
values set to
$\omega_c=$ 1, 2, 4, 8, and 16 -- see subfigures ~\ref{fg_model_mean}e-\ref{fg_model_mean}-f. 
These results demonstrate that solver \lssOrelE\
exhibits the best asymptotic average-case performance with
$cntProbe$ of $650.07 * 1.1435^L$.
The second and third best results are achieved with $\omega_c=16$ and 4 while the
worst results are achieved with $\omega_c=1$. For example, consider the case of $L=127$:
by changing $\omega_c = 1$  to $\omega_c = 8$, the mean value of
$cntProbe$  decreases from $2^{34.64}$ to $2^{33.83}$, a decrease by a factor of 1.76.

Since experiments show that the effectiveness of \lssOrel\ is best for the 
walk segment coefficient value of $\omega_c = 8$, we shall use \lssOrelE\
as the reference solver for comparisons with all other solver configurations
in the remainder of this paper. 

\begin{figure*}[t!]
\centering
\subfloat[Results for $L=71$ up to $L=105$.]{
\begin{minipage}{0.45\textwidth}
\includegraphics[width=0.99\textwidth]{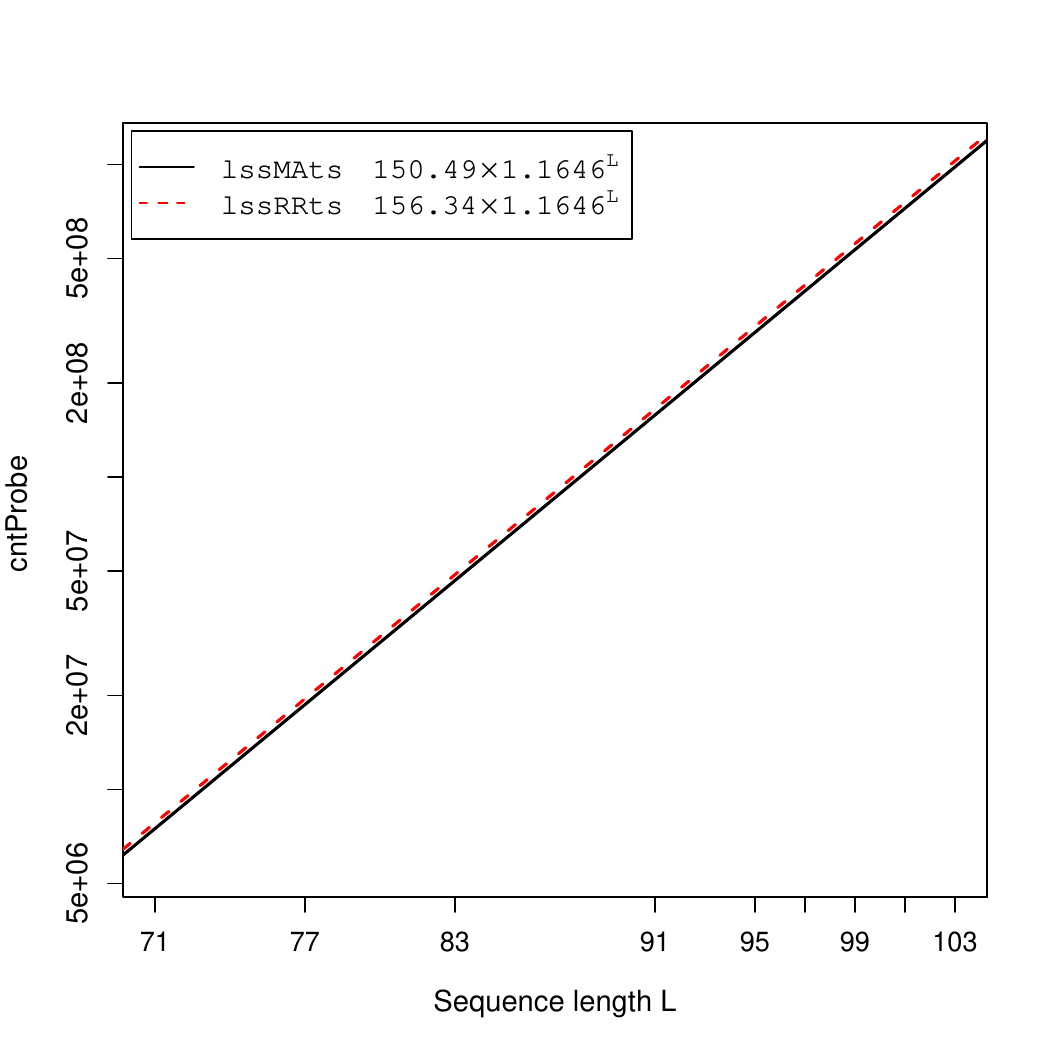}
\label{fg_comparison_lssRRts_vs_lssMAts1}
\end{minipage}
}
\subfloat[Results for $L=107$ up to $L=127$.]{
\begin{minipage}{0.45\textwidth}
\includegraphics[width=0.99\textwidth]{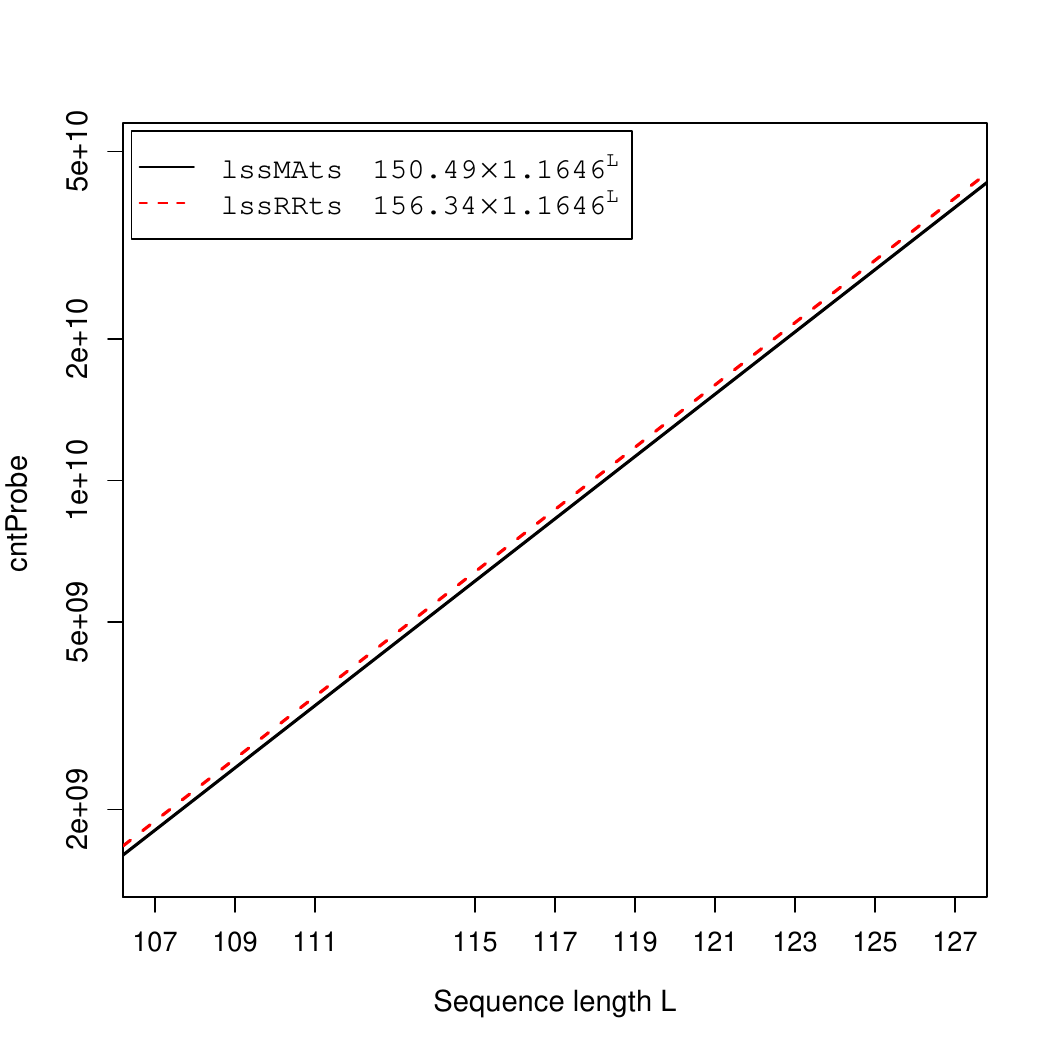}
\label{fg_comparison_lssRRts_vs_lssMAts2}
\end{minipage}
}
\vspace*{2ex}
\caption{Comparison between solvers \lssMAts\ and \lssRRts. 
        We observe that the asymptotic average-case performance of these two solvers are statistically equivalent.
        Thus, for the labs problem, the evolutionary component within \lssMAts\ is not effective.}
\label{fg_comparison_lssRRts_vs_lssMAts}
\end{figure*}


\noindent
{\bf (5) Comparisons of \lssOrelU\ and \lssOrelE.}
The main difference between \lssOrelU\ and \lssOrelE\ is the walk segment length and consequently, the memory usage.
See Figure~\ref{fg_unlimited_walk} for experiments with 14 hardest-to-solve instances, ranging from $L$ = 41 to $L$ = 105. Results
show (a) $cntProbe$, (b) $runtime$, (c) $speed$, and (d) memory usage. 
The solver speed   
is defined as the number of function evaluations (probes) per
second. For $L<71$ $runtime$ is close to 0 and the speed cannot estimated 
accurately, hence results are shown for $L \geq 71$ only.
The  memory usage in Figure
\ref{fg_unlimited_walk_d} is not an average value, it is the maximum memory usage 
observed for one of the 100 samples.

When observing $cntProbe$ alone, the solver \lssOrelU\ has a slight advantage over
\lssOrelE\ -- which we would expect. However, as $L$ increases, this advantage 
decreases for $runtime$ -- due to the increased reduction in speed observed for \lssOrelU.
A significant factor in this speed reduction for \lssOrelU\
is the increasing memory requirement for \lssOrelU, inducing an increased 
probability of a hash collision to maintain a self-avoiding walk under a fixed memory limit. As shown in the graph, the memory required by 
\lssOrelU\ increases with the  instance size $L$ while \lssOrelE\ requires a
constant amount of memory, about 1.8 MB in our case.

What we learned from these experiments is that the solver such as 
\lssOrelU\ cannot deliver solutions under a single self-avoiding walk segment when the required walk length exceeds the available memory
constraints of the solver -- a compromise solver such as 
\lssOrelE\ is needed for solving larger instance sizes. 

\par\vspace*{1ex}
\noindent
{\bf (6) Comparisons of \lssMAts\ and \lssRRts.}
The solver \lssRRts\ is a derivative of \lssMAts; 
asymptotic comparison of the two solvers is expected to reveal
whether or not the initialization of tabu search by the evolutionary component within \lssMAts\
significantly improves the solver performance in comparison with \lssRRts\
where tabu search is initialized with a random binary sequence.

\begin{figure*}[t!]
\centering
\vspace*{-7ex}
\subfloat[$cntProbe$ for $L=71$ up to $L=105$.]{
\begin{minipage}{0.40\textwidth}
\includegraphics[width=0.99\textwidth]{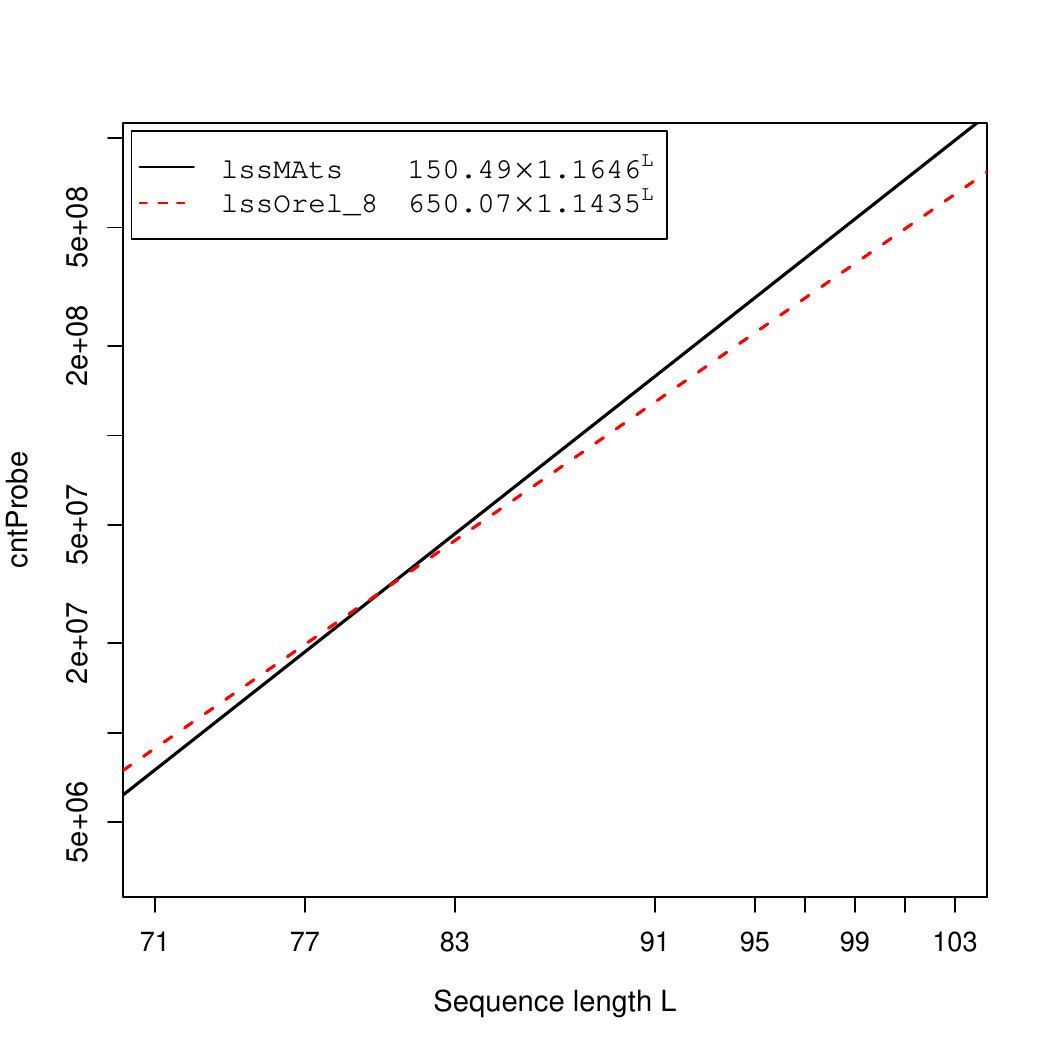}
\label{fg_comparison_lssOrel_vs_lssMAts_a}
\vspace*{-2ex}
\end{minipage}
}
\subfloat[$cntProbe$ for $L=107$ up to $L=127$.]{
\begin{minipage}{0.40\textwidth}
\includegraphics[width=0.99\textwidth]{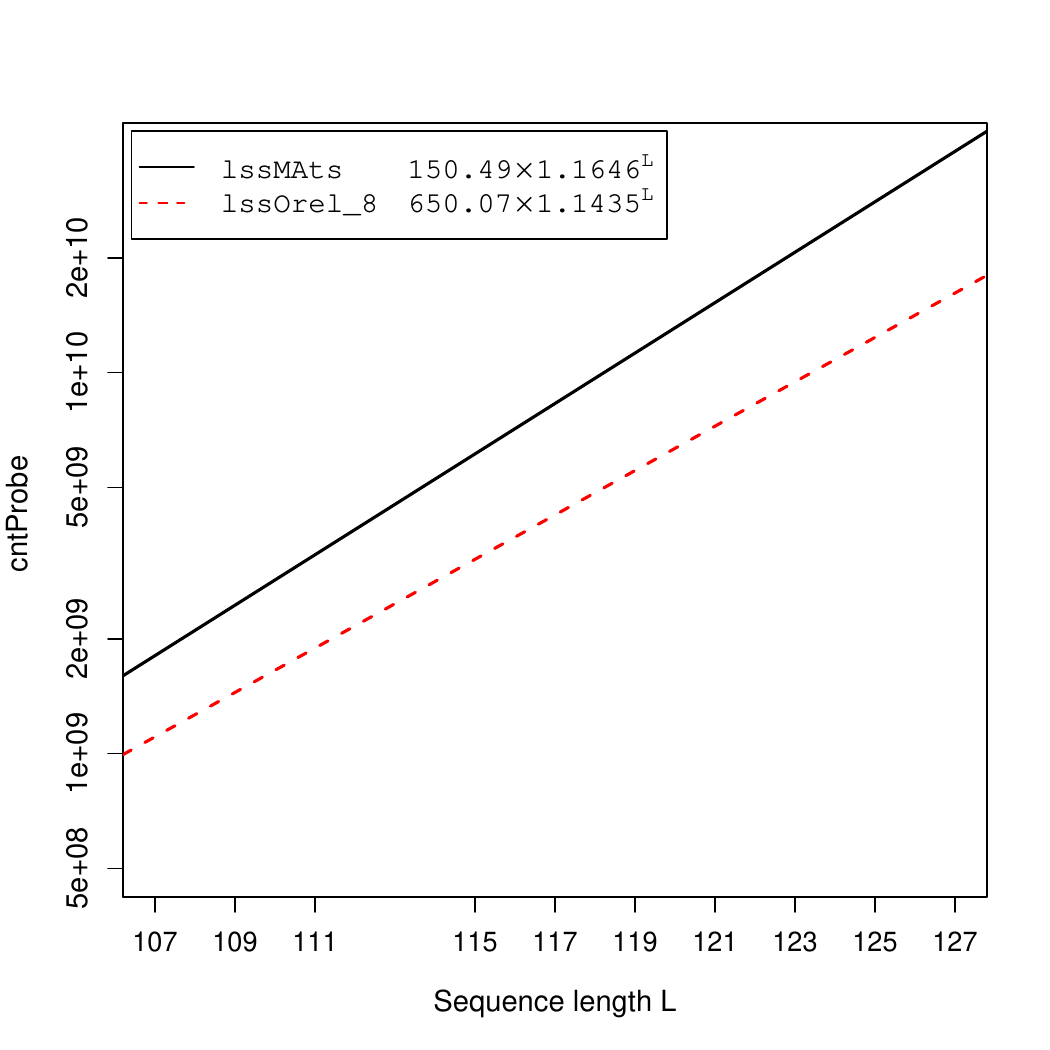}
\label{fg_comparison_lssOrel_vs_lssMAts_b}
\vspace*{-2ex}
\end{minipage}
}

\vspace*{-5ex}
\subfloat[$runtime$ in seconds for $L=107$ up to $L=127$.]{
\begin{minipage}{0.40\textwidth}
\includegraphics[width=0.99\textwidth]{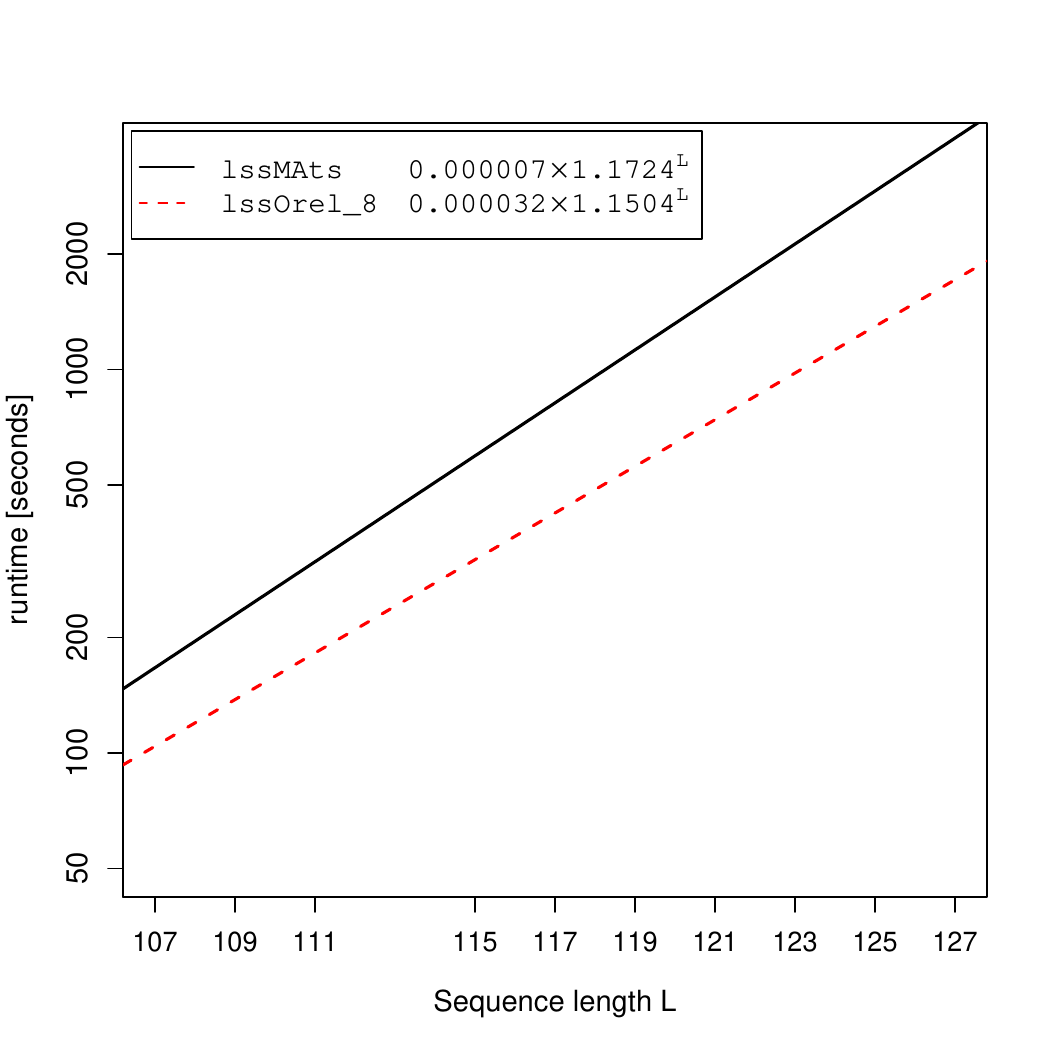}
\label{fg_comparison_lssOrel_vs_lssMAts_c}
\vspace*{-2ex}
\end{minipage}
}
\subfloat[$speed$ for $L=107$ up to $L=127$.]{
\begin{minipage}{0.40\textwidth}
\includegraphics[width=0.99\textwidth]{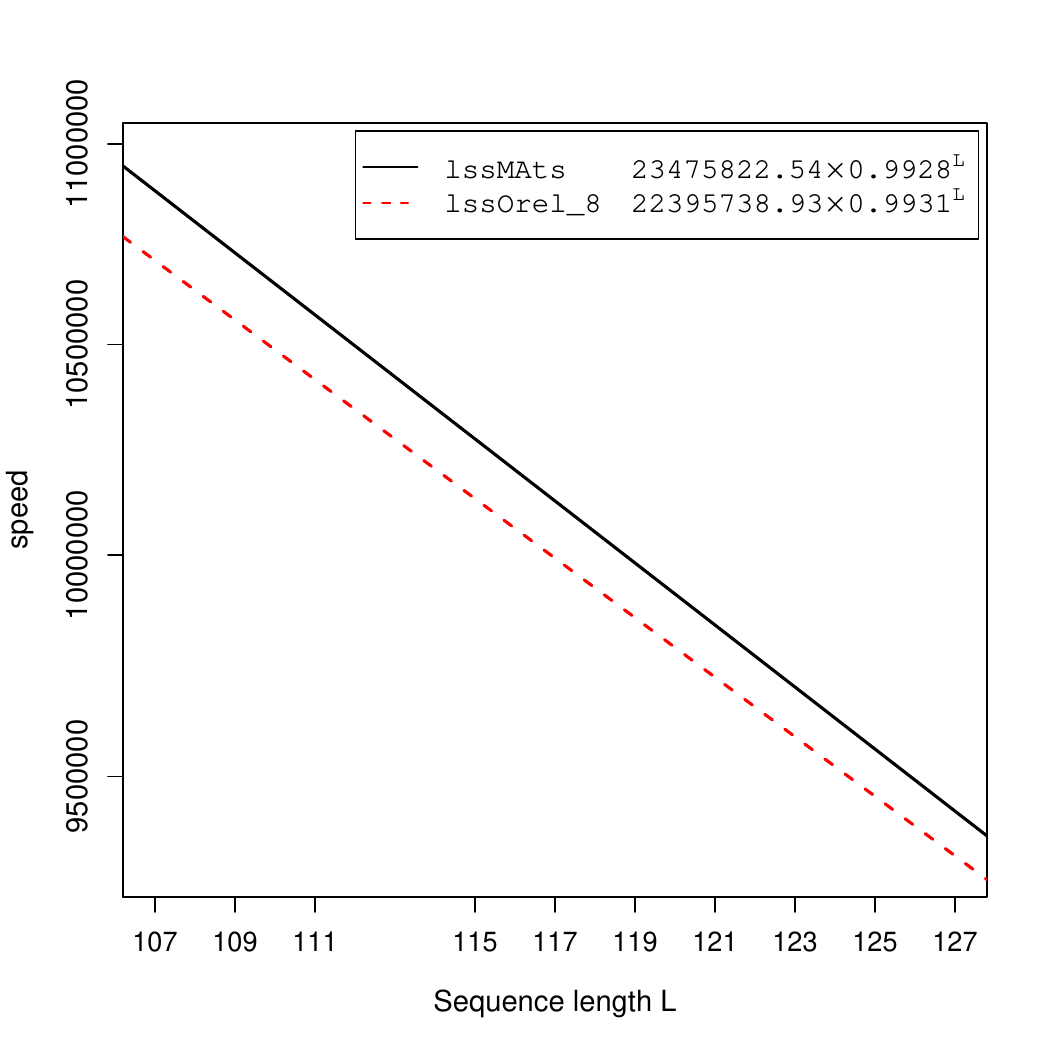}
\label{fg_comparison_lssOrel_vs_lssMAts_d}
\vspace*{-2ex}
\end{minipage}
}

\vspace*{-5ex}
\subfloat[\lssMAts\ versus \lssMAtsE\ only.]{
\begin{minipage}{0.40\textwidth}
\includegraphics[width=0.99\textwidth]{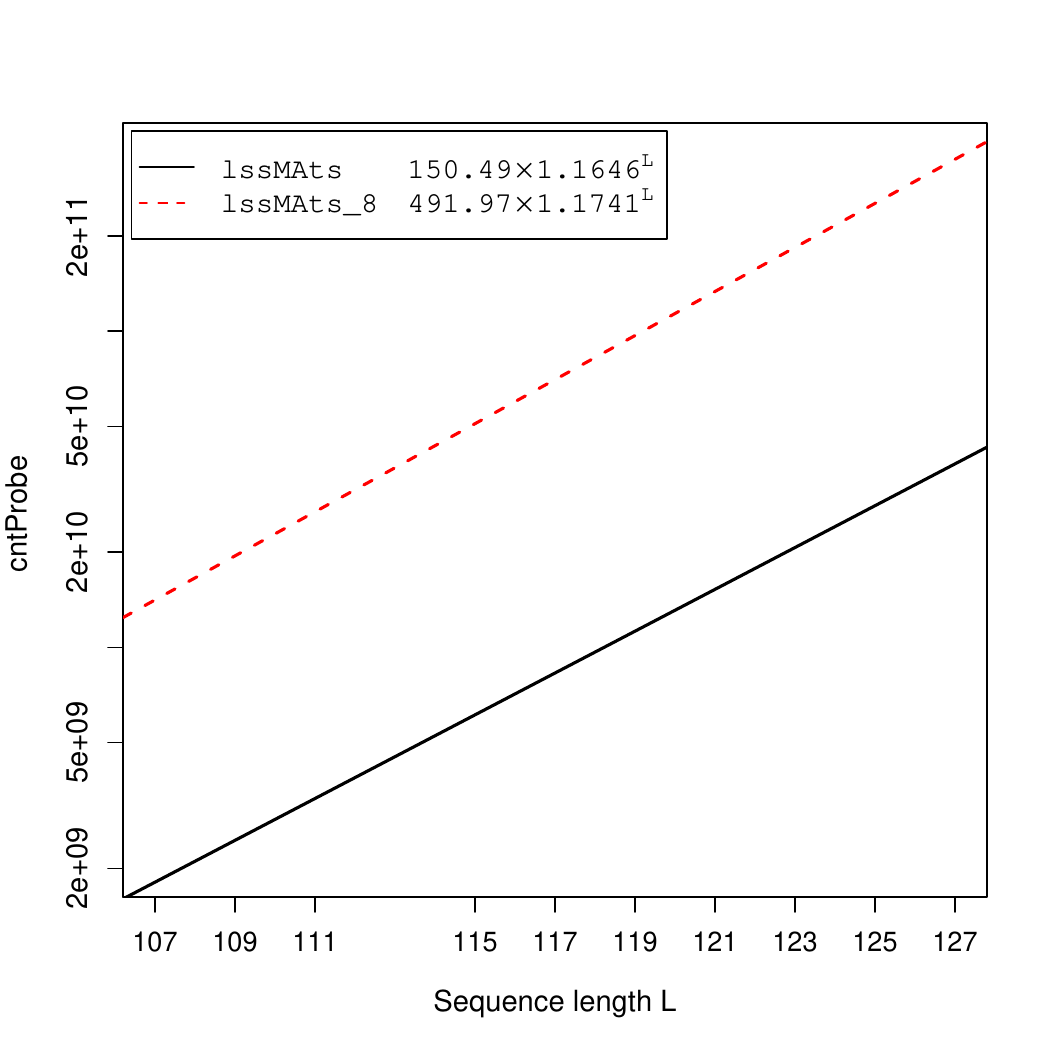}
\label{fg_comparison_lssOrel_vs_lssMAts_e}
\vspace*{-2ex}
\end{minipage}
}
\subfloat[\lssMAts\ versus \lssOrelE\ and others.]{
\begin{minipage}{0.40\textwidth}
\includegraphics[width=0.99\textwidth]{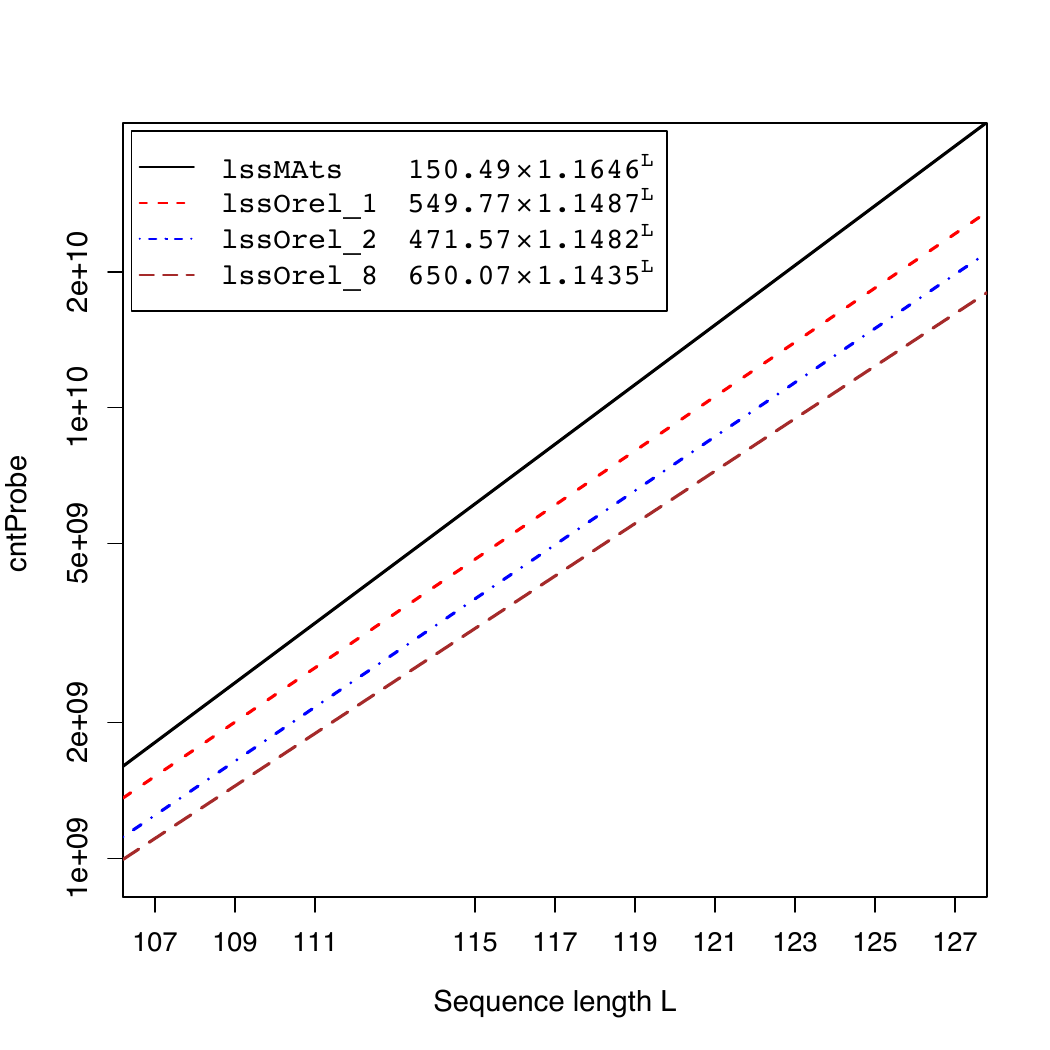}
\label{fg_comparison_lssOrel_vs_lssMAts_f}
\vspace*{-2ex}
\end{minipage}
}
\caption{Asymptotic average-case performance comparison between two solvers: \lssOrelE\ and \lssMAts. 
While \lssMAts\ dominates \lssOrelE\ in terms of {\em speed} performance,
\lssOrelE\ exhibits a significantly better 
{\em cntProbe} performance which results in a significantly better {\em runtime} performance.
}
\label{fg_comparison_lssOrel_vs_lssMAts}
\end{figure*}

We ran experiments with the secondary group of the hardest-to-solve instances
(Eq.~\ref{eq_L_secd}, see also Table~\ref{tb_L_asymptotic}).
The settings of \lssMAts\ are the same as described in~\cite{Lib-OPUS-labs-2009-ASC-Gallardo-memetic} and also
shown in Figure~\ref{alg_lssMAts_lssRRts}. The solver \lssRRts\ has only one control parameter: the tabu search walk
length, set in the same way as  \lssMAts. 
Results are shown in Figure~\ref{fg_comparison_lssRRts_vs_lssMAts}. 
We conclude that the asymptotic average-case performance of these two solvers are statistically equivalent. For the range $71 \leq L \leq 127$ 
we find $cntProbe$ as $150.49 * 1.1646^L$ for \lssMAts\ and  $156.34 * 1.1646^L$ for \lssRRts. Thus, for the \labs\ problem, 
the evolutionary component within \lssMAts\ is not effective.

\par\vspace*{1ex}\noindent
{\bf (7) Comparisons of \lssOrelE\ and \lssMAts.}
We ran two sets of experiments to compare the two solvers.
With the first set, we measure the asymptotic average-case performance,  with hardest-to-solve
instances from the secondary group in Eq.~\ref{eq_L_secd}.
For the second set, we select 7 instances that belong to the tertiary list in ~Eq. \ref{eq_L_tertiary}
and analyze solvabilities and hit ratios observed and predicted for the two solvers.

Results from first set are shown in Figure~\ref{fg_comparison_lssOrel_vs_lssMAts},
which itself consists of  {\em six subfigures}.
The average values of $cntProbe$ required by
each solver to reach the best-known upper bound, are shown in Figures~\ref{fg_comparison_lssOrel_vs_lssMAts_a} 
and~\ref{fg_comparison_lssOrel_vs_lssMAts_b}. We can conclude by inspection that the solver \lssOrelE\ dominates
\lssMAts\ in terms of $cntProbe$. Notably, for $L > 107$, the gap in average values of $cntProbe$
between \lssOrelE\ and \lssMAts\ is statistically significant and also continues to increase with increasing value of $L$.

In Figure~\ref{fg_comparison_lssOrel_vs_lssMAts_c} we also observe a statistically significant and increasing gap in 
average values of $runtime$ between the two solvers.
However, in Figure~\ref{fg_comparison_lssOrel_vs_lssMAts_d}
the observed speed of \lssOrelE\ is below the observed speed of \lssMAts\ -- with the gap slowly reducing as $L$ increases.
Apparently,  solver \lssOrelE\ overcomes its speed disadvantage by significantly better $cntProbe$ performance.
For example, in
the case of $L=109$, the difference between mean values of $runtime$ is 74 seconds and  for $L=127$,
this difference increases to 1555 seconds. In conclusion, the solver of choice
for the remainder of this paper is \lssOrelE, we settle on its mean runtime model (in unit of seconds) as
\vspace*{-0.3ex}
\begin{equation}
\overline{m}(L) = 0.000032*1.1504^L
\label{eq_runtime_seconds}
\end{equation}

A meticulous reader may notice that performance differences between the two solvers
can also be attributed to differences in walk length segments between random restarts.
By default, the solver \lssMAts\ selects the walk segment length $\omega_{lmt}$ randomly from the interval 
$[\frac{L}{2}, \frac{3L}{2}]$ whereas for \lssMAts\_$8$ and \lssOrelE\ 
we keep the walk segment length constant at $\omega_{lmt} = 8*\frac{L+1}{2}$.

In Figure~\ref{fg_comparison_lssOrel_vs_lssMAts_e}, we compare the performance of \lssMAts\  and \lssMAts\_$8$: the solver based on tabu search
that extends the walk segment length to $\omega_{lmt} = 8*\frac{L+1}{2}$
is at a significant disadvantage.

In Figure~\ref{fg_comparison_lssOrel_vs_lssMAts_f}, 
we compare \lssMAts\ with \lssOrel\ for values $\omega_c = 1, 2, 8$. 
We observe significant differences between \lssMAts\ and all version of \lssOrel.
Even the version of \lssOrel\_1 with $\omega_{lmt} = \frac{L+1}{2}$, performs better
than \lssMAts,  where on the average, $\omega_{lmt}$ is larger than $\frac{L+1}{2}$.
In conclusion, we keep \lssMAts\ as the default solver based on tabu search and \lssOrelE\ as the default solver based on self-avoiding walk segments.

\begin{table*}[t]
  \caption{Predictions versus observations from experiments with \lssOrelE\ and \lssMAts,
  under the constraint of runtime limit of 4 days for each of the selected values of $L$.
  The observed mean represents the sample mean based on the value of observed solvability
  (the sum total of $cntProbe$ of each instance  exhibits a gamma distribution). The observed hitRatio value of 100\% 
  signifies that none of the solutions have been censored. The model mean values are computed
  from the two predictors based on empirical data described in Figure~\ref{fg_comparison_lssOrel_vs_lssMAts}. 
  The values of predicted solvability, computed from Eq.~\ref{eq_solvP_ser}, represent the value of 
  $cntProbe$ to reach hitRatio of 100\% with probability of 0.99 -- provided (1) the solver
  has been scheduled on a single CPU invoke on each instance serially, and (2) the solution
  produced by the solver has not been censored.
}
\label{tb_lssOrel_lssMAts_cntProbe_solvability}
\centering
\begin{footnotesize}
\begin{tabular}{ c || c c | c c | r || c c | c c | r}
   & \multicolumn{5}{c||}{\lssOrelE}& \multicolumn{5}{c}{\lssMAts} \\
   \hline
 ~ & model~ & observed & predicted & observed & & model~ & observed & predicted & observed \\[-0.45ex]
 L & mean$^*$ & mean & solvability$^\dagger$ & solvability & $\mathit{solvP}^\ddagger$  & mean$^+$ & mean & solvability$^\dagger$ & solvability & $\mathit{solvP}^\ddagger$ \\
\hline
&&&&&&&&& \\[-2ex] 
  115 & 3.236e+09 & 2.858e+09 & 4.037e+11 & 2.858e+11 & 100 & 6.135e+09 & 6.132e+09 & 7.652e+11 & 6.131e+11 & 100 \\
  121 & 7.236e+09 & 6.539e+09 & 9.025e+11 & 6.539e+11 & 100 & 1.530e+10 & 1.335e+10 & 1.909e+12 & 1.334e+12 & 100 \\
  127 & 1.617e+10 & 1.479e+10 & 2.017e+12 & 1.479e+12 & 100 & 3.819e+10 & 2.997e+10 & 4.763e+12 & 2.997e+12 & 100 \\
  141 & 1.057e+11 & 5.339e+10 & 1.318e+13 & 5.339e+12 & 100 & 3.224e+11 & 1.611e+11 & 4.021e+13 & 1.610e+13 & 100 \\
  151 & 4.042e+11 & 3.874e+11 & 5.041e+13 & 3.874e+13 & 95 & 1.479e+12 & 7.363e+11 & 1.845e+14 & 7.363e+13 & 80 \\
  161 & 1.545e+12 & 7.033e+11 & 1.927e+14 & 7.033e+13 & 76 & 6.791e+12 & 1.200e+12 & 8.470e+14 & 1.200e+14 & 44 \\
  181 & 2.257e+13 & 1.221e+12 & 2.816e+15 & 1.221e+14 & 6 & 1.430e+14 & 1.338e+12 & 1.784e+16 & 1.338e+14 & 1\\
 \hline
  & \multicolumn{4}{@{~}c@{~}}{} & & \multicolumn{4}{@{~}c@{~}}{} \\[-2.3ex]
  & \multicolumn{4}{@{~}c@{~}}{${\rm ^*cntProbe(\lssOrelE)} = 650.07 * 1.1435^L$} & & \multicolumn{4}{@{~}c@{~}}{${\rm ^+cntProbe(\lssMAts)} =  150.49 * 1.1646^L$} \\
  \hline
  \multicolumn{11}{@{~}l@{~}}{} \\[-1.3ex]
  \end{tabular}\\
$^\dagger$ using Eq.~\ref{eq_solvP_ser} with $N = 100, p = 0.99$~~;~~$^\ddagger$ using Eq.~\ref{eq_hitO} 
  \end{footnotesize}
\end{table*}

\begin{table*}[th!]
\centering
\caption{New best-known solutions provided by \lssOrelE. Solution coordinates of length $\frac{L+1}{2}$ are shown in run length encoded notation. Rules of skew-symmetry must be applied to expand each cooodinate to its full length $L$.}
\label{tb_new_records}
\begin{scriptsize}
\begin{tabular}{@{}c | c c p{15.0cm} }
L & F & E & solution coordinate (in canonic form, e.g. for $L=181$, 11 represents a run of eleven 0's) \\
\hline
181 & 8.9316 & 1834 & 
11,1,2,1,4,4,2,3,2,1,1,3,2,2,2,1,2,2,1,2,3,4,1,2,1,1,4,1,2,1,1,2,3,1,1,6,1,1,4,1,1\\
201 & 8.4876 & 2380 & 
9,1,2,2,1,2,1,2,1,6,8,2,1,6,1,1,2,1,2,5,1,1,1,2,1,2,1,1,3,3,5,1,1,2,6,2,3,2,2,2,2,1\\
215 & 8.5888 & 2691 & 
4,3,1,1,3,4,1,1,1,1,1,4,1,1,1,1,1,2,1,2,2,1,2,1,1,2,5,1,2,1,3,3,2,2,1,3,1,1,1,4,2,2,1,4,1,1,2,2,1,3,1,1,1,1,4,4,3\\
221 & 8.8544 & 2758 & 
7,11,1,2,2,1,2,1,2,2,1,2,6,6,1,1,2,1,2,1,2,1,4,1,1,1,2,1,7,2,2,1,3,2,1,2,1,2,1,2,1,1,4,3,1,1,2,2,3\\
241 & 8.0668 & 3600 & 
2,1,1,2,1,2,2,1,1,1,1,3,1,2,1,5,1,2,1,5,1,1,2,2,3,2,1,3,5,2,1,5,1,4,1,1,2,3,1,2,4,1,4,3,2,1,3,1,1,2,1,3,1,1,3,2,1,2,1,1,1,1,1,1\\
249 & 8.1323 & 3812 & 
4,1,2,1,2,1,1,2,2,1,4,1,2,1,4,1,3,6,2,1,4,1,2,7,3,1,2,1,2,1,2,5,2,1,2,1,2,1,12,1,3,5,1,1,1,4,1,2,3,1,1,2,1,2,2\\
259 & 8.0918 & 4145 & 
3,3,1,3,2,3,2,1,4,1,3,5,4,1,4,1,1,1,1,2,1,4,2,3,3,3,3,2,1,1,1,1,2,3,2,6,3,1,1,4,1,2,2,1,3,4,1,2,1,5,1,5,1,3,1,1,1,2\\
261 & 7.8517 & 4338 & 
2,2,2,1,1,1,1,2,1,1,1,4,1,2,1,2,1,1,4,1,1,1,3,2,2,1,1,1,1,1,2,2,1,3,1,2,1,4,1,1,1,1,1,2,3,1,1,1,2,1,1,2,3,1,1,1,2,2,1,5,1,1,1,2,1,1,2, 1,2,2,1,1,1,8,2,2,3,2\\
271 & 7.5386 & 4871 & 
3,3,1,1,1,1,3,1,1,4,2,1,3,3,2,1,1,1,1,1,1,1,1,2,3,2,3,1,3,3,2,1,1,1,2,4,4,1,2,4,4,1,1,1,1,2,2,1,1,1,2,1,2,1,1,3,1,4,3,2,3,1,1,1,1,1,3, 6,1,2,2,2\\
281 & 7.5058 & 5260 & 
15,3,3,10,2,3,2,2,1,4,2,3,2,2,3,2,1,4,4,2,1,5,3,2,4,1,1,1,2,3,2,4,2,1,3,3,1,1,7,2,1,3,2,1,1,1,1,1,2,1,1,1,2,1,2,1\\
283 & 7.5088 & 5333 & 
4,1,1,11,1,1,1,1,9,1,3,3,1,2,1,2,1,1,1,2,2,3,2,4,1,1,1,1,2,1,3,2,2,2,1,1,2,1,1,1,4,2,3,3,2,2,5,3,1,2,1,3,2,1,1,2,1,1,3,2,1,2,1,1,1,1, 2,1,2,1,1,2\\
301 & 7.4827 & 6054 & 
3,3,3,3,3,3,4,2,3,5,1,1,2,5,3,1,2,1,1,1,1,1,1,2,1,3,1,3,2,3,6,1,1,1,1,1,2,3,2,7,6,2,1,1,1,1,3,3,1,3,1,1,1,1,1,3,1,2,1,3,1,2,2,1,1,1,1, 1,1,2,1,2,2,2,1\\
303 & 7.2462 & 6335 & 
2,1,1,2,2,1,2,2,2,1,1,1,1,1,1,1,1,2,1,1,2,2,1,1,1,1,1,1,1,2,1,2,1,2,2,2,3,1,1,1,1,1,2,3,1,1,2,1,1,1,4,1,2,1,2,2,2,1,1,1,1,1,1,2,1,2,3, 
2,2,1,1,1,1,4,1,1,1,2,1,1,1,1,2,2,1,6,2,1,1,3,3,1,1,1,4,2,1,1,1\\
341 & 6.9397 & 8378 & 
2,4,3,1,1,2,2,2,1,2,1,2,1,3,1,2,2,1,6,5,2,2,1,1,3,1,3,3,1,1,1,4,3,2,1,1,2,1,1,3,1,2,1,1,1,1,1,1,5,5,1,2,1,2,4,1,1,3,5,1,1,1,1,3,1,1,1,
4,1,1,2,3,1,1,3,3,1,2,1,3,2,4,1,1,1,1,1,2,1,1\\
381 & 7.0893 & 10238 & 
5,2,1,2,1,6,2,1,7,2,2,2,1,2,2,1,2,1,2,1,2,1,2,2,2,2,1,7,6,2,1,2,6,2,1,2,5,5,2,1,7,4,9,1,1,2,2,3,6,1,1,2,1,2,1,1,2,1,2,2,5,1,1,1,5,1,2,
1,3,1,3,1,1,7,1,1,1,4,1\\
401 & 6.7632 & 11888 & 
2,4,1,2,4,1,5,1,1,2,1,2,1,1,2,5,6,2,4,5,1,1,2,3,2,2,4,1,2,3,1,4,1,2,3,4,1,3,2,1,2,1,1,1,3,2,1,1,1,2,3,4,3,1,5,2,1,4,2,5,1,1,1,1,1,1,1,
1,1,1,2,3,1,2,1,3,2,2,2,3,1,6,1,1,1,1,1,1,5,1,1,1,2,1,2,1,3\\
\end{tabular}
\end{scriptsize}
\end{table*}

Results from the second set of experiments,
based on seven instances from the tertiary list in Eq.~\ref{eq_L_tertiary},
are shown in Table~\ref{tb_lssOrel_lssMAts_cntProbe_solvability}.
Here we compare
asymptotic predictions for $cntProbe$, calculated under the first set of experiments in Figure~\ref{fg_comparison_lssOrel_vs_lssMAts}
versus the observed mean and observed solvability (defined as the sum of total of $cntProbe$,
exhibiting a gamma distribution).
There are a number of important observations that can be inferred from this set of experiments:
(1) as long as the hit ratio stays at 100\% (for all instance sizes up to $L$=141), 
the value differences between the model mean and the observed mean (and the asymptotic solvability and the observed solvability) are relatively small for {\em both} solvers,
the differences increases significantly as the hit ratio reduces to 6\% and 1\% respectively;
(2) for each instance, the asymptotic predictions represent the upper bound 
on the observed values (in this set of experiments);
(3) for each instance, \lssOrelE\ significantly outperforms \lssMAts.\\

\par\vspace*{-2ex}\noindent
{\bf (8) Comparisons with best known merit factors.}
The new best-known merit factors returned by \lssOrelE\ for all tertiary group instances that are greater
than 160 and their {\em canonic solution} coordinates are shown in Table~\ref{tb_new_records}. For brevity, we list coordinates to represent
{\em the first} $\frac{L+1}{2}$  binary
symbols in the run-length notation. For example, 
the solution coordinate for L = 221 is listed as 7,11,1,2,2,... 
The value of 7 implies a run of seven 0's,
followed by a run of eleven 1's, etc. Note these solutions are in the canonic form:
each solution begins with a run of at least two 0's.

\begin{figure*}[t!]
\vspace*{-5ex}
\centering
\begin{minipage}{0.36\textwidth}
  \begin{small}
  (a) Mean value runtime predictions  
  versus the observed runtime means  with the solver \lssOrelE:
  $\overline{m}(L) = 0.000032*1.1504^L$ (Eq.~\ref{eq_runtime_seconds})
  versus $\overline{m}_{obs}(L)$.
  For values of $L \le 141$,
  none of the $N=100$ runs are censored within 
  the time limit of $t_{lmt} = 4$  days (345600 seconds),
  hence the observed number of hits is {\em hitO = 100}
  and the reported values of observed runtime means  are within the confidence
  interval defined in Eq.~\ref{eq_confidenceInterval}.
  Relative to the predicted value of $\overline{m}(L)$, the
  ratio of $\overline{m}_{obs}(L)/\overline{m}(L) > 1$
  is an indicator of the instance solvability as well as the level of the landscape 
  {frustration}~\cite{Lib-OPUS-labs-2014-NatNano-frustrated-energy-landscape};
  a  characteristic  of the 
  the \labs\ problem. All instances with more than a single canonic solution ($C_L > 1$)
  have been excluded from this prediction model. For $L = 141$, $C_{141} = 2$.
 \end{small}
\end{minipage}
\begin{minipage}{0.45\textwidth}
  \centering
  \begin{small}
 \begin{tabular}{c|c c c c c}
 $L$& $\overline{m}(L)$ & $\overline{m}_{obs}(L)$ & $\Delta(\rm{seconds})$ & ratio & $C_{L}$ \\
 \hline
 83 &   3  &   12 &     9 &  4.00 & 1\\
 91 &   11 &   7  &    -4 & -0.63 & 1\\
 95 &   19 &   30 &    11 &  1.58 & 1\\
 97 &   25 &   28 &     3 &  1.12 & 1\\
 99 &   33 &   14 &   -19 & -0.42 & 1\\
101 &   44 &   34 &   -10 & -0.77 & 1\\
103 &   59 &   48 &   -11 & -0.81 & 1\\
105 &   78 &   69 &    -9 & -0.88 & 1\\
107 &  103 &  446 &   343 &  4.33 & 1\\
109 &  137 &   83 &   -54 & -0.61 & 1\\
111 &  181 &  154 &   -27 & -0.85 & 1\\
115 &  318 &  279 &   -39 & -0.87 & 1\\
117 &  421 &  299 &  -122 & -0.71 & 1\\
119 &  557 &  355 &  -202 & -0.63 & 1\\
121 &  737 &  673 &   -64 & -0.91 & 1\\
123 &  976 & 2181 &  1205 &  2.23 & 1\\
125 & 1291 & 1427 &   136 &  1.11 & 1\\
127 & 1709 & 1598 &  -111 & -0.94 & 1
 \end{tabular}
 \end{small}
\end{minipage}

\begin{minipage}{0.49\textwidth}
\includegraphics[width=0.99\textwidth]{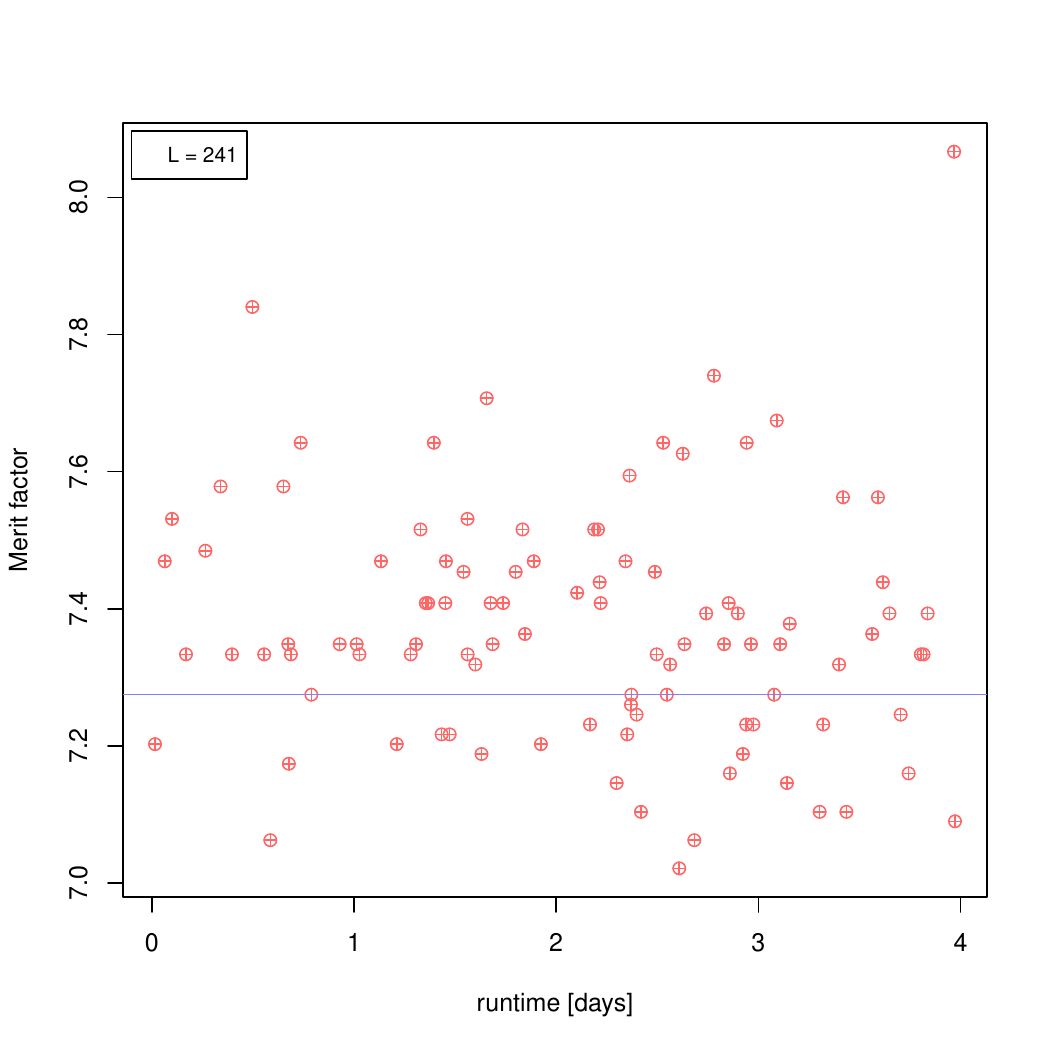}
\end{minipage}
\begin{minipage}{0.49\textwidth}
\includegraphics[width=0.99\textwidth]{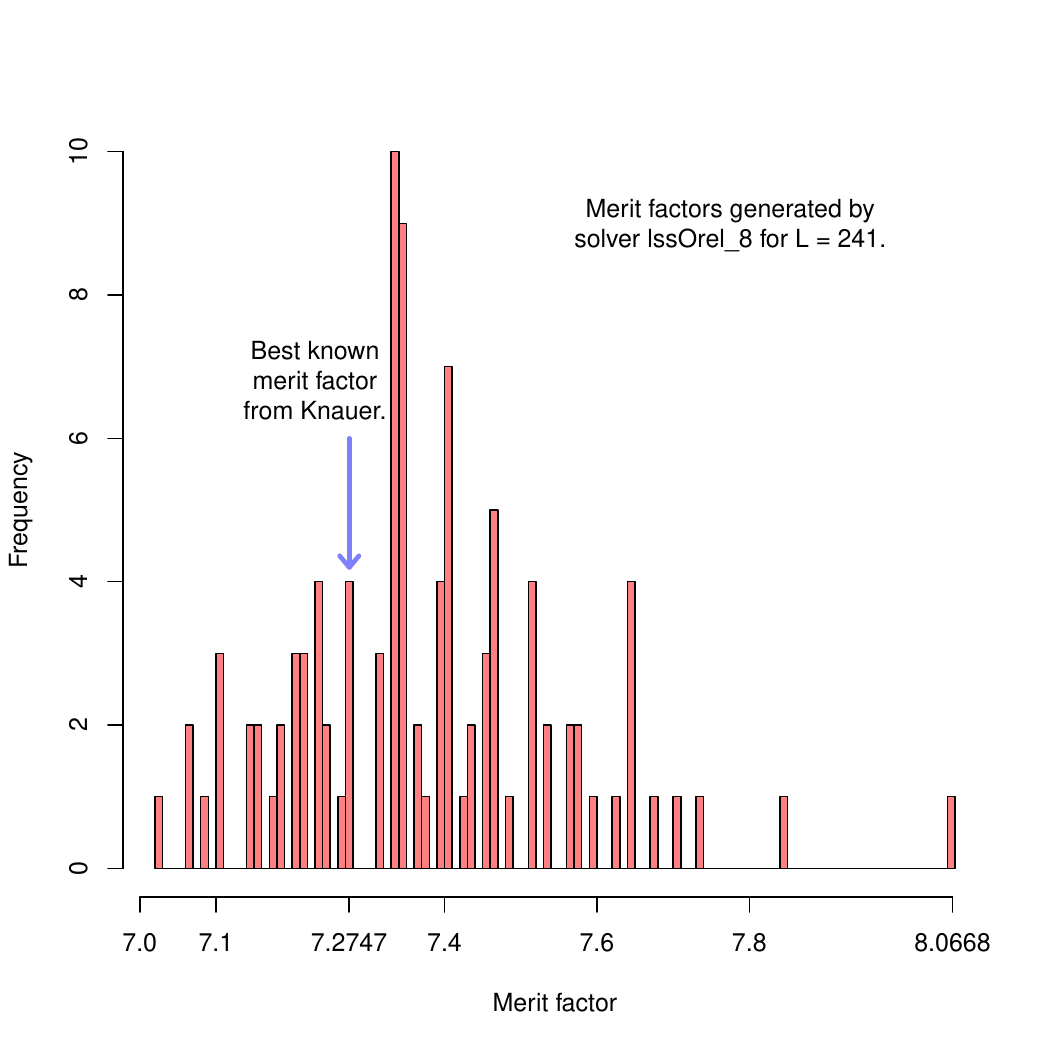}
\end{minipage}

\begin{minipage}{0.99\textwidth}
\centering
{\small (b) $L=241$: a runtime scatter plot of best merit factors
and a histogram of merit factor frequencies}
\end{minipage}

\caption{
Two empirical views illustrating the \labs\ problem:
(a) the asymptotic runtime predictions and observations,
(b) best merit factors for $L$ = 241 in a special-case experiment with \lssOrelE\
running on 100 instances for a total of 4 days.
}
\label{fg_histogram}
\end{figure*}

For additional empirical views that illustrate 
a number of the characteristics of the \lssOrelE, see
Figure~\ref{fg_histogram}. In Figure~\ref{fg_histogram}a, we observe the quality of
runtime predictions in comparisons with the observed \lssOrelE\ runtimes.
In Figure~\ref{fg_histogram}b,
we use a large instance of $L=241$ to demonstrate the variability in the
quality of solutions reported by  \lssOrelE:  the  
runtime scatter plot  and the histogram of 
best merit factors based on the sample size of 100.
In this experiment, the only stopping criterion is the \lssOrelE\ runtime limit of 4 days.
The best-known
merit factor for $L=241$ has been 7.2747~\cite{Lib-OPUS2-git_labs-Boskovic}; 
represented with the blue line in the histogram.
Observations of most interest in this figure include:
(1) the point in the scatter graph on the extreme left, where the merit factor of 7.2 is reached near the very start of the experiment and {\em is not improved in 4 days}, and
(2) the point on the extreme right, where the new best-known  merit factor of 8.0668
is reached just before the end of the 4 day experiment.
Overall, \lssOrelE\ found 69 solutions with a merit factor better than 7.2747.
Furthermore, the histogram 
shows that with the sample size of 100, there is only 1 solution with the best merit factor of 8.0668, and that there are now a
total of 24 unique merit factor values that exceed the value 7.2747,
currently reported as the best known value~\cite{Lib-OPUS2-git_labs-Boskovic}. 
More computational resources, better solvers, or both
are required to reach solutions with merit factor that will most likely exceed 
the value of 8.0668 for $L=241$.

\par
\begin{figure*}[t!]
  \vspace*{-10ex}
  \centering
  \subfloat[merit factors based on constructions from Legendre sequences]{
  \begin{minipage}{0.49\textwidth}
  \includegraphics[width=0.99\textwidth]{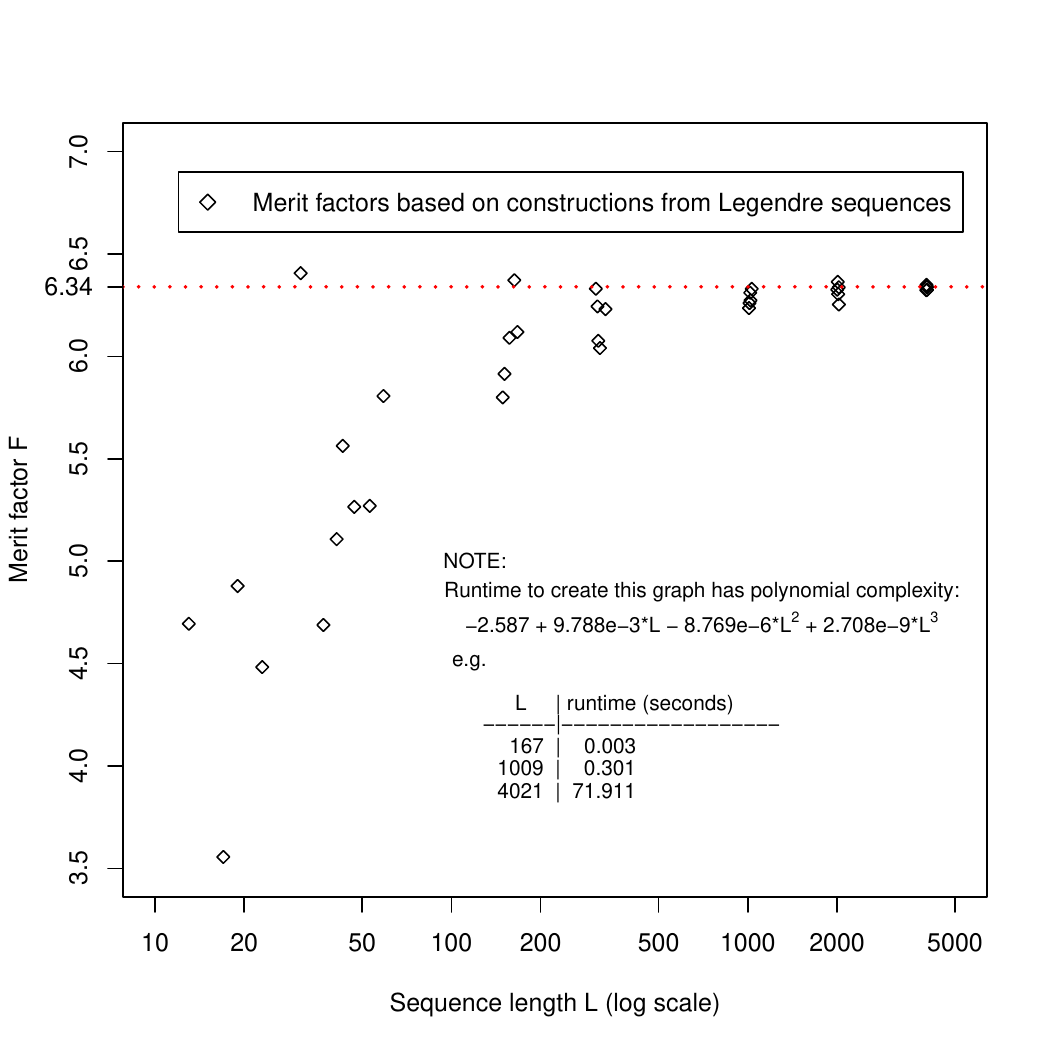}
  \label{fg_asymptotics_a}
  \vspace*{-1ex}
  \end{minipage}
  }
  \subfloat[asymptotes and merit factors based a stochastic solver]{
  \begin{minipage}{0.49\textwidth}
  \includegraphics[width=0.99\textwidth]{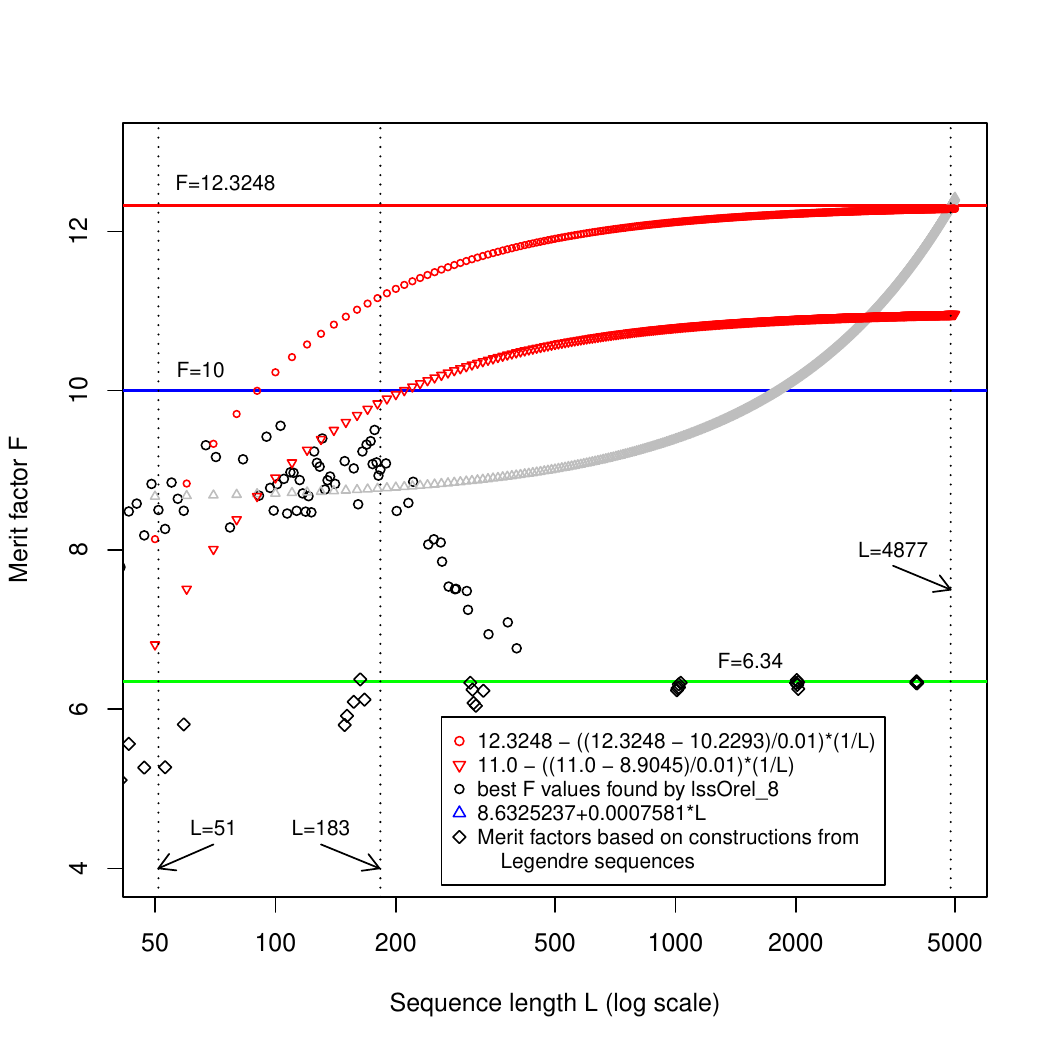}
  \label{fg_asymptotics_b}
  \vspace*{-1ex}
  \end{minipage}
  }
  \caption{
We contrast two views of merit factor asymptotics as $L$ increases towards the value of 5000. 
In (a), the merit factor values are based on constructions from Legendre sequences, using a technique similar 
to~\cite{Lib-OPUS-labs-2004-IEEE_TIT-Borwein,Lib-OPUS-labs-2004-IEEE_TIT-Parker-Legendre}.
Parameters $r$ and $t$ are taken in increments of $1/L$:
for $L <   500$, $r \in [0,0.5], t \in [0,0.1]$; 
for $L \ge 500$, $r \in [0.2,0.24], t \in [0.055,0.063]$.
Notably, the observed
merit factor variability is decreasing rapidly as $L$ increases.
For the prime values of (13, 17, 19, 23, 31, 37, 41, 43, 47, 53, 59), values of $F$ range from 3.55556 to 6.40667. 
For the prime values of (149, 151, 157, 163, 167), values of $F$ range from 5.79981 to 6.37224. 
For the prime values of (1009, 1013, 1019, 1021, 1031), values of $F$ range from 6.23938 to 6.33041 and runtime for each value of $L$ is less than 1 CPU second. 
For the prime values of (4003, 4007, 4013, 4019, 4021), values of $F$ range from 6.32348 to 6.34917 and runtime for each value of $L$ is $\approx 72$ CPU seconds.
The runtime complexity to construct these sequences is $O(L^3)$.
%
%
In (b), the linear model  $8.6325237 + 0.0007571*L$ `predicts' to cross the conjectured asymptotic value of 12.3248 at $L = 4877$. 
This model is based on extrapolation of
{\em observed merit factor values} in Figure~\ref{fg_Deviation-vs-SequenceLength}d:  for $L > 50$ and $L \leq 183$, the values of $F$ range from 8.2618 to 9.5577.
The crossover value of $L = 4877$ may at this point
be pessimistic. On the other hand,
there is a rigorous interpretation for the slope provided by Bernasconi in Figure~\ref{fg_Deviation-vs-SequenceLength}b,
see Figure 5 and Equations 22-24 in~\cite{Lib-OPUS-labs-1987-JourPhys-Bernasconi}.
By interpreting the slope at 12.3248 in Figure~\ref{fg_Deviation-vs-SequenceLength}b
as (12.3248 - 10.2293)/0.01), we create a non-linear predictor $b_U = 12.3248 - ((12.3248 - 10.2293)/0.01)*(1/L)$ 
as an upper bound and $b_L = 11.0 - ((11.0 - 8.9045)/0.01)*(1/L)$ as its lower bound counterpart.
For a reality check, recall the runtime asymptotics for the better of the two solvers, \lssOrelE, extrapolated
from~Figure~\ref{fg_comparison_lssRRts_vs_lssMAts}c: $0.000032*1.1504^L$ (Eq.~\ref{eq_runtime_seconds}), implying
the mean runtime to solve for `the best merit factor' for $L = 161$ approaches 
2.32 days (on a single CPU) and is increasing rapidly: 
28.9 days for  $L = 179$, 629.9 days for $L = 201$, and 467 years for $L=241$.
  }
  \label{fg_asymptotics}
\end{figure*}

\vspace*{1ex}\noindent
{\bf (9) Challenges for the next generation of labs solvers.}
We conclude the section with Figures~\ref{fg_merit_factors_and_solvability} and~\ref{fg_asymptotics}. 
Both of these figures summarize the most important findings in this paper as well as 
the challenges for the next generation of labs solvers.

The table in Figure \ref{fg_merit_factors_and_solvability-a} compares 
merit factors obtained with \lssOrelE\
with the best-known merit factors 
reported in the literature.
Notably, \lssOrelE\ always finds a solution that has equal or better merit factor than those reported earlier.
The merit factors where the best-known solutions were not skew-symmetric are marked with *. All these solutions have been
improved by \lssOrelE\ and {\em all the best-known solutions for odd instance sizes greater than 100 are now skew-symmetric}.
This is not unexpected; skew-symmetry significantly reduces the problem size and 
the solver has a better chance of finding new and better solutions for larger instances.

\begin{figure*}
\centering 
\vspace*{-4ex}
\subfloat[]{
\begin{minipage}{0.99\textwidth}
\renewcommand{\arraystretch}{0.9}
\centering
\begin{footnotesize}
\begin{tabular}{c|c c c c c c c c c | c c c }
	{\bf L} & \cite{Lib-OPUS-labs-1985-Phillips-Beenker} & \cite{Lib-OPUS-labs-1990-IEEE_TIT-Golay-skewsym} & 
	\cite{Lib-OPUS-labs-1993-Diploma-Reinholz-genetic-alg} & \cite{Lib-OPUS-labs-2007-Prestwich-local-search} & \cite{Lib-OPUS-labs-1998-IEEE_EC-Militzer} &
	\cite{Lib-OPUS2-labs-2014-homepage-Knauer} & \cite{Lib-OPUS-labs-2008-LMSLNS-Borwein}& \cite{Lib-OPUS-labs-2009-ASC-Gallardo-memetic} & 
	{\bf \lssOrelE} & $\mathit{hitO}$  & $C_{L}$ & $\Delta (E)$\\
	\hline \\[-1.8ex]
	107 & 6.53 & {\bf 8.46} & {\bf 8.46} & 8.36 & {\bf 8.46} & {\bf 8.4557} & - & {\bf 8.4557} & {\bf 8.4557} & 100 & 1 & 0 \\
	109 & 6.15 & {\bf 8.97} & {\bf 8.97} & {\bf 8.97} & {\bf 8.97} & {\bf 8.9736} & - & {\bf 8.9736} & {\bf 8.9736} & 100 & 1 & 0 \\
	111 & 6.02 & {\bf 8.97} & {\bf 8.97} & {\bf 8.97} & {\bf 8.97} & {\bf 8.9672} & - & {\bf 8.9672} & {\bf 8.9672} & 100 & 1 &  0\\
	113 & 6.33 & {\bf 8.49} & {\bf 8.49} & {\bf 8.49} & {\bf 8.49} & {\bf 8.4900} & - & {\bf 8.4900} & {\bf 8.4900} & 100 & 2 & 0\\
	115 & 6.40 & {\bf 8.88} & 8.60 & {\bf 8.88} & {\bf 8.88} & {\bf 8.8758} & - &  {\bf 8.8758} & {\bf 8.8758} & 100 & 1 & 0\\
	117 & 6.42 & {\bf 8.71} & 8.12 & {\bf 8.71} & {\bf 8.71} & {\bf 8.7080} & - & {\bf 8.7080} & {\bf 8.7080} & 100 & 1 & 0\\
	119 & 6.01 & - & 7.67 & {\bf 8.48} & 8.02 & {\bf 8.4796} & - & {\bf 8.4796} & {\bf 8.4796} & 100 & 1 & 0\\
	121 & 6.61 & - & {\bf 8.67} & - & {\bf 8.67} & {\bf 8.6736} & - &  {\bf 8.6736} & {\bf 8.6736} & 100 & 1 & 0\\
	141 & 6.01 & - & 7.45 & - & {\bf 8.83} & {\bf 8.8282} & - & {\bf 8.8282} & {\bf 8.8282} & 100 & 2 & 0\\
	149 & - & - & - & - & - & {\bf 9.1137} &  {\bf 9.1137} & - & {\bf 9.1137} & 95 & 1 & 0\\
	157 & - & - & - & - & - & {\bf 9.0223} &  {\bf 9.0223} & - & {\bf 9.0223} & 98 & 1 & 0\\
	161 & 6.02 & - & 6.89 & - & 8.39 & 8.5266 & - & {\bf 8.5718} & {\bf 8.5718} & 76 & 2 & 0\\
	165 & - & - & - & - & - & {\bf 9.2351} &  {\bf 9.2351} & - & {\bf 9.2351} & 26   & 1 & 0\\       
	169 & - & - & - & - & - & {\bf 9.3215} &  {\bf 9.3215} & - & {\bf 9.3215} & 24   & 1 & 0\\
	173 & - & - & - & - & - & 9.3179 &  {\bf 9.3645} & - & {\bf 9.3645} & 28 & 1 & 0\\
	175 & - & - & - & - & - & 8.9078 &  {\bf 9.0768} & - & {\bf 9.0768} & 12 & 1 & 0\\
	177 & - & - & - & - & - & 8.6640 &  {\bf 9.5052} & - & {\bf 9.5052} & 10 & 1 & 0\\
	179 & - & - & - & - & - & 8.4452 &  {\bf 9.0974} & - & {\bf 9.0974} & 6 & 1 & 0\\
	181 & 7.70 & - & 6.77 & - & 7.75 & 8.6304 & - & 7.7194 & {\bf 8.9316} & 6 & 1 & 64 \\
	183 & - & - & - & - & - & 8.3932 &  {\bf 9.0073} & - & {\bf 9.0073} & 5 & 2 & 0 \\
	189 & - & - & - & - & - & {\bf 9.0847} &  {\bf 9.0847} & - & {\bf 9.0847} & 1 & 1 & 0\\
	201 & - & - & 6.29 & - & 7.46 & 8.2116 & - &  7.6633 & {\bf 8.4876} & 1 & 1 & 80\\
	215 & - & - & - & - & - & ~$8.1641^{*}$ & - & - & {\bf 8.5888} & 1 & 1 & 140\\
	221 & - & - & - & - & - & 7.6171 & - & - & {\bf 8.8544} & 1 & 1 & 448\\
	241 & - & - & - & - & - & 7.2747 & - &  - & {\bf 8.0668} & 1 & 1 & 392\\
	249 & - & - & - & - & - & ~$7.2431^{*}$ & - & - & {\bf 8.1323} & 1 & 1 & 468\\
	259 & - & - & - & - & - & ~$7.1287^{*}$ & - & - & {\bf 8.0918} & 1 & 1 & 560\\
	261 & - & - & - & - & - & ~$7.1108^{*}$ & - & - & {\bf 7.8517} & 1 & 1 & 452\\
	271 & - & - & - & - & - & ~$7.0037^{*}$ & - & - & {\bf 7.5386} & 1 & 1 & 372\\
	281 & - & - & - & - & - & 7.0957 & - & - & {\bf 7.5058} & 1 & 1 & 304\\
	283 & - & - & - & - & - & ~$7.0291^{*}$ & - & - & {\bf  7.5088} & 1 & 1 & 364\\
	301 & - & - & - & - & - & - & - & - & {\bf  7.4827} & 1 & 1 & - \\
	303 & - & - & - & - & - & 7.1115 & - & - & {\bf 7.2462} & 1 & 1 & 120 \\
	341 & - & - & - & - & - & - & - & - & {\bf 6.9397} & 1 & 1 & - \\
	381 & - & - & - & - & - & - & - & - & {\bf 7.0893} & 1 & 1 & - \\
	401 & - & - & - & - & - & - & - & - & {\bf 6.7632} & 1 & 1 & -\\
\end{tabular}
\end{footnotesize}
\label{fg_merit_factors_and_solvability-a}
\end{minipage}
}

\subfloat[]{
\begin{minipage}{0.49\textwidth}
\begin{footnotesize}
 \begin{tabular}{@{~}r@{~}|l l | l l}
  \multirow{2}{*}{L} & \multicolumn{2}{c|}{runtime [years]} & \multicolumn{2}{@{~}c@{~}}{100 $\times$ runtime [years]} \\
 & \multicolumn{1}{c}{BB} 
 & \multicolumn{1}{c|}{ratio$^\dagger$} 
 & \multicolumn{1}{c}{\lssOrelE} 
 & \multicolumn{1}{c}{\lssMAts} \\
  \hline
75 & 1.1590e-04~\, & 3.1194e+03 & 3.7154e-06$^*$ & 3.3649e-06$^+$ \\
77 & 2.8983e-04~\, & 5.7999e+03 & 4.9971e-06~\, & 4.7695e-06~\,\\
79 & 5.6726e-04~\, & 8.7172e+03 & 6.5074e-06$^*$ & 6.3574e-06$^+$ \\
81 & 8.9805e-04~\, & 1.0428e+04 & 8.6120e-06$^*$ & 8.7384e-06$^+$ \\
83 & 1.1310e-03~\, & 2.8420e+03 & 3.9795e-05~\, & 8.5217e-06~\,\\
85 & 2.3780e-03~\, & 1.5766e+04 & 1.5083e-05$^*$ & 1.6510e-05$^+$ \\
87 & 4.5392e-03~\, & 2.2739e+04 & 1.9962e-05$^*$ & 2.2693e-05$^+$ \\
89 & 9.0476e-03~\, & 3.4249e+04 & 5.3345e-06 & 1.0364e-05 \\
115 & 1.5758e+01$^*$ & 1.7756e+06 & 8.8751e-04~\, & 1.8324e-03~\,\\
121 & 9.0011e+01$^*$ & 4.2144e+06 & 2.1358e-03~\, & 4.2444e-03~\,\\
127 & 5.1414e+02$^*$ & 1.0145e+07 & 5.0680e-03~\, & 1.0521e-02~\,\\
141 & 2.9986e+04$^*$ & 7.7784e+07 & 3.8551e-02~\,  & 1.2189e-01~\,   \\
151 & 5.4732e+05$^*$ & 3.4972e+08 & 1.5650e-01$^*$ & 5.9805e-01$^+$ \\
161 & 9.9897e+06$^*$ & 1.5723e+09 & 6.3534e-01$^*$ & 2.9342e+00$^+$ \\
181 & 3.3280e+09$^*$ & 3.1783e+10 & 1.0471e+01$^*$ & 7.0634e+01$^+$ \\
201 & 1.1087e+12$^*$ & 6.4245e+11& 1.7257e+02$^*$ & 1.7003e+03$^+$ \\
241 & 1.2304e+17$^*$ & 2.6322e+14 & 4.6744e+04$^*$ & 9.8259e+05$^+$ \\
\hline
\multicolumn{5}{l}{ }  \\[-1.5ex]
\multicolumn{5}{l}{The unmarked values have been observed experimentally, }  \\
\multicolumn{5}{l}{the values marked with $^*$ and $^+$ are based on predictions below: }  \\
\multicolumn{5}{l}{BB = 1.55311e-06*$1.3370^L/(3600*24*366)$ }  \\
\multicolumn{5}{l}{$^\dagger$~ratio~=~runtime(BB)/runtime(\lssOrelE) }  \\
\multicolumn{5}{l}{$^*~\lssOrelE = 100*(0.000032*1.1504^L/(3600*24*366))$} \\
\multicolumn{5}{l}{$^+~\lssMAts~~~~  = 100*(0.000007*1.1724^L/(3600*24*366))$} \\
\end{tabular}
\end{footnotesize}
\label{fg_merit_factors_and_solvability-b}
\end{minipage}
}
\subfloat[]{
\begin{minipage}{0.49\textwidth}
\vspace*{-1.1ex}
\hspace*{+0.5em}
\includegraphics[width=1\textwidth]{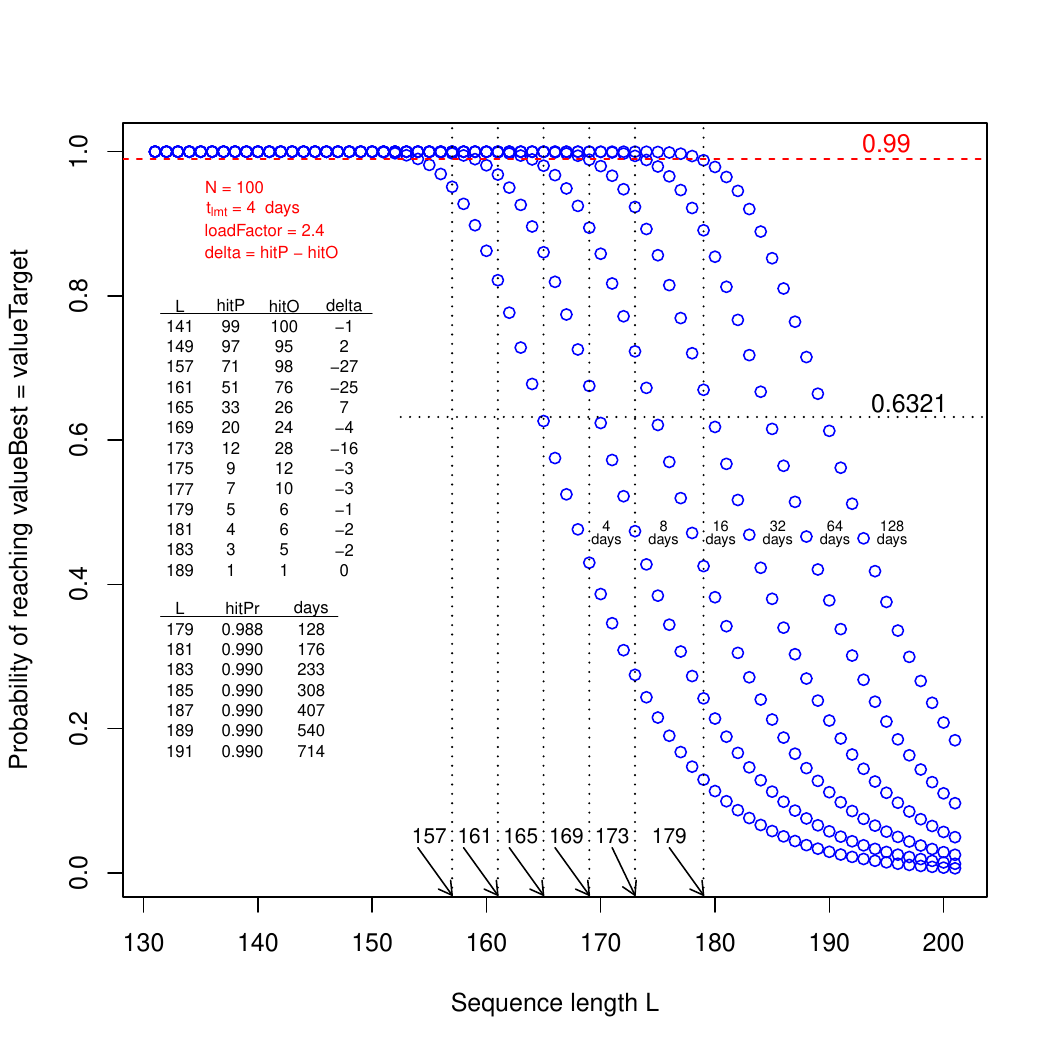}
\label{fg_merit_factors_and_solvability-c}
\vspace*{-4.5ex}
\end{minipage}
}
\vspace*{-1.1ex}
\caption{
(a) Comparisons of the best-known merit factors reported in the literature and 
the best-known merit factors obtained by \lssOrelE. 
(b) Observed/predicted asymptotic performance of state-of-the-art branch-and-bound \labs\ solver
under skew-symmetry \cite{Lib-OPUS-labs-2013-arxiv-Prestwich-branch-and-bound-odd} versus the observed/predicted
runtime of two stochastic solvers under skew-symmetry: \lssOrelE\ and \lssMAts.
(c) A hit ratio model $\mathit{hitP}_r$ for the solver \lssOrelE\
predicts the probability of reaching {\em uncensored} {\em valueBest = valueTarget} solutions on a grid of $N$ independent processors,
given 
(1) the runtime limit in days of $t_{lmt} \in (4, 8, 16,...,128)$, 
(2) the ideal value of processor load factor $\mathit{loadFactor} = 1$, and
(3) $\overline{m}(L) = 0.000032*1.1504^L/(3600*24)$. 
In other words, when $\mathit{hitP}_r \ge 0.99$, we predict that
at most 1\% of the $N$ instances may be censored by the solver.
The table summary under the headline of $\mathit{loadFactor} = 2.4$
displays a remarkable agreement with 
the predicted number of hits ($\mathit{hitP}$) versus the
observed number of hits ($\mathit{hitO}$);
the empirical value of $\mathit{loadFactor} = 2.4$ is associated with~\cite{2014-Web-SLING}.
Small deviations in the column {\tt delta} 
are of the same order as the ones explained in Figure~\ref{fg_histogram}-a.
For formulas about $\mathit{hitO}$, $\mathit{hitP}$, $\mathit{hitP_r}$, and
$\overline{m}(L)$ see Eqs.~\ref{eq_hitO}, \ref{eq_hitP_r}, \ref{eq_hitP}, \ref{eq_runtime_seconds}.
}
\label{fg_merit_factors_and_solvability}
\end{figure*}

The table also lists additional columns:
the observed number of hits $\mathit{hitO}$ as defined in Eq.~\ref{eq_hitO}, 
the cardinality of the canonic solutions $C_L$ as defined in Section~\ref{sec_notation},
and the energy level difference $\Delta (E)$ of the improved solution with respect to
the best-known previous solution. The value of $\Delta (E)$ 
conveys the significance of a given merit factor improvement.
For example, for $L=221$, the solver \lssOrelE\ reports the merit factor of 8.8544 which represents a reduction of 448 energy level
(under the constraint of skew-symmetry)
with respect to the merit factor of 7.6171 reported by Knauer~\cite{Lib-OPUS2-git_labs-Boskovic} (without the constraint of skew-symmetry). 
It is instructive to observe how the observed number of hits $\mathit{hitO}$ relates to the
the cardinality of canonic solutions $C_L$: $C_L \le 2$ and remains as $C_L = 1$ for most of instances where
$\mathit{hitO} > 1$, even when $\mathit{hitO} \gg 1$.  
%

As the size of the \labs\ problem increases, the monotonically decreasing number of observed hits
$\mathit{hitO}$ 
illustrates not only the limitation of \lssOrelE\, it also suggests the need for
massively parallel computational resources so that we can continue to maintain 
the observable hit ratio at 100\% with $N \ge 100$. We argue that pursuing this strategy,
we have the best chance of finding solutions with merit factors approaching the 
postulated limit of 12.3248.

The table in Figure~\ref{fg_merit_factors_and_solvability}b
shows that
the computational complexity of the new branch-and-bound 
solver~\cite{Lib-OPUS-labs-2013-arxiv-Prestwich-branch-and-bound-odd}, now limited
to odd values of $L$ under skew-symmetry, 
is $O(1.3370^L)$. However, the B\&B solver scales poorly and the stochastic solvers are the only viable alternative.
Both \lssMAts\ and \lssOrelE\
stochastic solvers can find the same optimum solutions
with significantly less computational effort, even when comparing
a single run with the branch and bound solver with the runtime for 100 repeated runs of each stochastic solver.

The hit ratio model $\mathit{hitP_r}$ in Figure~\ref{fg_merit_factors_and_solvability}c
predicts, for the solver \lssOrelE,
the probability of reaching {\em uncensored} {\em valueBest = valueTarget} solutions on a grid of $N$ independent processors,
given (1) the runtime limit in days of $t_{lmt} \in (4, 8, 16,...,128)$, 
(2) the ideal value of processor load factor $\mathit{loadFactor} = 1$,
and (3) $\overline{m}(L) = 0.000032*1.1504^L/(3600*24)$). 
The agreement with  
the predicted number of hits, $\mathit{hitP}$, versus the
observed number of hits, $\mathit{hitO}$ in the
small table summary under empirically determined 
$\mathit{loadFactor} = 2.4$ under~\cite{2014-Web-SLING} 
is as remarkable as is the 
agreement with the  mean value runtime predictions  versus the observed runtime means
for the solver \lssOrelE\ and values of $L \le 141$, i.e. the table shown
in Figure~\ref{fg_histogram}a.

In Figure~\ref{fg_asymptotics} we contrast two views of merit factor asymptotics as $L$ increases towards the value of 5000. 
In Figure~\ref{fg_asymptotics}a, the merit factors rely on Legendre sequences, using a construction technique similar 
to~\cite{Lib-OPUS-labs-2004-IEEE_TIT-Borwein,Lib-OPUS-labs-2004-IEEE_TIT-Parker-Legendre}.
A few short and mostly
very long sequences, and with merit factors hovering around 6.34,
can be  computed in polynomial runtime $O(L^3)$. The best merit factor under this construction, 6.40667, has been found for $L=31$.
On the other hand, as shown in Figure~\ref{fg_asymptotics}b, 
all merit factors found by \lssOrelE\ are well above 6.34, some with merit factors larger than 9.0, and all 
below 10.0.
In Figure~\ref{fg_BB-vs-others-asymptotes} we show that 
even to find the binary sequence of length $L=573$ with merit factor of at least 6.34, the 
average runtime for the current generation of stochastic solvers such as \lssOrelE\
is around 32 years. 
Nevertheless, the trend of merit factors achieved with \lssOrelE\ 
in Figure~\ref{fg_asymptotics}b points in the right direction -- as long as we 
continue to find {\em uncensored solutions}
with progressively increasing merit factors.
Currently, sequences that would converge closer
to the conjectured asymptotic value of 12.3248 are yet to be discovered.
In order to find better merit factors as $L$ increases we
need both: new approaches to design better solvers and 
a significant increase in computational resources.

\section{Discussion}
\label{sec_introduction}
\noindent
This paper focuses on the stochastic solvers for the \labs\ problem. Stochastic solvers can not guarantee optimal solutions. We can compare the performance
of each solver only by measuring
the memory footprint, the walk length, the probe count, and the runtime until it returns a solution coordinate with {\em the best known value} 
({\em BKV}) -- which may or may not be an optimum value. When {\em BKV} is not improved, only repeated after a large number of independent experiments, a statistician can make not only a {\em reliable} statement about the average-case performance of the specific solver but also about the probability of ever finding a better solution with the given solver. 

The branch-and-bound solvers do guarantee optimal solutions, but only for instance of size up to $L=66$ without the use of skew-symmetry and up to 
$L=119$~\cite{Lib-OPUS-labs-2016-Packebusch} with the use of skew-symmetry.
In contrast, the new stochastic solver \lssOrel, also using skew-symmetry, returned solutions with {\em BKVs} that match {\em all} of the exact solutions reported by branch-and-bound solvers -- not only in a single experiment and in a small fraction of runtime required by the branch-and-bound solver, but also in at least 100 independent experiments.
Moreover, \lssOrel\ now reports  {\em BKVs} for skew-symmetric instances 
$119 < L \le 401$ on a grid of 100 processors with a runtime limit of 4 days. The number of repeated  {\em BKVs} drops from 100 to 95, 76, 6, 1 starting with instances $L \ge 151$.

The analysis of the self-avoiding walk reveals that on smaller instances the solver
\lssOrelU, where the length of the walk segment is kept at `unlimited',
has an advantage over the nominal solver where
the walk segment length is limited. However, as $L$ increases, 
the advantage of \lssOrelU\ decreases in terms of $runtime$.
The reason for the reduced efficiency of the solver is the {\em runtime} cost of memory management, necessary to maintain the self-avoiding walk. 
The instance with $L=105$ hits the memory restriction of 8 GB on our processor.
The experiments show
that with \lssOrelE, where the length of the walk segment is limited to $8 * \frac{L+1}{2}$, 
we get the best asymptotic average case
performance: cntProbe is $650.07*1.1435^L$ and the memory footprint is constant at about 1.8 MB.

The rigorous asymptotic experiments with solvers \lssMAts\ and \lssRRts, both using the tabu search, show that
(1) the evolutionary
component within \lssMAts\ is not effective and (2), the solver \lssOrelE\ significantly outperform \lssMAts.
The same asymptotic performance testing methodology, including the platform-independent performance comparisons, can be applied to engineering
the next generation of \labs\ solvers.

\OMIT{
The new best-known merit factors were also obtained using \lssOrelE\ for all tested instances that are greater than $L=160$.
We also take into account the constructive method that can produce sequences in polynomial time with merit factor of approximately 
6.34 and these methods are suitable for very large instance sizes. While our stochastic solver is useful for instance sizes
up to approximately $L=500$.
}

\OMIT{
On Nov 26, 2015, at 1:21 PM, Borko Bo?kovi? <Borko.Boskovic@um.si> wrote:

V prilogi so vsi modeli.

na kratko:

rulerOrder m=9 - 14
samleSize = 100
walkSegmCoef = 2
walkSegmLmt = walkSegmCoef * rulerLength

glede na order (m):
model(walklength) = 0.00073 ? 5.3138^m
model(runtime) = 0.000000000065 ? 6.6941^m

glede na length:
model(walklength) = 27.38208*1.0127^length 
model(runtime) = 0.000018883248*1.1206^length 

for labs:
walkLength(lssOrel_U) = 6.29*1.1498^L
runtime(lssOrel_8) = 0.000032?1.1504^L,
}
\section{Conclusions And Future Work}
\label{sec_conclusions}
\noindent
This paper introduces a new stochastic solver and demonstrates its merits by following a rigorous methodology of experimental design.
We now have models that predict not only the asymptotic runtime performance of this solver,
we also have similar models for alternative state-of-the-art 
solvers~\cite{Lib-OPUS-labs-2009-ASC-Gallardo-memetic}. Moreover, we have shown why the new {\em self-avoiding walk search} solver \lssOrel\ dominates the solver \lssMAts\ (memetic/tabu search) and why the solvers \lssRRts\ (tabu search only) and \lssMAts\  are equivalent -- at least 
when applied to the \labs\ problem. Despite their superficial similarities, the self-avoiding walk search and the long-established tabu search are not equivalent.

We borrowed the notion of self-avoiding walk from chemists and physicists. In the follow-up work, we are generalizing these stochastic walks -- on directed vertex-weighted graphs -- as being Hamiltonian as well 
as Eulerian~\cite{Lib-OPUS2-ebook-markov-2016-Brglez}. For example, the Hamiltonian walk illustrated in
Figure~\ref{fg_walks_restarts_vs_saw}b reaches the target vertex in 17 steps. However, by 
considering edges as bidirectional and switching to an Eulerian walk  after the
step 7, we can reach the target vertex in 11 steps only. Work in progress includes an exploration
of new walk strategies to improve the current \labs\ solver; such strategies are also showing
promising results in domains of other combinatorial problems, ranging from 
{\em optimum Golomb ruler} or {\tt ogr} problem, minimum vertex/set cover, linear ordering, 
protein folding, and beyond. A version of the new walk strategies is particularly
effective in solving the {\tt ogr} problem~\cite{Lib-OPUS2-ebook-ogr-2016-Brglez}; the new solver
already dominates 
a state-of-the-art memetic solver~\cite{Lib-OPUS-ogr-2007-Springer-Cotta-Golomb_Ruler-Memetic}.

The invariants that characterize \lssOrel\ are the models
for the {\em average} {\em cntProbe},  {\em walkLength}, and {\em runtime}. 
These invariants are standards that
should not only be met but also improved by the new \labs\ solver. Both {\em cntProbe} and {\em walkLength}
facilitate platform-independent performance comparisons
with other solvers. Thus:
\begin{eqnarray}
 {\mathit cntProbe(\lssOrelE)}   &=& 650.07   * 1.1435^L \\ \nonumber
 {\mathit walkLength(\lssOrelE)} &=& 35.51    * 1.1321^L \\ 
 {\mathit runtime(\lssOrelE)}    &=& 0.000032 * 1.1504^L   \nonumber
\end{eqnarray}

\begin{table}[t!]
\caption{
Predictions, based on Eq.~\ref{eq_forN}, for the required number of processors $N$ running concurrently for 5 years (5*365 days)
in order to get at least 100
repeated hits of the best known merit factor with the solver \lssOrelE.
In this table, the runtime for each processor is reported in days:\\
\hspace*{3em}${\mathit runtime(\lssOrelE)} = (0.000032 * 1.1504^L)/(3600*24)$
}
\label{tb_predicting_N}
  \centering
  \begin{footnotesize}
  \vspace*{-1ex}
 \begin{tabular}{c|c c}
 $L$& $\mathit runtime$ & $\mathit N$ \\
 \hline
199 &  4.759e+02 &  1.030e+02\\
216 &  5.152e+03 &  3.360e+02\\
246 &  3.447e+05 &  1.893e+04\\
283 &  6.149e+07 &  3.369e+06\\
333 &  6.781e+10 &  3.715e+09\\
 \end{tabular}
 \end{footnotesize}
\vspace*{-3ex}
\end{table}
For a bigger picture about the hardness of 
of the \labs\ problem, we contrast it with the {\tt ogr}  problem, given that our
 experiments with {\tt ogr} solvers in~\cite{Lib-OPUS2-ebook-ogr-2016-Brglez}
show that the asymptotic runtime complexity of a stochastic {\tt ogr} solver is significantly lower than the 
complexity of \lssOrelE. 
Consider the results in Table~\ref{tb_predicting_N}.
We use the Eq.~\ref{eq_forN} to predict 
the required number of processors $N$ running concurrently for 5 years (5*365 days)
in order to get at least 100
repeated hits of the best known merit factor with the solver \lssOrelE.
In the example that follows  
Eq.~\ref{eq_forN} we take $L = 179$, a runtime limit of 4 days
(under the {\em loadFactor} of 2.4), find
$\mathit{hitP_r} = 0.05607801$ and get the prediction of $N = 1784$ processors 
that we should run concurrently
in order to achieve at least $N_c = 100$ hits (repeated best-known value solutions).
In Table~\ref{tb_predicting_N} we choose a runtime limit 
of 5 years (5*365 days with a {\em loadFactor} of 1) and
assign a subset of values of $L$ associated with the known optimum Golomb rulers.
The choice of the runtime limit of 5 years is related to the {\em waiting time},
under {\em massively parallel computational effort}, that elapsed
before finding the optimal ogr solution for $L=553$ in
2014~\cite{WEB-Project-Golomb_Rulers,WEB-Wikipedia-Golomb_Rulers}.

Given results in Table~\ref{tb_predicting_N}, finding the near-optimum values for 
the \labs\ problem with $L > 246$ may not be an option unless we devise a faster solver. 
Revisiting the 100-processor, 4-day experiment with $L=241$ in 
Figure~\ref{fg_histogram}, we find that under a runtime limit of 5 years we need to run in parallel at least
9425 processors. This number increases to 46923 processors if the runtime limit is 1 year.

In conclusion, the paper reports a number of 
important computational milestones.
Using the grid environment with only 100 processors and a runtime limit of only 4 days, \lssOrel\ 
either re-confirmed or improved the best merit factors reported in the literature. For some instances 
the improvement is huge. For  $L=259$, the best
merit factor was improved from 7.1287 to 8.0918 which represents a reduction 
of 560 energy levels, from 4705 to 4145. All of the best known solutions for instances with $L_{odd} > 100$ 
are now skew-symmetric. 
\par\vspace*{-1.5ex}

\section*{Acknowledgments}
\par\vspace*{-0.9ex}\noindent
This work was supported in part by the Slovenian Research Agency under program
P2-0041: Computer Systems, Methodologies, and Intelligent Services.\\

\noindent
We gratefully acknowledge the cooperation of Jos\'{e} E. Gallardo, Carlos Cotta and Antonio J. Fern\'{a}ndez 
in providing the source code of solvers \lMAts\ and \lssMAts.

We also thank for the generous advice and encouragement offered by Dr.~Jacob Bernasconi.
The comments from anonymous reviewers helped to clarify a number of points in the paper.
For the presentation of results, we were greatly influenced by
the features offered in 
R~\cite{WEB-Project-R} and the arguments in~\cite{Lib-OPUS-book-2006-Tufte-Beautiful_Evidence}.

\section*{Source code release}
\noindent
For links to 
(1) comprehensive tables of {\em best-known-value solutions},
(2) the number of {\em unique}  solutions in {\em canonic form} and the solutions themselves, 
(3) the source code of relevant solvers, and 
(4) the {\tt xBed} statistical testing environment, customized for the  \labs\ problem,
see~\cite{Lib-OPUS2-git_labs-Boskovic}.

In addition,
a crowd-sourcing server prototype to facilitate experimentation and push the
frontiers on finding new {\em best-known values, BKVs,}  for the 
\labs\ problem  
is under construction. 
Researchers will be able to download the \labs\  solver
\lssOrelE\ with the data set and simply {\em run the solver on their host} in the browser. Upon finding the solution (or a breakthrough value that improves the current {\em BKV}), the browser will return solution details to the server and start 
a new run on the local host either with the current or the new {\em BKV}. For an interactive \labs\ puzzle and the timeline
of the crowd-sourcing server prototype availability~\cite{Lib-OPUS2-git_labs-Boskovic}.


\begin{thebibliography}{10}
\expandafter\ifx\csname url\endcsname\relax
  \def\url#1{\texttt{#1}}\fi
\expandafter\ifx\csname urlprefix\endcsname\relax\def\urlprefix{URL }\fi
\expandafter\ifx\csname href\endcsname\relax
  \def\href#1#2{#2} \def\path#1{#1}\fi

\bibitem{Lib-OPUS-labs-1977-IEEE_TIT-Golay-sieves}
M.~J. Golay, Sieves for low autocorrelation binary sequences, IEEE Transactions
  on Information Theory 23 (1977) 43--51. \doi{10.1109/TIT.1977.1055653}.

\bibitem{Lib-OPUS-labs-1982-IEEE_TIT-Golay-merit-proof}
M.~J. Golay, The merit factor of long low autocorrelation binary sequences,
  IEEE Transactions on Information Theory IT-28~(3) (1982) 543--549.
  \doi{10.1109/TIT.1982.1056505}.

\bibitem{Lib-OPUS-labs-1990-IEEE_TIT-Golay-skewsym}
M.~J. Golay, D.~B. Harris, A new search for skewsymmetric binary sequences with
  optimal merit factors, IEEE Transactions on Information Theory 36~(5) (1990)
  1163--1166. \doi{10.1109/18.57219}.

\bibitem{Lib-OPUS-labs-1972-Golay}
M.~Golay, {A class of finite binary sequences with alternate auto-correlation
  values equal to zero}, IEEE Trans. Inf. Theor. 18~(3) (1972) 449--450.
  \doi{10.1109/TIT.1972.1054797}.

\bibitem{Lib-OPUS-labs-1985-Phillips-Beenker}
G.~Beenker, T.~Claasen, P.~Hermens, {Binary sequences with a maximally flat
  amplitude spectrum}, Philips J. Res. vol. 40 (1985) 289--304.

\bibitem{Lib-OPUS-labs-2000-Pasha}
I.~A. Pasha, P.~S. Moharir, N.~S. Rao, Bi-alphabetic pulse compression radar
  signal design, Sadhana 25~(5) (2000) 481--488.
  \doi{10.1007/BF02703629}.

\bibitem{Lib-OPUS-labs-1987-JourPhys-Bernasconi}
J.~Bernasconi, {Low autocorrelation binary sequences: statistical mechanics and
  configuration space analysis}, J. Phys. 48 (1987) 559--567.
\doi{10.1051/jphys:01987004804055900}.

\bibitem{Lib-OPUS-labs-1968-littlewood}
J.~Littlewood, Some problems in real and complex analysis, Heath mathematical
  monographs, D. C. Heath, 1968.

\bibitem{Lib-OPUS-labs-2002-Borwein}
P.~Borwein, Computational Excursions in Analysis and Number Theory,
  Springer-Verlag, 2002. \doi{10.1007/978-0-387-21652-2}.

\bibitem{Lib-OPUS-labs-1996-Stadler}
P.~F. Stadler, Landscapes and their correlation functions, Journal of
  Mathematical Chemistry 20~(1) (1996) 1--45. \doi{10.1007/BF01165154}.

\bibitem{Lib-OPUS-labs-2016-Schmidt}
K.-U. Schmidt, J.~Willms, Barker sequences of odd length, Designs, Codes and
  Cryptography 80~(2) (2016) 409--414. \doi{10.1007/s10623-015-0104-4}.

\bibitem{Lib-OPUS-labs-2016-Packebusch}
T.~Packebusch, S.~Mertens, Low autocorrelation binary sequences, Journal of
  Physics A: Mathematical and Theoretical 49~(16) (2016) 165001.
  \doi{10.1088/1751-8113/49/16/165001}.
  
\bibitem{Lib-OPUS-labs-1994-Marinari}
E.~Marinari, G.~Parisi, F.~Ritort, Replica field theory for deterministic
  models: I. binary sequences with low autocorrelation, Journal of Physics A:
  Mathematical and General 27~(23) (1994) 7615--7645.
  \doi{10.1088/0305-4470/27/23/010}.

\bibitem{Lib-OPUS-labs-2004-Springer-Jedwab-survey}
J.~Jedwab, A Survey of the Merit Factor Problem for Binary Sequences, in: T.~Helleseth, D.~Sarwate, H.-Y.
  Song, K.~Yang (Eds.), Sequences and Their Applications - SETA 2004, Vol. 3486
  of Lecture Notes in Computer Science, Springer Berlin Heidelberg, 2005, pp.
  30--55. \doi{10.1007/11423461\_2}.

\bibitem{Lib-OPUS-labs-2010-Elsevier-Ukil-theory}
A.~Ukil, Low autocorrelation binary sequences: Number theory-based analysis for minimum
  energy level, Barker codes, Digit. Signal Process. 20~(2) (2010) 483--495.
  \doi{10.1016/j.dsp.2009.08.003}.

\bibitem{BernasconiMail}
J.~Bernasconi, personal communication (2015).

\bibitem{Lib-OPUS-labs-2003-GECCO-Goldberg-periodic}
M.~Pelikan, D.~E. Goldberg,
  HHierarchical BOA Solves Ising Spin Glasses and MAXSAT, in: Proceedings of the 2003
  international conference on Genetic and evolutionary computation: PartII,
  GECCO'03, Springer-Verlag, Berlin, Heidelberg, 2003, pp. 1271--1282.
  \doi{10.1007/3-540-45110-2\_3}.

\bibitem{Lib-OPUS-labs-1996-JPhysA-Mertens-BB_solutions}
S.~Mertens, Exhaustive search for low-autocorrelation binary sequences, Journal
  of Physics A: Mathematical and General 29 (1996) 473--481, The sequences for
  $48 < L \le 60$ have been found with an improved implementation due to Heiko
  Bauke. \doi{10.1088/0305-4470/29/18/005}. All values are available at
  \url{http://www-e.uni-magdeburg.de/mertens/research/labs/open.dat}.

\bibitem{Lib-OPUS2-git_labs-Boskovic}
B.~Bo\v{s}kovi\'{c}, F.~Brglez, J.~Brest, {A GitHub Archive for Solvers and
  Solutions of the {\tt labs} problem}, {For updates, see \newline
  \url{https://github.com/borkob/git_labs}} (January 2016).

\bibitem{Lib-OPUS-labs-2013-arxiv-Prestwich-branch-and-bound-odd}
S.~D. Prestwich, {Improved Branch-and-Bound for Low Autocorrelation Binary
  Sequences}, \href{http://arxiv.org/abs/1305.6187}{http://arxiv.org/abs/1305.6187}.

\bibitem{Lib-OPUS-labs-1992-Optimization-deGroot}
C.~D. Groot, D.~W\"{u}rtz, K.~H. Hoffmann, Low autocorrelation binary
  sequences: exact enumeration and optimization by evolutionary strategies,
  Optimization 23 (1992) 369--384. \doi{10.1080/02331939208843771}.

\bibitem{Lib-OPUS-labs-1993-Diploma-Reinholz-genetic-alg}
A.~Reinholz, {Ein paralleler genetischer Algorithmus zur Optimierung der
  binaren Autokorrelations-Funktion}, Master's thesis, Diplom Thesis,
  Universitaet Bonn (October 1993).

\bibitem{Lib-OPUS-labs-1996-PRL-Dittes}
F.-M. Dittes, Optimization on rugged landscape: A new general purpose monte
  carlo approach, Physical Review Letters 76 (1996) 4651--4655.
  \doi{10.1103/PhysRevLett.76.4651}.

\bibitem{Lib-OPUS-labs-1998-IEEE_EC-Militzer}
B.~Militzer, M.~Zamparelli, D.~Beule, Evolutionary search for low
  autocorrelated binary sequences, IEEE Transactions on Evolutionary
  Computation 2~(1) (1998) 34--39. \doi{10.1109/4235.728212}.

\bibitem{Lib-OPUS2-labs-2003-FEA-Brglez-short}
F.~Brglez, X.~Y. Li, M.~F. Stallmann, B.~Militzer, {Reliable Cost Predictions
  for Finding Optimal Solutions to LABS Problem: Evolutionary and Alternative
  Algorithms}, in: {Fifth Int. Workshop on Frontiers in Evolutionary Algorithms
  (FEA2003)}, 2003,
  \url{http://militzer.berkeley.edu/papers/2003-FEA-Brglez-posted.pdf}.

\bibitem{Lib-OPUS2-labs-2004-InfoSciences-Brglez-InPress}
F.~Brglez, X.~Y. Li, M.~F. Stallmann, B.~Militzer, {Evolutionary and
  Alternative Algorithms: Reliable Cost Predictions for Finding Optimal
  Solutions to the LABS Problem}, 
  \url{http://militzer.berkeley.edu/papers/2003-InfoSciences-accp-FB.pdf}.

\bibitem{Lib-OPUS-labs-2007-Prestwich-local-search}
S.~D. Prestwich, {Exploiting relaxation in local search for LABS}, Annals of Operations Research 156~(1)
  (2007) 129--141.\doi{10.1007/s10479-007-0226-9}.

\bibitem{Lib-OPUS-labs-2008-CP-Halim-tabu}
S.~Halim, R.~H. Yap, F.~Halim,
  Engineering stochastic local search for the low autocorrelation binary sequence problem, in:
  Proceedings of the 14th international conference on Principles and Practice
  of Constraint Programming, CP '08, Springer-Verlag, Berlin, Heidelberg, 2008,
  pp. 640--645. \doi{10.1007/978-3-540-85958-1\_57}

\bibitem{Lib-OPUS-labs-2008-LMSLNS-Borwein}
P.~Borwein, R.~Ferguson, J.~Knauer, The merit factor problem, in: London
  Mathematical Society Lecture Note Series, Vol. 352, 2008, pp. 52--70.
  \url{http://www.cecm.sfu.ca/personal/pborwein/PAPERS/P218.pdf}.

\bibitem{Lib-OPUS-labs-2009-ASC-Gallardo-memetic}
J.~E. Gallardo, C.~Cotta, A.~J. Fern\'{a}ndez,
  Finding low autocorrelation binary sequences with memetic algorithms, Appl. Soft Comput.
  9~(4) (2009) 1252--1262. \doi{10.1016/j.asoc.2009.03.005}.

\bibitem{Lib-OPUS-labs-1983-IEEE_TIT-Golay-merit-legendre}
M.~J. Golay, The merit factor of legendre sequences, IEEE Transactions on
  Information Theory IT-29~(6) (1983) 934--936. \doi{10.1109/TIT.1983.1056744}.

\bibitem{Lib-OPUS-labs-1991-IEEE_TIT-Jensen-cyclic-difference_sets}
J.~Jensen, H.~Jensen, T.~Hoholdt, The merit factor of binary sequences related
  to difference sets, Information Theory, IEEE Transactions on 37~(3) (1991)
  617--626. \doi{10.1109/18.79917}.

\bibitem{Lib-OPUS-labs-2004-IEEE_TIT-Borwein}
P.~Borwein, K.-K. Choi, J.~Jedwab, Binary sequences with merit factor greater
  than 6.34, Information Theory, IEEE Transactions on 50~(12) (2004)
  3234--3249. \doi{10.1109/TIT.2004.838341}.

\bibitem{Lib-OPUS-labs-2004-IEEE_TIT-Parker-Legendre}
R.~A. Kristiansen, M.~G. Parker, Binary sequences with merit factor $>$ 6.3,
  Information Theory, IEEE Transactions on 50~(12) (2004) 3385--3899.
  \doi{10.1109/TIT.2004.838343}.

\bibitem{Lib-OPUS-labs-2013-Elsevier-Jedwab-merit-factor}
J.~Jedwab, D.~J. Katz, K.-U. Schmidt,
  Advances in the Merit Factor Problem for Binary Sequences, J. Comb. Theory Ser. A 120~(4) (2013)
  882--906. \doi{10.1016/j.jcta.2013.01.010}.
  
\bibitem{Lib-OPUS-labs-2013-AdvMath-Jedwab-Littlewood_Polynomials}
J.~Jedwab, D.~J. Katz, K.-U. Schmidt,
  Littlewood polynomials with small L$^4$ norm, Advances in Mathematics 241 (2013) 127 -- 136.
  \doi{10.1016/j.aim.2013.03.015}.

\bibitem{Lib-OPUS-labs-2011-IEEE_TIT-Baden-Jacobi}
J.~Baden, Efficient optimization of the merit factor of long binary sequences,
  Information Theory, IEEE Transactions on 57~(12) (2011) 8084--8094.
  \doi{10.1109/TIT.2011.2164778}.

\bibitem{Lib-OPUS2-dice-2011-EV-Brglez}
F.~Brglez, {Of n-dimensional Dice, Combinatorial Optimization, and Reproducible
  Research: An Introduction}, {Eletrotehni\v{s}ki Vestnik 78(4) 2011: pp.
  181--192, English Edition, \url{http://ev.fe.uni-lj.si/4-2011/Brglez.pdf} ; An
  invited talk at 2011-ERK, Portoroz, Slovenia, Sept. 2011~} 78~(4) (2011)
  181--192.

\bibitem{Lib-OPUS2-walk-2013-MIDEM-Brglez}
F.~Brglez, {Self-Avoiding Walks across n-Dimensional Dice and Combinatorial
  Optimization: An Introduction}, {Informacije MIDEM, 44 (1) (2014), pp. 53-68,
  English Edition \url{http://www.midem-drustvo.si/journal/} ; also at
  \url{http://arxiv.org/abs/1309.7508} ; An invited talk at 2013-MIDEM, Sept. 2013,
  Kranjska Gora, Slovenia~} 44~(1) (2014) 53--68.

\newpage

\bibitem{Lib-OPUS-walk-self-avoiding-2014-Wikipedia}
Self-avoiding walk (2014) [cited 2014].
\newline\url{http://en.wikipedia.org/wiki/Self-avoiding\_walk}

\bibitem{Lib-OPUS-walk-2011-EMS-Slade-self-avoiding}
G.~Slade, {The self-avoiding walk: A brief survey}, in: J.~Blath, P.~Imkeller,
  S.~R{\oe}lly (Eds.), Surveys in Stochastic Processes, European Mathematical
  Society, 2011, pp. 181--199. \newline \url{https://www.math.ubc.ca/~slade/spa_proceedings.pdf}.

\bibitem{Lib-OPUS-Book-2013-Birkhauser-Madras-The-Self-Avoiding-Walk}
N.~Madras, G.~Slade, {The Self-Avoiding Walk}, {Modern Birkh\"{a}user Classics}, 2013. \doi{10.1007/978-1-4614-6025-1}.

\bibitem{Lib-OPUS-walk-2010-StatPhysics-Clisby-pivot-alg-self-avoiding}
N.~Clisby, {Efficient Implementation of the Pivot Algorithm for Self-avoiding
  Walks}, Journal of Statistical Physics 140 (2010) 349--392.
  \doi{10.1007/s10955-010-9994-8}.

\bibitem{Lib-OPUS2-ebook-xBed-2016-Brglez}
F.~Brglez, J.~Brest, B.~Bo\v{s}kovi\'{c}, {xBed: An Open Environment for Design
  and Experiments with Combinatorial Solvers}, A preprint available from the
  authors.

\bibitem{Lib-OPUS-statistics-1995-MathStat-Sen-censoring}
{Pranab Kumar Sen}, {Censoring in Theory and Practice: Statistical Perspectives
  and Controversies}, Institute of Mathematical Statistics, Analysis of
  Censored Data 27 (1995) 177--192. \doi{10.1214/lnms/1215452220}.
  
\bibitem{2014-Web-SLING}
{SLING - Slovenian Initiative for National Grid}, \newline
See \url{http://www.sling.si/sling/} for an overview (2014).
  
\bibitem{1980-Macmillan-Olkin}
I.~Olkin, L.~J. Gleser, C.~Derman, {Probability Models and Applications},
  {Macmillan Publishing Co., Inc.}, 1980.
  
\bibitem{WEB-Project-R}
{The R Project for Statistical Computing}, \newline
\url{http://www.r-project.org/} (Nov 2015).

\bibitem{2013-Web-NCSU-HPC}
{A dedicated 'sam cluster' of 22 processors.}, NCSU High Performance Computing
  Services. See also \url{http://www.ncsu.edu/itd/hpc/} for an overview (2013).

  \vfill
\begin{footnotesize}
\begin{tabular}[b!]{p{17.8cm}}
 ~ \\
\hline
 \multicolumn{1}{|p{17.8cm}|}{
 \footnotesize This version, created on \thedate, is an electronic reprint of the original article published in the Applied Soft Computing Journal, 2017, Vol. 56, pp. 262--285,
 \href{http://dx.doi.org/10.1016/j.asoc.2017.02.024}{doi:10.1016/j.asoc.2017.02.024}. This reprint differs from the original in pagination and typographic detail.} \\
 \hline
\end{tabular}
\end{footnotesize}

\newpage
  
\bibitem{Lib-OPUS2-labs-2014-homepage-Knauer}
{LABS Problem: 2002 Merit Factor Records posted by Knauer}, {Now re-posted
  under \url{https://github.com/borkob/git_labs} } (2016).

  
\bibitem{Lib-OPUS-labs-2014-NatNano-frustrated-energy-landscape}
J.~Trastoy, M.~Malnou, C.~U.~R. Bernard, N.~Bergeal, G.~Faini, J.~Lesueur,
  J.~Briatico, J.~E. Villegas,
  Freezing and thawing of artificial ice by thermal switching of geometric frustration in magnetic flux
  lattices, {Nat Nano} 9~(9) (2014) 710--715. \doi{10.1038/nnano.2014.158}

\bibitem{Lib-OPUS2-ebook-markov-2016-Brglez}
{F. Brglez}, {On Stochastic Combinatorial Optimization and Markov Chains: from
  Walks with Self-Loops to Self-Avoiding Walks}, A preprint available from the
  author.

\bibitem{Lib-OPUS2-ebook-ogr-2016-Brglez}
F.~Brglez, B.~Bo\v{s}kovi\'{c}, J.~Brest, {On Asymptotic Complexity of the
  Optimum Golomb Ruler Problem: From Established Stochastic Methods to
  Self-Avoiding Walks}, A preprint available from the authors.

\bibitem{Lib-OPUS-ogr-2007-Springer-Cotta-Golomb_Ruler-Memetic}
C.~Cotta, .~Dotu, A.~J. Fernandez, P.~V. Hentenryck, {A Memetic Approach to
  Golomb Rulers}, in: {T. P. Runarsson et al} (Ed.), {Parallel Problem Solving
  from Nature}, {LNCS}, {Springer}, 2007, pp. {252--261}. \doi{10.1007/11844297_26}.

\bibitem{WEB-Project-Golomb_Rulers}
{Golomb Rulers Project}, \newline \url{http://www.distributed.net/OGR} (2016).

\bibitem{WEB-Wikipedia-Golomb_Rulers}
{Golomb Ruler Wikipedia}, \newline \url{http://en.wikipedia.org/wiki/Golomb\_ruler}
  (2016).

\bibitem{Lib-OPUS-book-2006-Tufte-Beautiful_Evidence}
E.~R. Tufte, {Beautiful Evidence}, {Graphics Press LLC}, 2006.

\end{thebibliography}
\end{document}